\input{psfig}
\documentclass[]{aa}
\usepackage{graphicx}
\usepackage{deluxetable}
\usepackage{aalongtable}
\begin{document}
\title{RACE-OC Project:\\ Rotation and variability in young stellar associations within 100 pc \thanks{The online Tables \ref{twa_lit}-\ref{abdor_lit} and on-line  Figs.\,\ref{twa_fig1}-\ref{abdor_fig6} are available in electronic form at the CDS via anonymous ftp to cdsarc.u-strasbg.fr (130.79.128.5)
or via http://cdsweb.u-strasbg.fr/cgi-bin/qcat?J/A+A/} \thanks{Based on the All Sky Automated Survey photometric data}}
\author{S.\,Messina\inst{1}
\and          S.\,Desidera\inst{2}
\and          M.\,Turatto\inst{1}
\and          A.\,C.\,Lanzafame\inst{1,3}
\and          E.\,F.\,Guinan\inst{4}
}
\offprints{Sergio Messina}
\institute{INAF-Catania Astrophysical Observatory, via S.Sofia, 78 I-95127 Catania, Italy \\
\email{sergio.messina@oact.inaf.it; massimo.turatto@oact.inaf.it}
\and   
INAF-Padova Astronomical Observatory, Vicolo dell'Osservatorio 5, I - 35122 Padova, Italy  \\
\email{silvano.desidera@oapd.inaf.it}
\and   
University of Catania, Dept. of Physics and Astronomy, via S.Sofia, 78 I-95127 Catania, Italy\\
\email{Alessandro.Lanzafame@oact.inaf.it}
\and   
Dept. of Astronomy and Astrophysics, Villanova University, Villanova, 19085, PA, USA\\
\email{edward.guinan@villanova.edu}
\\}

\date{}
\titlerunning{Rotation and variability in young associations}
\authorrunning{S.\,Messina et al.}
\abstract {Angular momentum and its interplay with magnetic fields represent a promising tool to probe the stellar internal structure and evolution of low-mass stars} 
{Our goal is to determine the rotational and magnetic-related activity properties of stars at different stages of  evolution. For this reason, we have focussed our attention primarily on members of clusters and young stellar associations of known ages. In this study, our targets are 6 young loose stellar associations within 100 pc and ages in the range 8-70 Myr: TW Hydrae ($\sim$8 Myr), $\beta$ Pictoris ($\sim$10 Myr), Tucana/Horologium, Columba, Carina ($\sim$30 Myr), and AB Doradus ($\sim$70 Myr). Additional data on $\alpha$ Persei and the Pleiades  from the literature is also considered.} {Rotational periods of stars showing
rotational modulation due to photospheric magnetic activity (i.e. starspots)
have been determined applying the Lomb-Scargle periodogram technique to
photometric time-series obtained by the All Sky Automated Survey (ASAS). The
magnetic activity level has been derived from the amplitude of the V
lightcurves. The statistical significance of the rotational evolution at
different ages has been inferred applying a two-sided Kolmogorov-Smirnov
test to subsequent age-bins. } {We detected the rotational modulation and measured the rotation periods of 93 stars for the first time, and confirmed the periods of 41 stars already known from the literature. For further 10 stars we revised the period determinations by other authors.  The sample was augmented with periods of 21 additional stars retrieved from the literature. 
In this way, for the first time we were able to determine  largest set of rotation periods  at ages of $\sim$8, $\sim$10 and $\sim$30 Myr, as well as increase by 150\% the number of known periodic members of AB Dor.} {The analysis of the rotation periods in young stellar associations, supplemented by Orion Nebula Cluster (ONC) and NGC\,2264 data from the literature, has allowed us to find that in the 0.6 - 1.2 M$_{\odot}$ range the most significant variations of the rotation period distribution are the spin-up between 9
and 30 Myr and the spin-down between 70 and 110 Myr.  Variations between 30 and 70 Myr are rather doubtful,
despite the median period indicates a significant spin-up. The photospheric activity level  is found to be correlated to rotation at ages greater than $\sim$70 Myr and to show some additional age dependence beside that related to rotation and mass.}
\maketitle
\rm

\section{Introduction}
Rotation is one of the basic stellar properties which undergoes dramatic changes along the whole stellar life.
Such changes depend both on the evolution of the internal structure - e.g., stellar radius contraction during pre main sequence (PMS) and its expansion during  post main sequence (Post MS) -
and on the presence and evolution of intense magnetic fields (Kawaler \cite{Kawaler88}; MacGregor \& Brenner \cite{Macgregor91}; 
Krishnamurthi et al. \cite{Krishnamurthi97}). Indeed, stellar magnetic fields play a fundamental role in the rotational history of late-type stars. 
During the PMS T-Tauri phase, they are responsible for the star-disk coupling  which maintains the star's rotation rate slow, in spite of the gravitational contraction (see, e.g., Scholz et al.  \cite{Scholz07}). 
During the MS and Post MS, they are responsible for the  angular momentum loss through magnetized stellar winds,
as well for  the redistribution of angular momentum through coupling processes between internal radiative zone 
and external convection zone (e.g., Barnes \cite{Barnes03}).
 Thus,  evolution of angular momentum and magnetic activity offer complementary diagnostics to study the
 mechanisms by which rotation and magnetic fields influence each other.\\ 
\indent 
Our knowledge on the rotation properties at different stellar ages is increasing thanks to a number of valuable projects either of decennial long-term monitoring of very young open clusters 
(see, e.g., Herbst \& Mundt \cite{Herbst05}; Herbst et al. \cite{Herbst07}) or of seasonal monitoring of intermediate-age open clusters (e.g., MONITOR, Hodgkin et al. \cite{Hodgkin06}; EXPLORE/OC Extrasolar Planet Occultation Research, von Braun et al. \cite{vonbraun05}). Differently than field stars, stars in open clusters form samples that are complete in mass and homogeneous in  
environmental conditions, initial chemical composition, age and interstellar reddening.
Such  stellar samples  allow us to accurately investigate the dependence on age and metallicity of  different stellar properties and of their mutual relationship.\\
\indent
However, much still needs to be done
since the number of studied open clusters, as well as the  number of periodic variables discovered in most clusters,  have not been large enough to fully constrain the various models proposed to describe the mechanisms that drive the angular
 momentum evolution. Specifically, the sequence of ages at which the angular momentum evolution has been studied  has still significant  gaps and
the sample of available periodic cluster members for a number of clusters is not as complete as necessary. Furthermore,  at most ages we have only one
representative cluster, a situation which does not allow us to investigate, e.g., the dependence on metallicity or on initial  environment conditions.\\
\indent 
RACE-OC, which stands for Rotation and  ACtivity Evolution in Open  Clusters, is a long-term
project aimed at studying the evolution  of the rotational properties and the magnetic activity of late-type members of stellar open
clusters (Messina \cite{Messina07}; Messina et al. \cite{Messina08}).
The RACE-OC targets are in stellar associations and open clusters with ages in the range from about 1 to about 600 Myr, for which no rotation and activity  investigations have been 
carried out so far. Top priority is given to the open clusters that  fill the gaps of the relationship among age, activity and rotation.
Nonetheless, we have also included clusters already extensively studied such as the very young Orion Nebula Cluster (Parihar et al. \cite{Parihar09}). 
The motivation behind this is  to  enrich further the sample of periodic rotational variables   and  to explore the long-term magnetic activity, e.g., to search for activity cycles and surface differential rotation (SDR), by making repeated observations of same clusters over several years.\\
\indent

\begin{table*}
\begin{tabular}{l l cccccccc}
\hline
Association & abbrev. & Age  & Distance & known   & late-type & periodic  & ASAS   &  literature & new period \\
            &        & (Myr)& (pc)     & members & members   & members   & period &  period     & (+revised) \\
\hline
TW Hydrae			& TWA	            & 8  & 48 & 36 & 29 & 23 & 12 & 11 & 4 (+0)\\
$\beta$ Pictoris 		& $\beta$ Pic 	& 10 & 31 & 51 & 32 & 27 & 25 & 2 & 10 (+1)\\
Tucana/Horologium 		&Tuc/Hor	& 30 & 48 & 54 & 35 & 29 & 27 & 2 & 19 (+5)\\
Columba 			       & Col 	& 30 & 82 & 35 & 23 & 19 & 19 & 0 & 15 (+2) \\
Carina 				& Car  	& 30 & 85 & 23 & 21 & 19 & 19 & 0 & 16 (+1) \\
AB Doradus 			& AB\,Dor		& 70$^a$ & 34 & 91 & 64 & 48 & 42 & 6 & 29 (+1) \\ 
\hline
\multicolumn{7}{l}{$^a$ Value taken from the literature but different from our estimate.}
\end{tabular}
\caption{Name, abbreviation, age, and mean distance of the nearby associations under study (Torres et al. 2008), together
with the number of known members; late-type (later than F) members selected for period search; total number of periodic members; periodic members discovered from ASAS photometry; periodic 
members with period adopted from the literature; new periods determined from this study (and periods revised by us  with respect to earlier literature values).
\label{tab-list}}
\end{table*}
In the present  study we considered stellar associations with  distances less  than 100 pc and
ages younger than about 100 Myr. In fact, while very few open clusters are within 100 pc, recent 
investigations successfully identified a number  of loose associations of nearby young stars (Zuckermann \& Song \cite{Zuckerman04}; Torres et al. \cite{Torres08}).
Like open clusters, the physical association among the members allows a much robust age determination than 
for isolated field stars. Furthermore, the brightness and the proximity to the Sun make it possible to carry out 
several complementary observations of individual objects that allow to put the rotational 
properties of these stars in a broader astrophysical context.
These observations include, e.g., high-resolution spectroscopy, trigonometric parallaxes, census of visual and
spectroscopic binaries, IR excess, searches for planets.\\
\indent
A knowledge of the rotational properties of very young stars, as a function of age and spectral type (=mass), is important for a number of issues:
a) some high-precision radial velocity studies of these targets are  on-going (Setiawan et al. \cite{setiawan08};  G\"unther \& Esposito \cite{Gunther07b}),
in spite of the challenge represented by the activity-induced radial velocity jitter.
As this is due to the occurrence of active regions on stellar surface, an independent determination
of rotational period is useful to disentangle radial velocity variations due to rotational modulation 
from those due to Keplerian motion (e.g., Lanza \cite{Lanza10a}); b) an accurate knowledge of the rotational properties of parent stars 
can illuminate how the star's angular momentum and planet formation influence each other.
The planet formation may alter significantly the rotational history of the parent stars and, conversely, 'anomalous' rotation may
reveal evidence of planet formation processes (Pont  2009; Lanza \cite{Lanza10b}); c)  the 
knowledge of rotation periods of young stars allows us to investigate the effect of rotation 
on the Lithium depletion (da Silva et al. 2009) and to establish a connection between rotation and Lithium on a basis firmer than using the projected rotational velocity
to estimate rotation; d)
finally, a comparison between the rotational properties of single stars and stars in binary systems can give some insight
on the effect of binarity in the early stages of the rotational evolution.\\
\indent
The nearby loose young stellar  associations we selected are: \object{TW Hydrae}, \object{$\beta$ Pictoris}, \object{Tucana/Horologium}, 
\object{Columba}, \object{Carina} and \object{AB Doradus}. All but the latter
have an age in the range between $\sim$8 and  $\sim$30 Myr, which is quite unexplored by
earlier rotational studies. In fact, to date no rotation period distribution was known in the age range from $\sim$ 
4 Myr (NGC\,2264; Lamm et al. \cite{Lamm04}) to $\sim$ 40 Myr  (IC\,4665; Scholz et al. \cite{Scholz09}).\\
\indent
This is an important age range in the rotational history of low-mass stars, when circumstellar disks dissipate and stars 
are free to spin their rotation up while they contract toward the zero age main sequence (ZAMS). This is also the age range of planet formation. Observations and theoretical studies of our planetary system (see Zuckerman \& Song \cite{Zuckerman04} and references therein) indicate that giant planets form in less than 10 Myr and Earth-like planets in less than 30 Myr. Thus, the study of these stars  allows to shed light on  formation and early evolution of planetary systems.
Indeed, nearby young stars are the prime targets for searches for planets with the direct imaging technique, as
planets are brighter at young ages (Burrows et al.~\cite{burrows97}). Several surveys already observed with the best state-of-art adaptive optics or space
instruments a number of members of young associations (e.g., Chauvin et al. \cite{Chauvin09}; Nielsen \& Close \cite{Nielsen09}). 
Efforts in this direction have recently lead to the first
planet discoveries  (e.g., Marois et al.~\cite{marois08}) and there are exciting perspectives for the use of future more sensitive instruments
that will be available within a few years (e.g., Beuzit et al. \cite{Beuzit08}).\\
\indent
In  Sect.\,2 we present the young loose associations considered in the present study. In Sect.\,3 we describe the  photometric data on which this study is based.
In Sect.\,4 we present the rotation period search. The results on the rotation period distributions and a discussion
in the context of angular momentum evolution are given in Sect.5 and 6. Sect.\,7 contains our conclusions.

\section{The sample}

The sample of our investigation is taken from the recent
compilations by  Zuckerman \&  Song (\cite{Zuckerman04}) and Torres et al. (\cite{Torres08}), that includes an updated
analysis of the membership of nearby associations younger than 100 Myr.

We selected the following associations that have a mean distance smaller
than 100 pc: {\object{TW Hydrae},  \object{$\beta$ Pictoris}, \object{Tucana/Horologium}, 
\object{Columba}, \object{Carina}\footnote{This 
association should not be confused with the
closer but older Carina-Near Moving Group identified by Zuckerman et al. (\cite{Zuckerman06})}, and
\object{AB Doradus}.
These associations are reported with ages in the range from $\sim$8 to $\sim$70 Myr.


The Torres et al. (2008) list of members is significantly more extended
than the previous ones, thanks to the availability
of the SACY (Search for Associations Containing Young stars) database of observation of young stars (Torres et al. \cite{Torres06}).
Only a very small number of objects have discrepant membership  with
respect to the previous investigations (Zuckermann \& Song 2004). 


Our initial target list included nearly 300 stars.
As the SACY sample was originally selected from bright ROSAT sources,
a large majority of the targets is represented by late-type stars and is
then  suitable for the photometric search of rotational modulations.
We excluded only a  fraction ($\sim$30\%) of stars with spectral types earlier than F9, 
since they are not expected to show any rotationally-induced variability due to
their shallow convective zones. We  note that a few members, although with unknown spectral type,
 were included in the search sample since their B$-$V colors were consistent with a late spectral type.
We further excluded stars fainter than V$\sim$13, as photometric errors  
of ASAS data, on which our study is based, become too large to allow a meaningful analysis.
After applying these selection criteria, we are left with 204 stars. 
The list of our target associations, together with age, mean distance, and number of known late-type members is reported in Table\,1.

The stars in young stellar associations have been studied extensively in the
past years especially in the context of the study of the
formation of planetary systems.
As a result, a significant fraction of the targets has accurate
spectroscopic characterisation. The search for  planetary companions
using direct imaging or radial velocities has also lead to a quite 
complete census of the binarity and to investigations on the presence of
circumstellar gaseous or dusty disks.
We exploited such resources in order to better understand trends
and correlations between the rotational periods and other properties.

Most of the spectroscopic observations (spectral types, projected
rotational velocity $v\sin i$) are from SACY database
(Torres et al. 2006). Information on binarity of the targets was taken from Torres et al. (2006, 2008),
and  Bonavita et al. (2010, in preparation). Additional bibliography
for individual targets is given in Appendix A.


\begin{figure*}
\begin{minipage}{18cm}
\includegraphics[scale = 0.7, trim = 30 0 0 80, clip, angle=90]{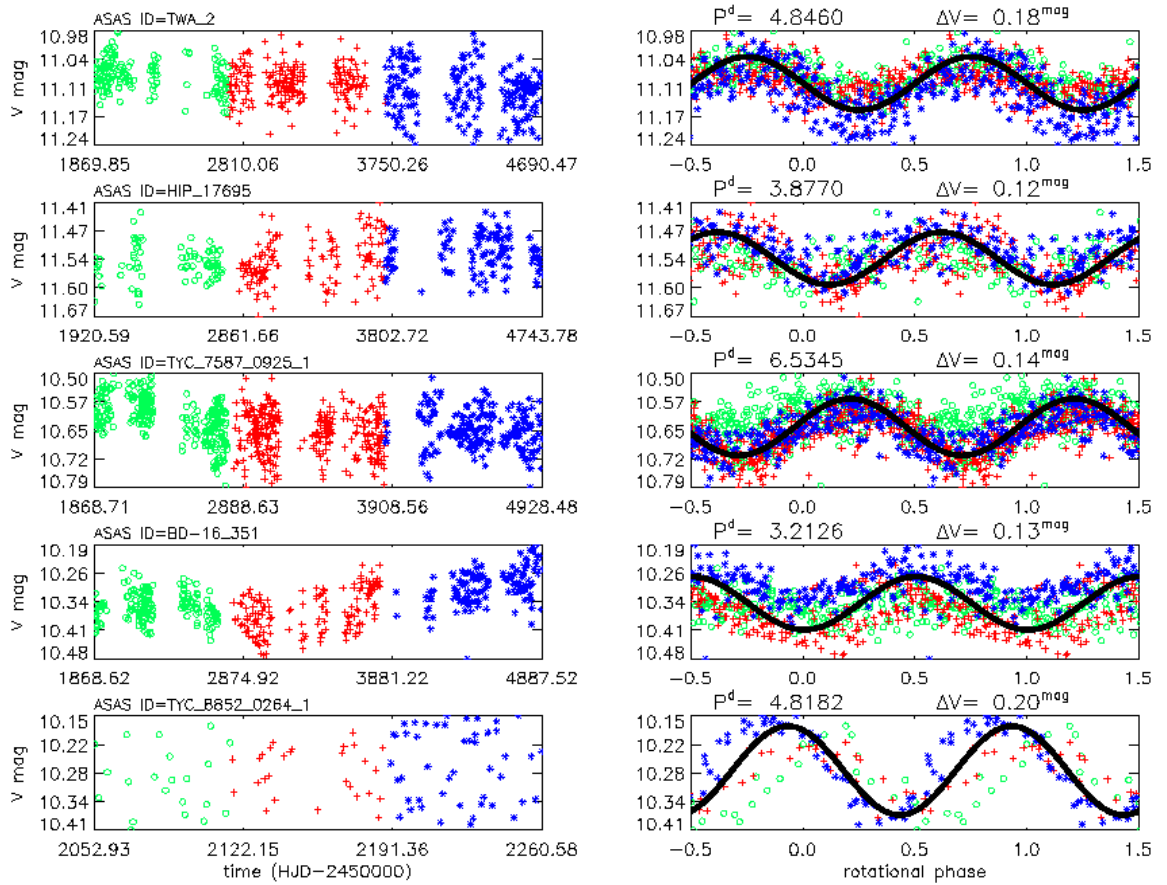}
\end{minipage}
\caption{\label{example} Representative example of light curves in our sample. V-band data time series in the left panels and  
phased light curve in the right panels together with the sinusoidal fit with the rotation period (thick solid line).
Different symbols and colors help identify  measurements collected in different time intervals.}
\end{figure*}

\section{Data}
\subsection{The ASAS photometry}
The All Sky Automated Survey (ASAS) is the major source of photometric observations on which the present analysis is based.
The  ASAS project started  in 1997 with the goal to photometrically monitor millions  of stars brighter
than 14 magnitude in  the V band and distributed all  over the sky at declinations $\delta < +28^{\circ}$, to investigate any kind of 
photometric variability (Pojmanski 1997; 2002).\\
\indent
Presently, ASAS is carried out by two observing stations. One is at Las Campanas Observatory, Chile (since 1997). It consists of two wide-field telescopes,  equipped with F200/2.8 Minolta telephoto lenses 
and  2K$\times$2K AP-10 CCD Apogee cameras, covering 8.8$\times$8.8 deg of the sky through the V and I filters.
The other station (the Northern Station) is  at Haleakala, Hawaii, Maui (since 2006), and  is equipped with two wide-field Nikon
F200/2.0  APO-G-10 telephoto lenses observing simultaneously in the standard
V and I filters. 
Data acquisition and processing is fully automated. The data reduction pipeline used to process ASAS 
data is described in detail by Pojmanski (1997). The linear scale at focal plane is 16 arcsec/pixel. The FWHM of stellar images is 1.3-1.8 pixels. Aperture photometry is used to extract stellar magnitudes
through 5 apertures (ranging from 1 to 3 pixels in radius, which corresponds to 16 to 48 arcsec).
Smaller apertures give better accuracy for brighter stars, whereas larger apertures for the fainter stars.
In the following analysis, we selected the magnitudes time series of each target by selecting the aperture 
giving the best photometric precision.
Due to the low spatial resolution, a check of the presence of any star close to the target star is crucial, 
especially for fainter stars for which  larger apertures (up to a 48 arcsec radius) are used to extract the stellar magnitudes.
All cases when nearby stars are not spatially resolved are discussed in Appendix A.
The astrometric calibration is currently based on the ACT catalog (Astrographic Catalog 2000 + Tycho, Urban et al. 1998) 
and achieves an accuracy around 3 arcsec.
Calibration to the standard system is based on the Tycho photometry (Perryman et al. 1997) and is accurate at about 0.05-mag level.
Default exposure time is 2 minutes in the I and 3 minutes in the V filter. The systems routinely secure from 160 to 200 frames per night in V and from 230 to 300 frames per night in I. At this rate, the telescopes can carry out photometry of the available sky in two filters in about 2 days.

\subsection{Data from the literature}
A few stars of our sample are not in the ASAS database, being located at declination $\delta > $ +28$^{\circ}$. 
We checked the bibliographical sources of all targets using 
 ADS (Astrophysical Data System) to see whether previous determinations of rotation period existed. 
A number of stars had the rotation period already determined within the ASAS survey and, therefore, they were found 
listed in the ASAS Catalogue of Variable Stars (ACVS). Nonetheless,  we made our period search also for these 
stars and, in a number of cases (10), we came to a different determination of the rotation period. Such cases are individually discussed.\\
\indent
In Tables\,\ref{twa_lit}-\ref{abdor_lit}\footnote{Available in the online material} we list  the following information  taken from the literature and used to discuss the results of our period search: target name; coordinates; V magnitude; B$-$V  and V$-$I colors; M$_V$  absolute  magnitude; distance; projected equatorial velocity; computed stellar mass and radius; spectral type; and notes on  membership.

\section{Photometry rotation period search}

We   have  used  the \bf  Lomb-Scargle periodogram \rm method to search for significant periodicities related to the stellar rotation
in the data time series. In  the following  sub-sections we briefly  describe our
procedure to determine the rotation period of our targets.

\begin{figure*}
\begin{minipage}{18cm}
\includegraphics[scale = 0.8, trim = 0 0 50 0, clip, angle=0]{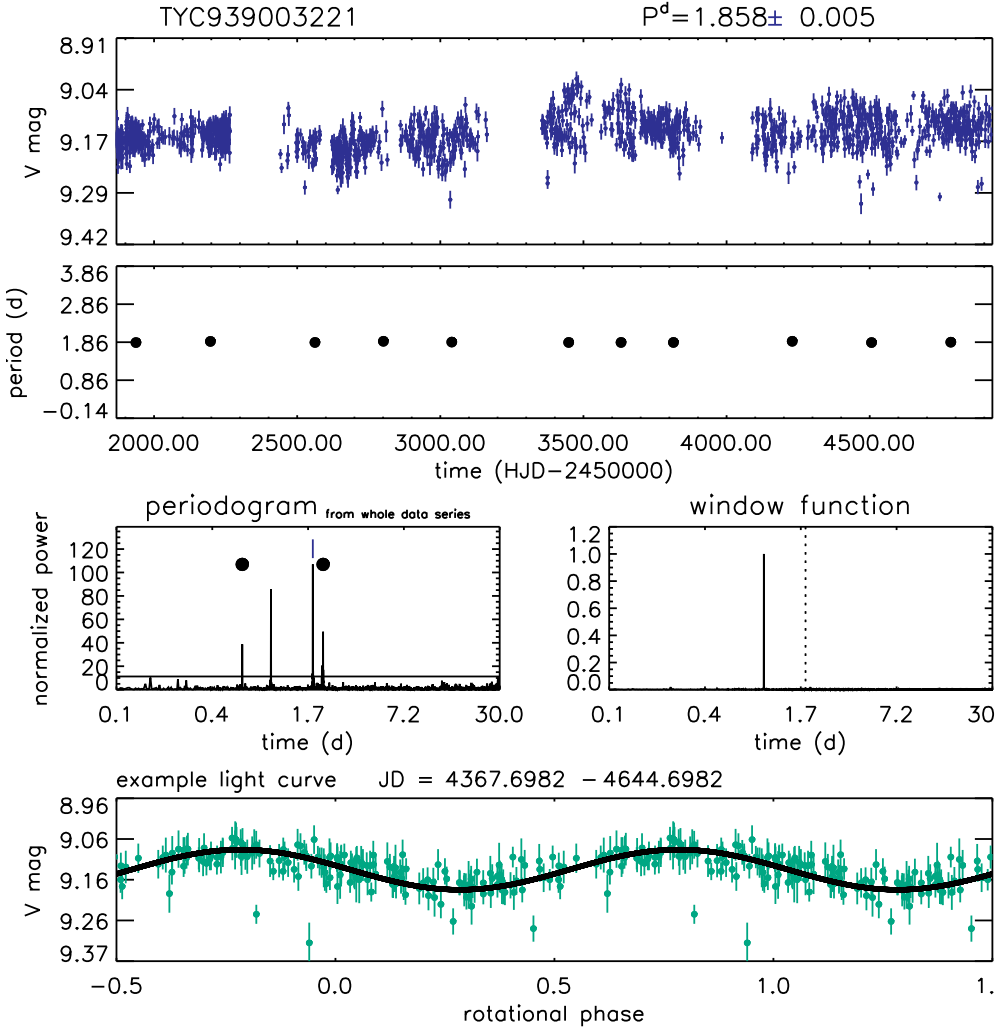}
\end{minipage}
\caption{\label{window} \it Top panel\rm: stellar V-magnitudes vs. time of TYC\,9390\,0322\,1. 
\it Second panel from top: \rm  rotation periods vs. time detected with a confidence level over 99\%. \it Third  panels (left): \rm
 periodogram with evidence of 4 peaks with confidence level larger than 99\%. Large bullets represents the peaks related to beat periods.   \it Third  panels (right): \rm
window function with evidence of a peak at about 1 day owing to the data sampling. The vertical dotted line indicates that
the P=1.8581d period is not affected by the window function peak.
 \it Bottom panel\rm: example light curve with data collected from MJD4367 to MJD4644 and phased with the P=1.8581d rotation period. The solid line is a sinusoidal fit.}
\end{figure*}

\subsection{Time series sectioning}
Since our analysis is focussed on solar and late-type stars, we expect to detect the stellar rotation period by analysing
the flux modulation induced by surface inhomogeneities unevenly distributed along the stellar
longitude. Such surface inhomogeneities can be either cool or hot spots arising from magnetic activity,
which is particularly efficient in stars with fast rotation (P$<$10-20 days) and deep outer convection zone
(spectral types from G to M). Indeed, the observed  variability is dominated by  phenomena 
that are manifested on different  time scales (see, e.g.,  Messina
et al.   2004).  The shortest time  scale, of the order  of seconds to
minutes, is related to  micro-flaring activity.  Its stochastic nature
increases the  level of  intrinsic noise in  the observed  flux time series.  
The  variability on time scales  from several hours  to days is
mostly  related to the  star's rotation. The  variability on
longer time scales, from months to years, is related to the growth and
decay of active regions (ARGD) as well as to the presence of starspot
cycles. These may be similar to the $\sim$11 yr sunspot cycle.\\
\indent
 Long-term monitoring of field stars (see, e.g., Messina \& Guinan \cite{Messina03}) shows that, 
because of  ARGD and surface differential rotation,  lightcurves'  amplitude and shape  change with typical timescales 
 of about 2-3 months, or even less for the fastest rotators (P$\sim$1 day). Such changes, if not taken into account,
can introduce aliases and  lead to incorrect results.
Therefore, a reasonable approach to the period search is to divide the complete 
data time series of each target (which is typically of about 8 yr in our case) into consecutive intervals not exceeding 
2 months and to carry out the period search in each interval separately. 
Following such an approach, we obtained  on average 10-15  intervals per target suitable for the period search.\\
\indent
Notwithstanding the 2-3 months timescale of lightcurve variation, Fourier analysis of long timespan series with sufficiently dense measurements can lead 
to a  period determination  with much higher confidence level and precision than the analysis of sectioned timeseries 
(e.g., Parihar et al. \cite{Parihar09}). 
Here we  anticipate that, without sectioning the data, we successfully detected the significant rotation periods in about
85\% of our periodic targets.\\
\indent
In Fig.\,\ref{example} we plot some representative example light curves together with the sinusoidal fit with the rotation period.
There are stars such as  TWA\,2 and HIP\,17695 whose amplitudes and phase of minima remain constant in time. Other stars, such as 
TYC\,7587\,0925\,1 and BD\,$-$16\,351, show constant phases of minima but variable amplitudes. 
In these  cases  a Fourier analysis of the complete time series without sectioning resulted in  very precise rotation period determinations.
On the other hand, stars like TYC\,8852\,0264\,1 have lightcurve phases of minima that change in less than two months. In such cases an
 accurate period determination requires
timeseries sectioning.
\rm
\indent

\subsection{Lomb-Scargle periodogram}

The  \bf Lomb-Scargle \rm technique (Press et al. \cite{Press92}; Scargle  \cite{Scargle82};
Horne \& Baliunas \cite{Horne86}) was  developed  to search  for
significant  periodicities  in unevenly  sampled  data. The algorithm calculates the normalized power
P$_{\rm N}$($\omega$) for a given  angular frequency $\omega = 2\pi\nu$. The
highest   peaks  in  the   calculated  power   spectrum  (periodogram)
correspond to  the candidate periodicities  in the analyzed  time series
data.  In order  to determine the significance level  of any candidate
periodic signal, the height of the corresponding power peak is related
with a false  alarm probability (FAP), that is  the probability that a
peak of given height is due to simply statistical variations, i.e.  to
Gaussian noise.  This method assumes  that each observed data point is
independent from the  others.  However, this is not  strictly true for
our time series data consisting of data that are generally 
collected  with a time sampling  much shorter than  both the
periodic  or the  irregular  intrinsic variability  timescales we  are
looking  for (P$^d$ = 0.1-30).  This correlation can have a significant impact  on the
period  determination  as it has  been   highlighted  by,  e.g.,  Herbst  \&
Wittenmyer (1996),  Stassun et  al. (1999), 
Rebull (2001),  Lamm  et al.   (\cite{Lamm04}). We decided  to determine the  FAP
in slightly different way than Scargle (\cite{Scargle82}) and Horne  \& Baliunas
(\cite{Horne86}), as discussed in the next sub-section, 
 to  overcome this  problem.

\subsection{False alarm probability}
Following the approach outlined by Herbst et al. (2002),  Monte Carlo simulations are used to determine the 
relationship between the normalized power and the FAP. Specifically, 
after dividing the data time series of each target into a number of  intervals, the data of each interval
\rm were randomized  by scrambling the day numbers of  the 
Julian Day (JD) while keeping photometric
magnitudes  and the  decimal part  of  the JD  unchanged. This  method
preserves  the same time sampling as in the  original  data set 
within the same night.
Then, we  applied a periodogram  analysis to
about 1000 "randomized'' data time series for each  time interval \rm and retained  the highest  power peak 
of each computed periodogram.
The FAP related to a given power  P$_{\rm N}$ is taken as the fraction of
randomised  light curves that  have the  highest power  peak exceeding
P$_{\rm N}$  which, in turn, is the probability that a peak of this height
is simply  due to statistical  variations, i.e. white noise.
As the rotation period, we selected  that 
corresponding to the highest power peak detected in the periodogram and with confidence level larger 
than 99\% (FAP$<0.01$), as computed from the mentioned simulations.  The same procedure was
repeated for each time interval and for all targets. \rm


\subsection{Alias detection}

To identify the true periodicities in the periodogram, it is
crucial to take into account that a few peaks, even with large power and high confidence levels, can be
 aliases arising from both the data sampling and the length of the time interval  during which the observations are collected.
In this respect, an inspection of the spectral window function  
helps to identify which peaks  in the periodogram may be alias.

In Fig.\,\ref{window} we plot, as an example, the ASAS photometric data time series of one of our targets (TYC\,9390\,0322\,1).
V-band magnitudes together with their uncertainties are plotted vs. the Heliocentric Julian Day (HJD) on the top panel.
The periodogram in the middle left panel shows the presence of 4 peaks with a large power exceeding the 99\% confidence level 
(solid horizontal line), but only one is related to the stellar rotation. 
If we look at the window function in the middle right panel, 
we find a major peak at about 1d which arises from the observation timing of about 1 day imposed by the rotation of the Earth 
and the fixed longitude of the observation site.
This inspection allows us to identify the 1-d peak in the periodogram (marked by a vertical dotted line) as an alias.
This peak is generally present in the periodogram of all the targets, being the observation timing similar for all the ASAS targets. 
The highest peak at P=1.858days is actually that one related to the stellar rotation period, whereas
the remaining two peaks (marked by bullets) arise from the convolution between the power spectrum
and the window function. These alias periods are beat periods (B) between the star's rotation period (P) and the data 
sampling and they obey to the relation
\begin{equation}
\frac{1}{B} = \frac{1}{P} \pm n \,\,\,\,(n=1,2,3,...)
\end{equation}
A method  to check whether secondary peaks are beat periods is to perform a prewhitening of the data time series
by fitting and removing a sinusoid with the star's rotation period  from the data.
After removing the primary frequency from the data time series and recomputing the periodogram, all the other peaks disappear, confirming
that they are  beat frequencies.

\subsection{Uncertainty in the rotation periods}

We followed the  method used by Lamm et al.  (\cite{Lamm04}) to compute the errors associated with the period determinations.
The uncertainty in the period can be written as
\begin{equation}
\Delta P = \frac{\delta \nu P^2}{2}
\end{equation}
where  $\delta\nu$ is  the finite  frequency resolution  of  the power
spectrum and is equal  to the full width at half maximum of  the main peak of
the window  function w($\nu$).   If the time  sampling is not  too non-uniform,
 which is the case  related to our observations, then $\delta\nu
\simeq 1/T$,   where T is the total  time span of  the observations. From
Eq.\,(2)  it  is clear that  the uncertainty in  the determined  period not
only depends  on the frequency resolution  (total time  span) but it is also
proportional to  the square of the  period. We also  computed the error
of  the period determinations  following the prescription suggested by Horne \&
Baliunas  (1986)  which  is  based on the formulation  given  by  Kovacs
(1981). The period uncertainty computed according
to  Eq.\,(2) was found  to be  a factor 5-10  larger than  the uncertainty
computed by the Horne  \& Baliunas
(\cite{Horne86})  technique. In this paper
we conservatively report the errors computed using Eq.\, (2) and, therefore, the precision in the 
periods could be better than that quoted in this paper.

\subsection{Data precision}
The precision of the ASAS photometry of the target stars under analysis is in the range
0.02-0.03 mag, as shown in Fig.\,\ref{accuracy}. We remind, as discussed in Sect.\,3.1, that we are using for each star the best aperture to extract the magnitude time series, which changes from star to star according to its magnitude.
Therefore,  we are comparing precision as determined from different apertures in Fig.\,\ref{accuracy}. This is the reason
for which we do not observe the typical trend of decreasing accuracy toward fainter stars in the magnitude range under analysis.
Each star's precision is computed by averaging  the uncertainty associated to the data points
of the complete time series. We plot also the amplitudes of the light curves of our targets. We see that all
stars for which we could determine the rotation period have a light curve amplitude at least a factor 2.5  larger than
the corresponding photometric accuracy. This circumstance permits 
the period search to detect high power peaks in the periodogram with large confidence level and, consequently,
to determine reliable rotation periods.
We found that the best photometric accuracy for our targets  was achieved by extracting the magnitude
with apertures from 15 to 30 arcsec. Using the ADS and SIMBAD databases, we have checked whether our target stars 
had nearby stars within the aperture radius  whose flux contribution may affect our analysis.
Such cases are mentioned in Appendix A, dedicated to the discussion of individual cases.

\begin{figure}
\begin{minipage}{10cm}
\includegraphics[scale = 0.5, trim = 0 0 0 0, clip, angle=0]{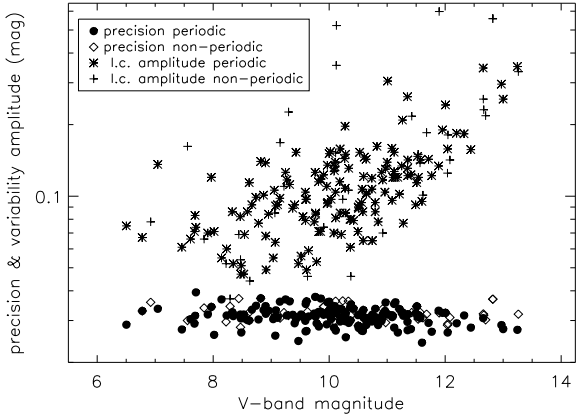}
\end{minipage}
\caption{\label{accuracy} Photometric precision and peak-to-peak variability amplitude  vs. V-band magnitude 
of target stars.}
\end{figure}

\section{Results}
We determined the rotation period of 144 out of the selected 204 stars. We determined the
rotation periods for the first time  for 93 of them. We confirmed the period determined by
other authors for 41 stars and revised the periods for 10 stars. Rotation
periods of further 21 stars were retrieved from the literature. We found non
periodic variability in 33 stars. The remaining 6 stars in the sample have
neither ASAS data nor periods reported in the literature. A summary for each
association is reported in Table\,\ref{tab-list}.\\
\indent
A comparison of our results with previous rotation period determinations was
possible for 51 stars (Fig.\,\ref{per_comp}). We confirm the results reported
in the literature in 41 cases.
Our periods differ in 10 cases from the periods reported either in the ACVS (in 9 cases)
or in the literature (only the case of HIP\,9892; Koen \& Eyer \cite{Koen02}).
A close inspection of our periodograms showed that in all 9 cases of disagreement with respect to ACVS
no power peak  at all exists at the period value reported in the ACVS
(see online Figs \,\ref{twa_fig1}-\ref{abdor_fig6}). Moreover, when we compute the rotation phases
using the ACVS period, in all, but the case of HIP\,12545, 
we obtain unconvincing light curves, that is with a high phase dispersion and
without any evident modulation.
On the contrary, our rotation periods were detected with a confidence level
greater than 99\% both in at least 8-10 time
intervals in which we divided the complete time-series (i.e. in over 60\% of
the time intervals) and in the periodogram computed without data sectioning.
The same holds for HIP\,9892 whose periodogram does not show any peak at the 
period reported by Koen \& Eyer (\cite{Koen02}).
We note that in 6 cases (TYC\,8852\,0264\,1, TYC\,8497\,0995\,1, TYC\,7026\,0325\,1, 
TYC\,7584\,1630\,1, TYC\,8160\,0958\,1, and HIP\,9892) the  period is twice our
value, which may be caused by the presence of two major spot groups located
at opposite stellar hemispheres. However, the light curves in this circumstance
should be double-peaked when they are phased with the long period,
and this is not observed.
In other 2 cases (TYC\,7617\,0549\,1 and TYC\,5907\,1244\,1), the ACVS period is
consistent with a beat period, according to Eq.(1). 
Finally, it is not possible to reconcile the
discrepant results with neither beat periods nor with spot groups at
opposite hemispheres  in just 2 cases (HIP\,12545 and HIP\,76768).\\
\indent
Note that our rotation period determinations include 10 stars in the TWA and
5 in the Tuc/Hor associations (flagged with an $a$ apex  in Tables\,\ref{twhya} and \ref{tucanae}) 
that were eliminated by Torres et al. (2008) from the high-probability member list, and two other stars, HIP\,84586 and V4046 Sgr, 
that are tidally locked binaries. These stars are expected to have a different rotational history than single stars, because of enhanced rotation rates 
from tidal synchronization..
All these 17 stars will not be considered in the following analysis on the rotation period distribution.\\
\begin{figure}
\begin{minipage}{10cm}
\includegraphics[scale = 0.5, trim = 0 0 0 0, clip, angle=0]{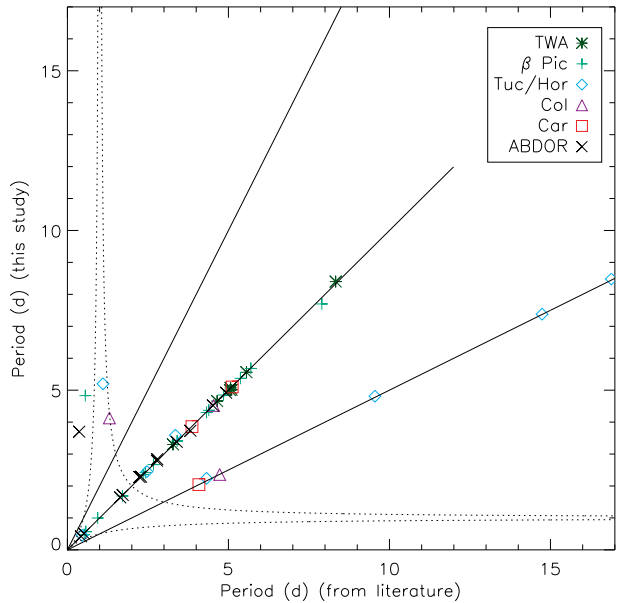}
\end{minipage}
\caption{\label{per_comp} Our rotation periods as a function of the periods derived from either ACVS or the literature. Straight solid lines represent
the loci where our periods are equal to, or they are a factor 0.5 and 2 larger than the literature values. 
The curved lines represent the loci of beat periods, according to Eq.\,(1).}
\end{figure}
The results of our period search are summarised in Tables\,\ref{twhya}-\ref{abdor}. To prevent
overestimating the maximum V-band light-curve amplitude ($\Delta V$$_{\rm max}$) we took
the difference between the median values of the upper and lower 15\% data points of the
timeseries section with the largest amplitude (see, e.g., Herbst et al. \cite{Herbst02}).
In the following analysis we will not use the brightest observed magnitude V$_{min}$, rather the 
V magnitudes (corrected for duplicity in the case of binary systems) taken from Torres et al. (\cite{Torres08}), and reported in the online 
Tables\,\ref{twa_lit}-\ref{abdor_lit}.
  The rotation periods, together with uncertainty and normalized power, determined in the individual timeseries sections, are listed in the online Table\,\ref{tab_period}. \\  \rm
\indent
The light curves of all stars for which the ASAS photometry allowed us to determine the rotation period are plotted in the on-line Figs.\,\ref{twa_fig1}-\ref{abdor_fig6}.\\
\indent
In Appendix A we report some detail on the nature (binarity and spectral classification) of individual targets and on the rotation 
periods when these were found in disagreement with previous determinations.
\begin{figure}
\begin{minipage}{10cm}
\centerline{
\psfig{file=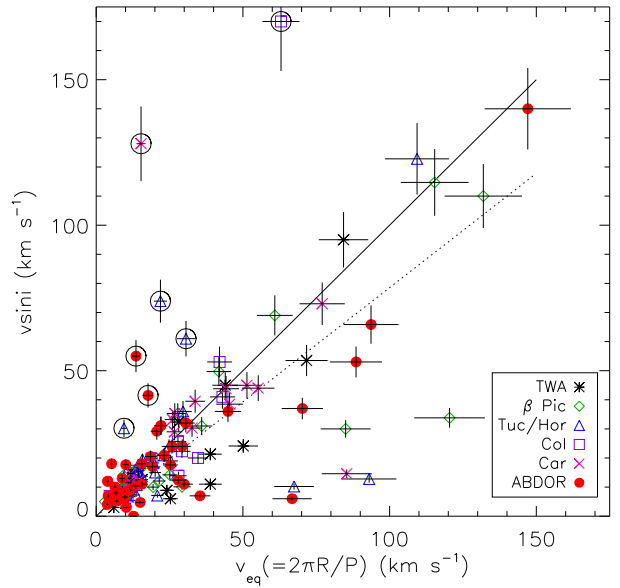,width=9cm,height=9cm,angle=0}
}
\end{minipage}
\caption{$v\sin i$ from the literature vs. equatorial velocity  v$_{\rm eq}$ = 2$\pi$R/P. The solid line marks $v\sin i$ = v$_{\rm eq}$, whereas the dotted line  $v\sin i$ = ($\pi$/4)v$_{\rm eq}$.\label{vsini} }
\end{figure}

\subsection{Vsin$i$ vs. equatorial velocity}

About 75\% of the periodic variables in our sample have known projected
equatorial velocities ($v\sin i$). We also derived stellar radii by comparing
the position of each target in the colour-magnitude diagram with the Baraffe
et al. (\cite{Baraffe98}) evolutionary tracks (see Sect. 6.1). When rotation period,
$v\sin i$, and stellar radius are known, it is possible to compare equatorial
velocity v$_{eq}$=2$\pi$ R /P and $v\sin i$ to check the consistency between the two
and derive the stellar inclination.\\
\indent
In Fig.\,\ref{vsini} we compare $v\sin i$ and v$_{eq}$, marking the loci of $v\sin i$=v$_{eq}$,
corresponding to equator-on orientation,  and  $v\sin i$=$\pi$/4 v$_{eq}$,
corresponding to a randomly orientated rotation axis distribution. The major
uncertainty of the equatorial velocity derives from the radius estimate. The
reported $v\sin i$ uncertainties are 10\% on average. Only 7 stars 
(flagged  with an apex  $c$ in Tables\,\ref{twhya}-\ref{abdor} and plotted with circled symbols in Fig.\,\ref{vsini})
have inconsistent $v\sin i$/v$_{eq}$ (i.e., much larger than unity).
Four out of seven stars (TYC\,9344\,0293\,1, TYC\,9529\,0340\,1, TYC\,7100\,2112\,1, and TYC\,8586\,2431\,1)
have only one measurement of $v\sin i$, whereas they have well established rotation periods.
It is important to carry out additional spectroscopic measurements to check the correctness 
of the $v\sin i$ value of these stars.
Three other stars (TYC\,8852\,0264\,1, TYC\,7059\,1111\,1, and TYC\,7598\,1488\,1) have
each 2-3 independent $v\sin i$ measurements, which give similar values within the errors.
These also have well established rotation periods which are confirmed by literature values.
The discrepancy in the case of  TYC\,7598\,1488\,1  may arise from an incorrect value of parallax.
In fact, Cutispoto (\cite{Cutispoto98b}) reports a photometric parallax larger than the one reported by Torres et al.  (\cite{Torres08})
which produces a larger stellar radius and would partly solve the disagreement of  v$_{eq}$ with $v\sin i$. 
Whereas in the other two cases the discrepancy requires further investigation to be addressed.\\
\indent
We have checked that the results of the following analysis on the rotation period distribution do not
change if the seven stars with inconsistent $v\sin i$/ v$_{eq}$ are either considered or not. One of these stars is anyway
excluded from the following analysis because it is a rejected member of $\beta$ Pictoris.
The average $v\sin i$ for each association are reported in Table\,\ref{tab-vsini}. This is obviously based
on the periodic variable sample excluding members with inconsistent $v\sin i$/ v$_{eq}$.
 In Table\,\ref{tab-vsini}, $r$ represents the correlation coefficient from the linear Pearson statistics, whereas the labels a and b give the significance level of the correlation coefficient. The significance level represents the probability of observing a value of the correlation coefficient larger than $r$ for a random sample having the same number of observations and degrees of freedom.\\
Taken at face values, mean inclinations are not consistent with the value expected for completely randomly orientated rotational axis distribution. However, an investigation on preferential orientations of rotational axis in young associations must take into account several observational biases and is outside the scope of this paper.\\
\rm

\begin{table}
\caption{Summary of results of the comparison between $v\sin i$ and equatorial velocity.\label{tab-vsini}}
\begin{tabular}{ccccc}
\hline
Association     &  \#		&	$<$sin$i$$>$$\pm$$\sigma$	&	r	&	significance	\\
			& stars	&                                             &              &   level                  \\
\hline
TWA			&	13	&	0.72$\pm$0.35		&	0.85		& 	a	\\
$\beta$ Pic	&	27	&	0.91$\pm$0.34		&	0.84		&	a	\\
Tuc/Hor		&	23	&	1.02$\pm$0.56		&	0.83		&	a	\\
Col			&	12	&	0.94$\pm$0.31		&	0.86		&	a	\\
Car			&	15	&	0.95$\pm$0.25		&	0.64		&	b	\\
AB\,DOR		&	44	&	1.02$\pm$0.71		&	0.77		& 	a	\\
all			&    134   &      0.95$\pm$0.50        & 	0.78		&	a	\\
\hline
\multicolumn{5}{l}{a: confidence level $>$ 99.999\%; b: confidence level = 99.995\%.}\\
\end{tabular}
\end{table}

\rm




\begin{table*}
\scriptsize
\caption{\label{twhya} {\bf TW Hydrae Association}. Summary of period searches: Star's name; ASAS number; rotation period and its uncertainty; fraction of timeseries sections in which the rotation period is detected with confidence level higher than 99\%; maximum light curve amplitude ($\Delta$V$_{\rm max}$ );  photometric accuracy ($\sigma_{\rm acc}$); brightest  V magnitude (V$_{min}$); B$-$V color; spectral type and note on binarity;  note on period, and variable star designation if it exists. Targets are listed according to increasing TWA number (and increasing ASAS number for the other associations).}
\centering
\begin{tabular}{llll@{\hspace{.1cm}}c@{\hspace{.1cm}}c@{\hspace{.1cm}}c@{\hspace{.1cm}}l@{\hspace{.1cm}}llll@{\hspace{.1cm}}c}

\hline
  name   &  ID$_{\rm ASAS}$   &P  &  $\Delta$P   & timeseries & $\Delta$V$_{\rm max}$ & $\sigma_{\rm acc}$ & V$_{min}$ & B$-$V & Sp.T. &  mult. & note & variable's  \\   
            &                                 & (d)  &  (d)      &   sections       & (mag)     & (mag)              & (mag) & (mag) &  & & & name   \\   
\hline
TW\,Hya 	&        110152-3442.3 &   2.80 &   0.04 & ...    &  0.25 &   0.031 &   11.00 &  0.97 & K6Ve & S 	&  P$_{\rm lit}$ &  \\
TWA\,2  	&        110914-3001.7 &   4.86 &   0.02 & 13/16  &  0.18 &   0.031 &   11.10 &  1.48 & M2Ve & V   	&  new &  \\
TWA\,3A 	& 	   111028-3731.9 &   ...  &  ...   & ...    & ...   &  ...    &   12.10 &  1.47 & M4Ve & V 	&  ...&  \\
TWA\,3B 	&   	   no-ASAS       &   ..   &  ...   & ...    & ...   &	 ...    &	13.07 &  1.52 & M4Ve & V 	&  ...&  \\
TWA\,4A  	&        112205-2446.7 &   2.52 &   0.20 & ...    &  0.09 &  ...    &    8.94 &  1.17 & K5V  & Q 	&  P$_{\rm lit}$ &  TV Crt\\
TWA\,4B  	&        112205-2446.7 &  14.8  &   0.3  &  9/14  &  0.09 &   0.032 &    8.92 &  1.17 & K5V  & Q 	&  new &  \\
TWA\,5 	&        113155-3436.5 &  0.776 &  0.001 &  2/15  &  0.22 &   0.031 &	11.56 &  1.47 & M2Ve & T 	&  P=P$_{\rm ACVS}$&   \\
TWA\,6$^a$ 	&        101829-3150.0 &  0.54  &   ...  &  ...   &  0.49 &  ...    &   10.88 &  1.31 & M0Ve & S	&  P$_{\rm lit}$&  \\
TWA\,7  	&        104230-3340.3 &  5.00$^b$  &  0.03  &  1/15  &  0.11 &   0.030 &   11.73 &  1.46 & M2Ve & S	&   P=P$_{\rm lit}$&  \\
TWA\,8A  	&        113241-2651.9 &  4.66  &   0.06 &  4/14  &  0.27 &   0.029 &   12.14 &  1.46 & M3Ve & V 	&  P=P$_{\rm lit}$ &  \\
TWA\,8B  	&        no-ASAS       &  0.78  & ...    &  ...   &  0.08 &   ...   &   15.20 &  1.10 & M5   & V	&  P$_{\rm lit}$ &  \\
TWA\,9A 	&        114824-3728.8 &  5.01  &   0.01 &  13/14 &  0.18 &   0.027 &   11.22 &  1.26 & K5V  & V 	&   P$\ne$P$_{\rm ACVS}$; P=P$_{\rm lit}$&  \\
TWA\,9B 	&        no-ASAS       &  3.980 &  ...   &  ...   &  0.08 &  ...    &   14.00 &  1.43 & M1V  & V	&  P$_{\rm lit}$ &  \\
TWA\,10 	&        123504-4136.6 &  8.4   &   0.1  &  2/14  &  0.33 &   0.027 &   13.04 &  1.43 & M2Ve & S 	&  P=P$_{\rm lit}$&   \\
TWA\,12$^a$ &        112105-3845.3 &  3.30  &  0.01  &  7/14  &  0.45 &   0.030 &   12.70 &  1.53 & M1Ve & S 	&   P=P$_{\rm ACVS}$=P$_{\rm lit}$&   \\	
TWA\,13A 	&        112117-3446.8        &  5.56  &  0.03  &  8/15 &  0.20 &   0.031 &   11.09 &  1.42 & M1Ve & V 	&  =1/2 P$_{ACVS}$;P=P$_{\rm lit}$&   \\
TWA\,13B 	&        112117-3446.8       &  5.35  &  0.03   & 4/12.   &  0.27 & 0.031.    &   11.96 &  1.47 & M5   & V 	&  P=P$_{\rm lit}$ &  \\
TWA\,14$^a$ &  	   111326-4523.7 &  0.63  &  ...   &  ...   &  0.11 &  ...    &   12.52 & 1.30$^1$ & M0  & S	& P$_{\rm lit}$&  \\
TWA\,15A$^a$&        123421-4815.2 &  0.65  &  ...   &  ...   &  0.13 &  ...    &   ...   & 1.45$^1$ & M2 & V 	& P$_{\rm lit}$ &  \\
TWA\,15B$^a$&        123420-4815.3 &  0.72  &  ...   &  ...   &  0.05 &  ...    &   ...   & 1.45$^1$ & M2 & V 	& P$_{\rm lit}$&  \\
TWA\,16$^a$ &        123456-4538.1 &   ...  &  ...   &  ...   &  ...  &  ...    &   12.69 &       &  M1.5  & B 	& ...&  \\ 
TWA\,17$^a$ &        132045-4611.6 &  0.69  &  ...   &  ...   &  0.12 &  ...    &   ...   & 1.20$^1$  & K5  & S  	& P$_{\rm lit}$&   \\
TWA\,18$^a$ &        132137-4421.9 &  1.11  &  ...   &  ...   &  0.08 &  0.030  &   13.12 & 1.30$^1$ & M0.5 & S	&     P$_{\rm lit}$ &  \\
TWA\,19AB$^a$	& 	   114721-4953.1 & ...    &  ...   &  ...	  &  ...  & ...     & 	...   & ...   & K7      & V 	& ...&  \\
TWA\,20 	&        123138-4559.0 & ...    &  ...   &  ...   &  ...  & ...     &   13.42 &  1.10 & M2 	& S	&  ...&  \\
TWA\,21$^a$	&        101315-5230.9 &  4.43  &   0.02 &  7/15  &  0.13 &   0.037 &    9.82 &  1.00 & K3Ve    & V 	&            new &  \\
TWA\,23$^e$ &        120727-3247.0 &   ...  & ...    &  ...   &  ...  &  ...    &   12.81 & 1.35$^1$ &  M1  & S 	&     ...     &   \\
TWA\,24$^a$ &	   120942-5854.8 &  0.680$^b$ & 0.004  &  6/12  & 0.08  &   0.030 &   10.28 & 0.94  &    K1Ve & S	& new&   \\
TWA\,25 	&        121531-3948.7 &  5.07  &   0.02 &  12/14 & 0.23  &   0.032 &   11.37 & 1.41  & M1Ve    & S 	&      P=P$_{\rm ACVS}$&   \\
\hline
\multicolumn{13}{l}{V=visual companion; S=single star; B=binary system; T= triple system; Q=quadruple system; P$_{\rm lit}$: period as given in the literature; }\\
\multicolumn{13}{l}{P$_{\rm ACVS}$: period in ACVS; $^a$ rejected member and excluded from rotation period distribution; $^b$ period undetected in the periodogram }\\
 \multicolumn{13}{l}{of the complete time series; $^e$ not included in the member list because of incomplete kinematic information; $^1$ inferred from spectral type.}\\
\end{tabular}

\end{table*}


\begin{table*}
\scriptsize
\caption{\label{beta_pic} As in Table\,\ref{twhya} for the {\bf $\beta$ Pictoris association.}}
\centering
\begin{tabular}{llll@{\hspace{.1cm}}c@{\hspace{.1cm}}c@{\hspace{.1cm}}c@{\hspace{.1cm}}l@{\hspace{.1cm}}llll@{\hspace{.1cm}}c}

\hline
  name   &  ID$_{\rm ASAS}$   &P  &  $\Delta$P   &timeseries  & $\Delta$V$_{\rm max}$ & $\sigma_{\rm acc}$ & V$_{min}$ & B$-$V & Sp.T. &  mult. & note & variable's  \\   
         &                    & (d)  &  (d)      &  sections        & (mag)     & (mag)              & (mag) & (mag) &  & & & name \\   
\hline
TYC\,1186\,706\,1		&        002335+2014.5	 &   7.7$^b$   &   0.3   &  1/6   & 0.09 &   0.028 &   10.80 &  1.4     &   K7.5V &	S  &	P=P$_{\rm lit}$ &  \\
HIP\,12545 			&        024126+0559.3   &   1.25  &   0.01  &  4/7   & 0.15 &   0.031 &   10.32 &  1.21    &   K6Ve & SB1 &  P=2P$_{\rm ACVS}$ &  \\
GJ\,3305 			&        043737-0229.5   &   6.10  &   ...   &  ...   & 0.03$^1$   & ...  &   10.59 &  1.45    &   M0.5 & T &    P$_{\rm lit}$ &  \\
HIP\,23200 			&        045935+0147.0   &   4.37  &   0.03  &  9/14  & 0.15 &   0.032 &   10.18 &  1.39 &    M0.5Ve & S & P=P$_{\rm ACVS}$=P$_{\rm lit}$ & V1005 Ori \\
HIP\,23309 			&        050047$-$5715.4 &   8.60  &   0.07  &  12/15 & 0.12 &   0.031 &   10.02 &  1.40 &       M0Ve & S &      new & \\
HIP\,23418 			&        050159+0959.0   &   6.42  &   0.04  &  1/10  & 0.12 &   0.034 &   11.57 &  1.54 &    M3V  & SB2+V &  new& \\
BD-21 1074 			&        050650-2135.1   &  13.3$^b$   &   0.2   &  4/14  & 0.07 &   0.034 &   9.94  &  1.52 &     M2 & T &       new&  \\
HIP\,29964 			&        061828$-$7202.7 &   2.67  &   0.01  &  12/15 & 0.12 &   0.032 &   9.78  &  1.13 &   K4Ve & S &    P=P$_{\rm ACVS}$ & AO~Men  \\
TWA\,22 	& 	   101727-5354.4 & ...    & ...    &  ...   &  ...  &  ...    &  ...    &  ...  &  M5     & S	& ...&  \\
HIP\,76629 			&        153857$-$5742.5 &   4.3   &   0.2   &  7/15  & 0.12 &   0.034 &   7.97  &  0.81 &   K0V & SB+V & P=P$_{\rm lit}$ & V343 Nor\\
HIP\,84586$^e$ 		&        171726$-$6657.1 &   1.688 &   0.003 &  8/13  & 0.07 &   0.034 &    6.77  &  0.76 & G5IV  & SB2+V &P=P$_{\rm lit}$ & V824 Ara  \\
TYC\,8728\,2262\,1  	&        172955$-$5415.8 &   1.819 &   0.007 &  8/16  & 0.11 &   0.036 &   9.61  &  0.85 &   K1V & S  &               new&  \\
TYC\,8742\,2065\,1  	&        174834$-$5306.7 &   2.61  &   0.02  &  8/16  & 0.09 &   0.033 &   8.95  &  0.83 &   K0IV & SB2+V   &    new & \\
HIP\,88399  		&	   180303-5138.9   &   ...   &   ...   &  ...   & ...  &  ...    &  12.50  & ...  &     M2+F6V & B & ...& \\
V4046\,Sgr$^e$		&        181411$-$3247.5 &   2.42  &   0.18  &  10/14 & 0.09 &   0.033 &   10.57 &  0.90 &    K5 &  SB2+V   &   P=P$_{\rm lit}$ & V4046 Sgr \\
UCAC2\,18035440 		&        181422-3246.2   &   ...   &   ...   &  ...   &  ... &   ...   &   12.78 & 1.36  &  M1Ve & SB? & ...& \\
HIP\,89829 			&        181952$-$2916.5 &   0.57 &   0.003 &  8/14 & 0.19 &   0.037 &  8.80 &  0.69 &    G1V & S   &       P=P$_{\rm ACVS}$ &  \\
TYC\,9077\,2489\,1      &        184537$-$6451.8 &   0.345  &   0.04  &   7/12 & 0.16 &   0.033 &   9.3 &  1.19 & K5Ve & SB2+V &	new & \\
TYC\,9073\,0762\,1  	&        184653$-$6210.6 &   5.37  &   0.04  &  10/15 & 0.33 &   0.032 &   12.08 &  1.46 &   M1Ve & S &         P=P$_{\rm ACVS}$  & \\
TYC\,7408\,0054\,1  	&        185044$-$3147.8 &   1.089 &   0.002 &  10/15 & 0.19 &   0.033 &   11.26 &  1.35 &     K8Ve & S &          new & \\
HIP\,92680	        	&        185306$-$5010.8 &   0.997 &   0.001 &  10/13 & 0.09 &   0.032 &    8.37 &  0.77 &   K8Ve & S &   P=P$_{\rm lit}$ & PZ Tel  \\
TYC\,6872\,1011\,1  	&        185804$-$2953.1 &   0.504 &   0.004 &   3/15 & 0.12 &   0.032 &   11.64 &  1.30 &     M0Ve & S &      new  & \\
TYC\,6878\,0195\,1  	&        191145$-$2604.2 &   5.65  &   0.05  &   7/12 & 0.09 &   0.032 &   10.33 &  1.1 &    K4Ve & V &        new &  \\
HIP\,102141AB		&        204151$-$3226.1 &   ...   &   ...   &   ...  & ...  &   ...   &   10.40 &  1.54 &   M4+M4.5 & B & &  AT Mic \\
HIP\,102409 		&        204510$-$3120.4 &   4.85  &   0.02  &   6/11 & 0.10 &   0.030 &   8.68  &  1.49 &  M1Ve & S &  P=P$_{\rm lit}$ & AU~Mic \\
TYC\,6349\,0200\,1  	&        205603$-$1710.9 &   3.41  &   0.05  &   7/10 & 0.11 &   0.028 &   10.55 &  1.22 &   K6Ve+M2 & V & P=P$_{\rm ACVS}$ & AZ~Cap \\
TYC\,2211\,1309\,1        &	   220042+2715.2   &   0.476 &   0.001 &    1/1 & 0.08 &   0.034 &   11.39 &  1.40 &  M0.0V  & S & P=P$_{\rm lit}$ & \\
TYC\,9340\,0437\,1  	&        224249$-$7142.3 &   4.48  &   0.03  &   6/12 & 0.16 &   0.030 &   10.65 &  1.35 &      K7Ve+K5V:e & B  &  new &  \\
HIP\,112312 		&        224458$-$3315.1 &   2.355 &   0.005 &   2/11 & 0.19 &   0.026 &   12.12 &  1.48 & M4IVe & V &  P=P$_{\rm ACVS}$ & WW~PsA \\
TYC\,5832\,0666\,1  	&        233231$-$1215.9 &   5.68  &   0.05  &   5/12 & 0.16 &   0.028 &   10.69 &  1.43 &   M0Ve & S &     P=P$_{\rm ACVS}$ & \\
HIP 11437     		&         no-ASAS        & 13.6928 &   ...   &   ...  & 0.03 & ...     &   10.62 &   1.21 & K8 & V & P$_{\rm lit}$ &  AG~Tri\\
HIP\,10679 			&         no-ASAS        &   ...   &   ...   &   ...  & ...  & ...     &    7.75 &   0.62 & G2V+F5V & B & ...& \\

\hline
\multicolumn{13}{l}{V=visual companion; S=single star; B=binary system; T= triple system; P$_{\rm lit}$: period as given in the literature;P$_{\rm ACVS}$ period in ACVS; }\\
\multicolumn{13}{l}{$^1$ derived from B-band light curve; $^b$ period undetected in the periodogram of the complete timeseries;}\\
\multicolumn{13}{l}{$e$ tidally-locked binary system, excluded from rotation period distribution.}\\

\end{tabular}

\end{table*}


\begin{table*}
\scriptsize
\caption{\label{tucanae}  As in Table\,\ref{twhya} for the {\bf Tucana/Horologium Association}.}
\centering
\begin{tabular}{llll@{\hspace{.1cm}}c@{\hspace{.1cm}}c@{\hspace{.1cm}}c@{\hspace{.1cm}}l@{\hspace{.1cm}}llll@{\hspace{.1cm}}c}

\hline
  name   &  ID$_{\rm ASAS}$   &P  &  $\Delta$P   & timeseries & $\Delta$V$_{\rm max}$  & $\sigma_{\rm acc}$ & V$_{min}$ & B$-$V & Sp.T. &  mult. & note & variable's  \\   
         &                    & (d)  &  (d)      &  sections       & (mag)     & (mag)              & (mag) & (mag) &  & &  & name \\   
\hline
HIP\,490 			& 	 000553-4145.2 &    ...   &  ...    & ...      &  ... &  ...    &  7.53  &  0.59 & G0V  & S & ...& \\
HIP\,1113 			&    	 001353-7441.3 &   3.72   &   0.01  &   13/17  & 0.09 &   0.034 &  8.79  &  0.74 & G8V  & S  &  new & \\
HIP\,1910 			&      002409-6211.1 &   1.750  &   0.003 &   11/20  & 0.13 &   0.031 &  11.42 &  1.40 & M0Ve & V &    new& \\
HIP\,1993 			&      002515-6130.8 &   4.35   &   0.02  &   11/19  & 0.15 &   0.031 &  11.43 &  1.35 & M0Ve & S     &                         new &  CT Tuc\\
HIP\,2729		&      003451-6155.0 &   0.37  &   0.002 &    ...   & 0.10 &   0.035 &   9.64 &  1.05 & K4Ve & S   &            P$_{\rm lit}$ & \\
TYC\,9351\,1110\,1	&      004220-7747.7 &   2.57   &   0.01  &   8/14   & 0.12 &   0.033 &  10.24 &  1.06 & K3Ve & S     &                       new & \\
HIP\,3556$^a$ 		& 	 004528-5137.5 &    ...   &   ...   &   ...    & ...  &   ...   &  11.91  & 1.48 & M3   & ...  & ...\\
TYC\,8852\,0264\,1$^a$ &  011315-6411.6 &   4.81$^c$   &   0.02  &   8/11   & 0.20 &   0.029 &  10.37 &  0.87 & K1V  & S     &    P=1/2P$_{\rm ACVS}$ & \\
HIP\,6485 			&      012321-5728.9 &   3.59   &   0.01  &   5/13   & 0.08 &   0.030 &   8.57 &  0.68 & G7V  & S     &                  =P$_{\rm lit}$ & \\
HIP\,6856 		      & 	 012809-5238.3 &    ...   &   ...   &   ...    & ...  &   ...   &   9.07 &  0.91 & K1V  & S  & & CC Phe\\
HIP\,9141 		      &      015749-2154.1 &   3.02   &   0.01  &   3/12   & 0.08 &   0.026 &   8.05 &  0.65 & G4V  & V &                       new & DK Cet\\
HIP\,9892 			&      020718-5311.9 &   2.24$^b$   &   0.03  &   7/13   & 0.06 &   0.030 &   8.63 &  0.65 & G7V  & SB1 &                  P =1/2P$_{\rm lit}$& \\
TYC\,8489\,1155\,1 	& 	 020732-5940.3 &   ...    &   ...   &   ...    & ...  &   ...   &  10.68 &  1.16 & K5Ve & V & ...& \\
RBS\,332$^a$  		&      023652-5203.1 &  1.538    &   0.001  &   6/15   & 0.24 &   0.032 &  12.05 &  1.48 & M2Ve & ...  &              new & \\
TYC\,8497\,0995\,1 	&      024233-5739.6 &   7.38   &   0.05  &   10/16  & 0.14 &   0.031 &  11.07 &  1.23 & K5Ve & S    &      P=1/2P$_{\rm ACVS}$ & \\
TYC\,8491\,0656\,1 	& 	 024147-5259.9 &  1.275$^b$   &   0.005 &    5/13  & 0.09 &   0.030 &  10.22 &  1.26 & K6Ve & V & new & \\
AF\,Hor 			& 	 024147-5259.5 &   ...    &   ...   &   ...    & ...  &    ...  &  12.21 &  1.49 & M2V  & V & ... & AF\,Hor\\
TYC\,7026\,0325\,1 	&      031909-3507.0 &   8.48   &    0.10 &   10/17  & 0.15 &   0.028 &  11.20 &  1.30 & K7Ve & S  &      P$\ne$P$_{\rm ACVS}$ & \\
TYC\,8060\,1673\,1 	&      033049-4555.9 &   3.74   &   0.04  &   7/15   & 0.09 &   0.027 &   9.63 &  0.95 & K3V  & S  &       new&  \\
TYC\,7574\,0803\,1 	&      033156-4359.2 &   2.94   &    0.01 &   12/16  & 0.15 &   0.031 &  10.98 &  1.30 & K6Ve & S  &                         new& \\
HIP\,16853 			&      033653-4957.5 &   ...    &   ...   &   ...    & ...  &   ...   &   7.63 &  0.59 & G2V  & SB &  ...        &              \\	
HIP\,21632 			&      043844-2702.0 &   4.25   &   0.02  &   ...    & 0.08 &   0.031 &   8.51 &  0.61 & G3V  & S   &  P$_{\rm lit}$   &   \\
TYC\,8083\,0455\,1 	& 	 044801-5041.4 &   8.46   &   0.05  &   4/15   & 0.12 &   0.030 &  11.53 &  1.35 & K7Ve & S & new & \\
TYC\,5907\,1244\,1		&      045249-1955.0 &   5.20   &   0.04  &   12/13  & 0.11 &   0.031 &   9.96 &  0.87 & ...  & SB2 &    P$\ne$P$_{\rm ACVS}$ &  \\
TYC\,5908\,230\,1 		&      045932-1917.7 &   4.06   &   0.02  &   11/17  & 0.15 &   0.030 &  10.68 &  1.20 & ...  & ...    &                        new& \\
BD-09\,1108 		&      051537-0930.8 &   2.72   &   0.01  &   5/12   & 0.11 &   0.032 &   9.87 &  0.67 & G5   & S  &                      new&  \\
TYC\,7048\,1453\,1 	&      051829-3001.5 &   1.70   &   0.02  &   2/15   & 0.09 &   0.024 &  11.65 &  1.27 & K4Ve & S     &                    new & \\
TYC\,7600\,0516\,1$^a$  &      053705-3932.4 &   2.45   &   0.01  &   7/15   & 0.10 &   0.031 &   9.59 &  0.80 & K1V(e) & S    &   P=P$_{\rm lit}$&  AT Col\\
TYC\,7065\,0879\,1$^a$  &      054234-3415.7 &   3.89$^b$   &   0.02  &   7/15   & 0.09 &   0.027 &  10.64 &  0.82 & K0V  & V    & new & \\
HIP\,105388 		&      212050-5302.0 &   3.36   &   0.01  &   7/15   & 0.10 &   0.033 &   8.71 &  0.72 & G7V  & S  &                    new & \\
HIP\,105404$^{a}$  	&      212100-5228.7 & 0.435334$^{b,d}$ & 0.000002 &  3/15   & 0.10 &   0.033 &   8.97 &  0.88 & G9V  & EB &     P=P$_{\rm ACVS}$ & BS~Ind  \\
HIP\,107345 		&      214430-6058.6 &   4.50   &   0.02  &   5/13   & 0.10 &   0.030 &  11.64 &  1.41 & M0Ve & S  &           new & \\
TYC\,9344\,0293\,1	&      232611-7323.8 &   1.32$^c$   &   0.02  &   8/12   & 0.13 &   0.033 &  11.95 &  1.39 & ...  & SB &                        new& \\
TYC\,9529\,0340\,1	&      232749-8613.3 &   2.31$^c$   &   0.01  &   11/13  & 0.11 &   0.036 &   9.37 &  0.69 & ...  & ... &                     new& \\
HIP\,116748AB 		&      233939-6911.8 &   2.85   &   0.02  &   5/12   & 0.06 &   0.031 &   8.26 &  0.70 & G6V+K3Ve & B &                       new & DS Tuc\\
\hline
\multicolumn{13}{l}{V=visual companion; S=single star; B=binary system; P$_{\rm lit}$: period as given  in the literature; P$_{\rm ACVS}$ period in ACVS;}\\
\multicolumn{13}{l}{$^a$ rejected member and excluded from rotation period distribution; $^b$ period undetected in the periodogram of the complete timeseries;}\\
\multicolumn{13}{l}{$^c$ $v\sin i$ inconsistent with v$_{\rm eq}$=2$\pi$R/P; $^d$ tidally-locked binary system.}\\
\end{tabular}
\end{table*}


\begin{table*}
\scriptsize
\caption{\label{columba}  As in Table\,\ref{twhya} for the {\bf Columba Association}. }
\centering
\begin{tabular}{llll@{\hspace{.1cm}}c@{\hspace{.1cm}}c@{\hspace{.1cm}}c@{\hspace{.1cm}}l@{\hspace{.1cm}}llll@{\hspace{.1cm}}c}

\hline
  name   &  ID$_{\rm ASAS}$   &P  &  $\Delta$P   & timeseries  & $\Delta$V$_{\rm max}$  & $\sigma_{\rm acc}$ & V$_{min}$ & B$-$V & Sp.T. &  mult. & note on& variable's  \\   
         &                    & (d)  &  (d)      &   sections       & (mag)     & (mag)              & (mag) & (mag) &  &  & period      & name  \\   
\hline

TYC\,8047\,0232\,1 &        015215-5219.6 &   2.40 &   0.01  &   5/14   & 0.11 &   0.028 &  10.88 &  0.95 &     K2V(e) & BD   &   new & \\
BD-16\,351         &        020136-1610.0 &   3.21 &   0.01  &   9/13   & 0.13 &   0.029 &  10.40 & 1.10$^1$ &  K5     & S   &   new &\\
TYC\,7558\,0655\,1 &        023032-4342.4 &   8.80 &   0.23  &   4/14   & 0.08 &   0.025 &  10.33 &  1.07 &     K5V(e) & S   &   new  &\\
HIP\,16413 	  	 &        033121-3031.0 &   2.25 &   0.01  &   4/14   & 0.07 &   0.028 &   9.86 &  0.64 &     G7IV   & B   &   new &\\
TYC5882\,1169\,1   &        040217-1521.5 &   3.78 &   0.04  &   7/15   & 0.07 &   0.028 &  10.12 &  1.01 &     K3/4   & S   &   new  &\\
HIP\,19775 	  	 &        041423-3819.0 &  2.345 &  0.008  &   3/12   & 0.07 &   0.025 &   9.10 &  0.58 &     G3V    & S   &   new & \\
TYC\,6457\,2731\,1 &        042110-2432.3 &    ... &   ...   &   ...    & ...  &   ...   &   9.42 &  0.62 &     G2V    & S   & ...    \\
TYC\,7584\,1630\,1 &        042149-4317.5 &  2.360 &  0.008 &    8/15   & 0.10 &   0.029 &  10.23 &  0.69 &      G7V    & S   &  P=1/2P$_{\rm ACVS}$&\\
TYC\,7044\,0535\,1 &        043451-3547.3 &   ...  &  ...    &   ...    & ...  &   ...   &  10.91 &  0.84 &     K1Ve   & S   &  ... &\\
TYC\,8077\,0657\,1 &        045153-4647.2 &  2.834 &  0.001  &   13/20  & 0.11 &   0.031 &   9.81 &  0.69 &     G5V    & B   &  new &\\
TYC\,8080\,1206\,1 &        045305-4844.6 &   4.52 &   0.03  &   9/18   & 0.15 &   0.029 &  10.79 &  0.87 &     K2V(e) & S   &   P=P$_{\rm ACVS}$ &\\
TYC\,8086\,0954\,1 & 	    052855-4535.0 &   4.50$^b$ & 0.05    &   4/15   & 0.11  & 0.030  &  11.45  &  0.86 &    K1V(e) & S   &   new  &\\
HIP\,25709 	  	 &        052924-3430.9 &   5.96 &   0.05  &   2/15   & 0.04 &   0.030 &   8.50 &  0.65 &     G3V    & SB2 &   new & \\
TYC\,7597\,0833\,1 & 	    054516-3836.8 &   ...  &   ...   &   ...    & ...  &   ...   &  10.95  & 0.70 &     G9V    & S   & ... &\\
TYC\,6502\,1188\,1 &        055021-2915.3 &   1.37$^b$ &   0.04  &   3/15   & 0.09 &   0.026 &   11.24 &  0.66 &    K0V(e) & S   &   new &\\
TYC\,8520\,0032\,1 &    055101-5238.2 &   1.203$^b$ &   0.005  &   8/14   & 0.10 &   0.031 &   10.60 &  0.75 &    G9IV   & S   &   new  &\\
TYC\,7617\,0549\,1 &  	    062607-4102.9 &   4.13 &   0.02  &   10/15  & 0.12 &   0.030 &   10.08 &  0.80 &    K0V    & S   & P$\ne$P$_{\rm ACVS}$ &\\
TYC\,8107\,1591\,1 &    062806-4826.9 &   1.293$^b$ &   0.002  &   8/15   & 0.11 &   0.032 &   11.01 &  0.65 &    G9V    & S   &   new  &\\
TYC\,7100\,2112\,1 &        065247-3636.3 &   0.83$^{b,c}$ &   0.002  &   11/15   & 0.12 &   0.035 &   11.18 &  0.62 &    K2V(e) & S   &   new &\\
TYC\,8118\,0871\,1 &        065623-4646.9 &   4.38 &   0.05  &   12/17  & 0.14 &   0.028 &   10.01 &  0.78 &    K0V(e) & S   &   new &\\
TYC\,7629\,2824\,1 &        070152-3922.1 &   1.335 &   0.002  &   9/12   & 0.14 &   0.032 &   11.32 &  0.63 &    G9V(e) & S   &   new &\\
AG\,Lep 	 	      & 	    no-ASAS         &  1.895 &   ...   &   ...    & 0.05 &   ...   &    9.62 &  0.60 &    G5V    & S   & P$_{\rm lit}$ & AG\,Lep \\
TYC\,5346-132-1    & 	    no-ASAS       &    ... &   ...   &   ...    & ...  &   ...   &    9.81 & 0.74 &     G7     & ... & ...\\
\hline
\multicolumn{12}{l}{ S=single star; B=binary system; BD= brown dwarf companion; $^1$: inferred from spectral type; }\\
\multicolumn{12}{l}{$^b$ period undetected in the periodogram of the complete timeseries;}\\
\multicolumn{12}{l}{$^c$: $v\sin i$ inconsistent with v$_{\rm eq}$=2$\pi$R/P; P$_{\rm lit}$: period as given in the literature; P$_{\rm ACVS}$ period in ACVS.}\\
\end{tabular}

\end{table*}


\begin{table*}
\scriptsize
\caption{\label{carina}  As in Table\,\ref{twhya} for the {\bf Carina Association}. }
\centering
\begin{tabular}{llll@{\hspace{.1cm}}c@{\hspace{.1cm}}c@{\hspace{.1cm}}c@{\hspace{.1cm}}l@{\hspace{.1cm}}llll@{\hspace{.1cm}}c}

\hline
  name   &  ID$_{\rm ASAS}$   &P  &  $\Delta$P   & timeseries & $\Delta$V$_{\rm max}$  & $\sigma_{\rm acc}$ & V$_{min}$ & B$-$V & Sp.T. &  mult. & note on& variable's  \\   
         &                    & (d)  &  (d)      &  sections        & (mag)     & (mag)              & (mag) & (mag) &  &  & period      & name  \\   
\hline
TYC\,9390\,0322\,1 &        055329-8156.9 &  1.858 &  0.005 &   11/12   & 0.14 &   0.036 &   9.20 &  0.79 &   K0V   &V &                        new &  \\
HIP\,30034 		 &	    061913-5803.3 &   3.85 &   0.01 &   6/10    & 0.10 &   0.031 &   9.33 &  0.86 &   K1V(e) & BD &        new  &  AB Pic\\
HIP\,32235 		 &	    064346-7158.6 &   3.83 &   0.01 &   9/12    & 0.09 &   0.031 &   9.11 &  0.70 &   G6V   & S      &          new  & \\
HIP\,33737 		 &	    070030-7941.8 &   5.10 &   0.03 &   11/12   & 0.14 &   0.034 &  10.16 &  0.91 &   K2V   & S       &            P=P$_{\rm ACVS}$ &  \\
TYC\,8559\,1016\,1 &        072124-5720.6 &   4.61 &   0.03 &   7/12    & 0.13 &   0.030 &  10.80 &  0.64 &   K0V   & V &                        new  &  \\
TYC\,8929\,0927\,1 &        082406-6334.1 &   0.79 &   0.02 &   9/12    & 0.09 &   0.037 &   9.89 &  0.63 &   G5V   & S  &                       new  &   \\
TYC\,8930\,0601\,1 &    084200-6218.4 &  1.224 &  0.004 &   10/14    & 0.16 &   0.035 &  11.04 &  0.80 &   K0V   & S  &                  new  &  \\
TYC\,8569\,1761\,1 &    084553-5327.5 &   1.28 &  0.003  &   10/16    & 0.07 &   0.033 &  10.46 &  0.63 &   G2V   & S      &               new  &   \\
TYC\,9395\,2139\,1 &        085005-7554.6 & 1.1435 &  0.005 &   11/14   & 0.10 &   0.032 &  10.62 &  0.76 &   G9V   & SB2? &                  new  &   \\
TYC\,8569\,3597\,1 &        085156-5355.9 & 1.942  &  0.005 &   10/14   & 0.09 &   0.033 &  10.84 &  0.69 &   G9V   & SB2 &                    new   &  \\
TYC\,8582\,3040\,1 & 	    085746-5408.6 & 1.94   &  0.005 &    8/14   & 0.16 &   0.030 &  11.71 & 0.88  &   K2IV(e) & S & new  &  \\
TYC\,8160\,0958\,1 &        085752-4941.8 &  2.043 &  0.001 &   8/15    & 0.15 &   0.027 &  10.55 &  0.73 &   G9V   & S   &             P=1/2P$_{\rm ACVS}$  &  \\
TYC\,8586\,2431\,1 &    085929-5446.8 &   3.64$^c$ &   0.01 &   9/14    & 0.11 &   0.033 &  10.16 &  0.66 &   G5IV  & S  &               new  &  \\
TYC\,8586\,0518\,1 &        090003-5538.4 &  0.916 &   0.36 &   8/12    & 0.11 &   0.031 &  10.84 &  0.68 &   G5V   & S     &                 new  &   \\
TYC\,8587\,1126\,1 &        090929-5538.4 &   0.77$^b$ &   0.003 &   2/12    & 0.05 &   0.033 &  10.24 &  0.73 &   G8V   & S         &               new  &   \\
TYC\,8587\,2290\,1 &        091317-5529.1 &   1.50 &   0.01 &   4/12    & 0.05 &   0.035 &  10.41 &  0.66 &   G5V(e) & S        &              new  &  \\
HIP\,46063 	   	 &        092335-6111.6 &   3.86 &   0.02 &   14/14   & 0.15 &   0.031 &  10.27 &  0.86 &   K1V(e) & S    &  P=P$_{\rm ACVS}$ &   V479~Car \\
TYC\,8584\,2682\,1 &        093226-5237.7 &    ... &   ...  &   ...     & ...  &    ...  &  10.86 &  0.76 &   G8V(e) & S  & ... &  \\
TYC\,8946\,1225\,1 &        094309-6313.1 &   1.70$^b$ &   0.01 &   2/11    & 0.07 &   0.034 &  10.40 &  0.68 &   G6V    &  S            &           new &    \\
HD\,107722 	   	 &        122329-7740.9 &   ...  &  ...   &   ...     &  ... &   ...   &  8.31  &  0.45 &   G6V    &  S          &   ...   &     \\
TYC\,8634\,1393\,1 &        114552-5520.8 &   5.35$^b$ &   0.04 &   3/12    & 0.05 &   0.030 & 10.24  &  1.01  &  K5Ve   & S  & new &  \\
\hline
\multicolumn{13}{l}{V=visual companion; BD= brown dwarf companion; S=single star; B=binary system; P$_{\rm lit}$: period as given in the literature; P$_{\rm ACVS}$ period in ACVS;}\\
\multicolumn{13}{l}{$^b$ period undetected in the periodogram of the complete timeseries; $^c$: $v\sin i$ inconsistent with v$_{\rm eq}$=2$\pi$R/P.}
\end{tabular}

\end{table*}


\begin{table*}
\scriptsize
\caption{\label{abdor}  As in Table\,\ref{twhya} for the {\bf AB Doradus Moving Group}.}
\centering
\begin{tabular}{llll@{\hspace{.1cm}}c@{\hspace{.1cm}}c@{\hspace{.1cm}}c@{\hspace{.1cm}}l@{\hspace{.1cm}}llll@{\hspace{.1cm}}c}

\hline
  name   &  ID$_{\rm ASAS}$   &P     &  $\Delta$P   &  timeseries & $\Delta$V$_{\rm max}$ & $\sigma_{\rm acc}$ & V$_{min}$ & B$-$V & Sp.T. &  mult. & note on& variable's  \\   
         &                    & (d)  &  (d)         &   sections       & (mag)     & (mag)              & (mag)     & (mag) &       &        & period      & name \\   
\hline
HIP\,5191			&        010626-1417.8 &   7.13 &   0.05 &   2/8    &  0.05 &   0.025 &   9.49 &  0.91 & K1V   & V &  new  & \\
HIP\,6276 			& 	   012032-1128.1 &    ... &   ...  &   ...    &  ...  &   ...   &   8.43 &  0.79 & G9V   & S & ...  &  \\
TYC\,8042\,1050\,1	&  	   021055-4604.0 &  1.116$^b$ &  0.001 &   7/12   &  0.15 &   0.029 &  11.48 &  0.91 & K3IVe & V & new   & \\
HIP\,10272			&        021216+2357.5 &   6.13 &   0.03 &   2/4    &  0.07 &   0.038 &   7.94 &  1.13 & K1    & B & new   &  \\
HIP\,13027 			& 	   024727+1922.3 &   ...  &   ...  &  ...     & ...   &  ...    &   6.90 &  0.68 & G0+G5 & B & ...  & \\
HIP\,14684 			&        030942-0934.8 &   5.46 &   0.08 &   2/10   & 0.07  &   0.026 &   8.51 &  0.81 & G0    & S &  new & IS Eri\\
HIP\,14809 			& 	   031114+2225.0 &   ...  &  ...   &   ...    & ...   &  ...    &   8.51 &  0.71 & G5    & V & ...  & \\
HIP\,17695 			&        034723-0158.3 &   3.87 &   0.01 &   4/7    & 0.15  &   0.031 &  11.60 &  1.51 & M3    & S &  new   & \\
TYC\,5899-0026\,1		&        045224-1649.4 &   ...  &   ...  &   ...    &  ...  &    ...  &  ...   &  ...  & M3    & S & ...  &  \\
HIP\,22738AB 		&        045331-5551.5 &  1.496 &   0.01 &   3/10   & 0.06  &   0.031 &  10.78 &  1.56 & M3Ve  & V & new  & \\
TYC\,7587\,0925\,1 	&        050230-3959.2 &   6.53 &   0.06 &   11/12  & 0.12  &   0.030 &  10.71 &  0.88 & K4V   & S & new   & \\
HD\,32981 			&        050628-1549.5 &  0.985$^b$ &   0.01 &   4/14   & 0.09  &   0.030 &  9.16  &  0.58 & F8V   & S & new   & \\
HIP\,25283 			&        052430-3858.2 &   9.34 &   0.08 &   7/12   & 0.07  &   0.028 &   9.06 &  1.25 & K6V   & S & new   & \\
HIP\,25647 		      &        052845-6526.9 & 0.5140 & 0.0003 &   2/10   & 0.13  &   0.033 &   6.74 &  0.83 & K0V   & Q & P=P$_{\rm lit}$   & AB Dor\\
TYC\,7059\,1111\,1 	&  	   052857-3328.3 &   2.29$^c$ &   0.01 &   8/14   & 0.11  &   0.027 &  10.61 &  1.06 & K3Ve  & S & P=P$_{\rm lit}$   & UX Col\\
TYC\,7064\,0839\,1 	&  	   053504-3417.9 &   7.82 &   0.18 &   1/14   & 0.15  &   0.026 &  11.83 &  1.08 & K4Ve  & S & new  & \\
HIP\,26373 			&        053657-4757.9 &   4.52 &   0.02 &   6/14   & 0.09  &   0.031 &   7.73 &  0.85 & K0+K6V   & V & P=P$_{\rm lit}$   &  UY Pic\\
HIP\,26401A			& 	   053713-4242.9 &   ...  &  ...   &   ...    &  ...  &  ..     &   9.55 &  0.67 & G7V+K1V & B & ...  & WX Col\\
CP-19\,878 			&        053923-1933.5 &   1.49 &   0.01 &   6/13   & 0.12  &   0.028 &  10.57 & 0.91$^1$ & K1V & S  & new &\\
TYC\,7605\,1429\,1 	&        054114-4118.0 &  2.75$^b$  &   0.02 &   3/14   & 0.15  &   0.030 &  12.29 &  0.87 & K4IVe  & S  & new  & \\ 
TYC\,6494\,1228\,1 	&        054413-2606.3 &   1.83 &   0.01 &   2/14   & 0.12  &   0.029 &  10.96 &  0.86 & K2Ve   & SB? & new  & \\
HIP\,27727 			&        055216-2839.4 &   2.84 &   0.01 &   9/14   & 0.08  &   0.033 &   9.12 &  0.63 & G3V & SB?  & P=P$_{\rm lit}$   & TZ Col \\
TYC\,7598\,1488\,1 	&        055751-3804.1 &   3.73$^c$ &   0.04 &   6/11   & 0.09  &   0.029 &   9.64 &  0.69 & G6V(e) & S  & P=P$_{\rm lit}$   & TY~Col \\
TYC\,7079\,0068\,1      &        060834-3402.9 &   3.38 &   0.01 &   9/12   & 0.14  &   0.031 &  10.36 &  0.79 & G9Ve   & S  & P=P$_{\rm ACVS}$  &  \\
TYC\,7084\,0794\,1 	&        060919-3549.5 &   1.717 & 0.004 &   6/12   & 0.11  &   0.031 &  11.13 &  1.69 & M1Ve   & S  & P=P$_{\rm ACVS}$   & \\
HIP\,30314 			&        062231-6013.1 &   ...  &   ...  &   ...    & ...   &   ...   &   6.54 &  0.61 & G1V    & V  & ...   & \\
HIP\,31711 		      &        063800-6132.0 &   2.60 &  ...   &   ...    & 0.05  &   ...   &   6.25 &  0.62 & G2V    & SB+V & P$_{\rm lit}$   & AK Pic\\
HIP\,31878 			&        063950-6128.7 &   9.06 &   0.08 &   4/10   & 0.07  &   0.032 &   9.78 &  1.26 & K7V(e) & S  & new  & \\
TYC\,7627\,2190\,1 	&        064119-3820.6 &   2.67$^b$ &   0.02 &   3/10   & 0.11  &   0.027 &  11.47 &  1.19 & K2Ve   & ...& new   & \\
UCAC2\,06727592 		&        064753-5713.5 &   6.05 &   0.05 &   5/11   & 0.11  &   0.032 &  11.54 & 0.90$^1$ & K4V & S  & new   & \\
TYC\,8558\,1148\,1 	&        071051-5736.8 &   2.94 &   0.01 &   4/11   & 0.07  &   0.031 &  10.53 &  0.68 & G2V    &  S & new   & \\
TYC\,1355-214-1 		&        072344+2025.0 &   2.79 &   0.02 &   6/8    & 0.16  &   0.036 &  10.08 &  1.15 & K3     &  S & P=P$_{\rm lit}$   & \\
HIP\,36108 			&        072618-4940.8 &   3.89 &   0.04 &   1/11   & 0.05  &   0.029 &   9.81 &  0.43 & G7V    & B  & new   & \\
HIP\,36349 		      &        072851-3014.8 &   1.642 &  0.006 & 15/15   & 0.15  &   0.032 &  10.01 &  1.44 & M1Ve   & B  & P=P$_{\rm lit}$   & V372 Pup \\
TYC\,9493\,0838\,1 	&        073100-8419.5 &   4.94 &   0.03 &   6/11   & 0.09  &   0.035 &  10.04 &  0.83 & G9V    & S  & new  & \\
HIP\,51317 			& 	   102856+0050.5 &  ...   &   ...  &   ...    & ...   &    ...  &   9.65 &  1.51 & ...    & ... & ...  & \\
HIP\,76768 			&        154028-1841.8 &   3.70 &   0.02 &   4/10   & 0.09  &   0.029 &  10.44 &  1.24 & K3/4V  & B   & P$\ne$P$_{\rm ACVS}$   & \\
HIP\,81084 			&        163342-0933.2 &   7.45 &   0.11 &   2/10   & 0.07  &   0.030 &  11.30 &  1.44 & M0.5   & S   & new  & \\
HIP\,82688 			& 	   165408-0420.4 &   ...  &   ...  &   ...    &  ...  &  ...    &  7.82  &  0.59 & G0     & ... & ...  & \\
TYC\,7379\,0279\,1 	&        172856-3244.0 &   ...  &   ...  &   ...    & ...   &  ...    &  10.40 &  0.93 & ...    & ... & ...   & \\
BD-13\,4687 		&        173746-1314.8 &   0.447 & 0.002 &  11/11   &  0.19 &   0.033 &  10.10 &  0.93 & K3/4IV & S  & new  & \\
HIP\,93375 			&        190106-2842.8 &   ...  & ...    &  ...     &  ...  &   ...   &  8.55  &  0.56 & G1V    & V   & ...   & \\
HIP\,94235 			&        191058-6016.3 &   2.24 &   0.01 &   2/10   & 0.04  &   0.031 &  8.37  &  0.59 & G1V    & V   & new   & \\
TYC0486-4943 		&        193304+0345.7 &   1.35$^b$ &   0.02 &   5/10   & 0.10  &   0.035 &  11.20 &  0.95 & K3V    & S   & new   & \\
HD\,189285 			&        195924-0432.1 &   4.85 &   0.03 &   2/12   & 0.06  &   0.031 &   9.54 &  0.71 & G5     & S   & new  & \\
TYC\,5164-567-1		&        200449-0239.3 &   4.68 &   0.06 &   5/7    & 0.10  &   0.032 &  10.17 &  1.00 & ...    & ... & new  & \\
HD\,199058 			& 	   205421+0902.4 &   ...  &   ...  &   ...    &  ...  &   ...   &   8.81 &  0.63 &  G5    & ... & ...  & \\
TYC1090-0543 		&        205428+0906.1 &   2.28 & 0.007  &    2/9   & 0.14  &   0.029 &  11.79 &  1.07 & ...    & ... & P=P$_{\rm lit}$   & \\
TYC\,6351\,0286\,1 	&        211305-1729.2 &   4.92 &   0.07 &   6/10   & 0.12  &   0.029 &  10.69 &  1.21 & K6Ve   & S   & P=P$_{\rm ACVS}$   & \\
HIP\,106231			&        213102+2320.1 &   0.42312$^b$ & 0.007 & 4/6    & 0.10  &   0.036 &   9.19 &  1.05 & K8     & S   & P=P$_{\rm ACVS}$=P$_{\rm lit}$   & LO~Peg \\
HIP\,107684			&        214849-3929.1 &   4.14 &   0.02 &   5/10   & 0.06  &   0.031 &   9.67 &  0.69 & G7V    & S   & new   & \\
HIP\,113579			&        230019-2609.2 &   2.169$^b$ &  0.007 &  1/10   & 0.06  &   0.027 &   7.49 &  0.65 & G5V    & B   & new   &  \\
HIP\,113597			&        230028-2618.7 &   ...   &  ...   &  ...    & ...   &   ...   &   9.63 &  1.34 & K7V    & S   & ...  &  \\
HIP\,114530 		&        231154-4508.0 &   5.17 &   0.02 &   4/10   & 0.07  &   0.030 &   8.82 & 0.80$^1$ & G8V & B   & new   & \\
HIP\,116910			&        234154-3558.7 &   2.294 &  0.007 &  6/10   & 0.11  &   0.029 &   9.45 &  0.71 & G8V    & S   & new   & \\
HIP\,118008  		& 	   235611-3903.1 &    ...  &  ...   &  ...    & ...   &   ...   &   8.22 &  0.97 & K3V    & S   & ...   & \\
PW\,And 			&   	   no-ASAS       &   1.762 &  ...   &  ...    & 0.13  &   ...   &   8.86 &  0.92 & K2V    & S   & P$_{\rm lit}$   & PW\,And  \\
HIP\,63742 			&	   no-ASAS       &   6.5400 & ...   &  ...    & 0.06  &   ...   &   7.73 &  0.85 & G5V    & B   & P$_{\rm lit}$   & PX Vir\\
HIP\,86346   		&	   no-ASAS       &   1.842 &  ...   &  ...    & 0.02  &    ...  &  10.29 &  1.23 & M0     & T   & P$_{\rm lit}$  & \\
HIP\,114066   		&	   no-ASAS       &   4.502 &  ...   &  ...    & 0.03  &   ...   &  10.87 &  1.44 & M0     & S   & P$_{\rm lit}$   & \\
HIP\,16563 			& 	   no-ASAS       &   1.430 & 0.006  &  ...    & ...   &   ...   &   8.15 &  0.80 & G5+M0  & B   & P$_{\rm lit}$   & V577 Per\\
HIP\,12635 			& 	   no-ASAS       &  ...    &   ...  &  ...    & ...   &   ...   &  10.21 &  0.88 & ...    & B & ...   & \\
HIP\,12638 		      &        no-ASAS       & ...     &   ...  &  ...    & ...   &   ...   &   8.70 &  0.71 &  G5    & B   & ...   & \\
HIP\,110526AB 		& 	   no-ASAS       & ...     &   ...  &  ...    & ...   &   ...   &  10.70 &  1.57 & M0     & B & ...   & \\
\hline
\multicolumn{13}{l}{V=visual companion; S=single star; B=binary system; QUAD=quadruple system; $^1$: inferred from spectral type;  }\\
\multicolumn{13}{l}{ $^c$: $v\sin i$ inconsistent with v$_{\rm eq}$=2$\pi$R/P; P$_{\rm lit}$: period as given in the literature; P$_{\rm ACVS}$ period in ACVS;}\\
\multicolumn{13}{l}{$^b$: period undetected in the periodogram of the complete timeseries.}\\
\end{tabular}

\end{table*}

\begin{figure*}[t,*h]
\begin{minipage}{18cm}
\centerline{
\psfig{file=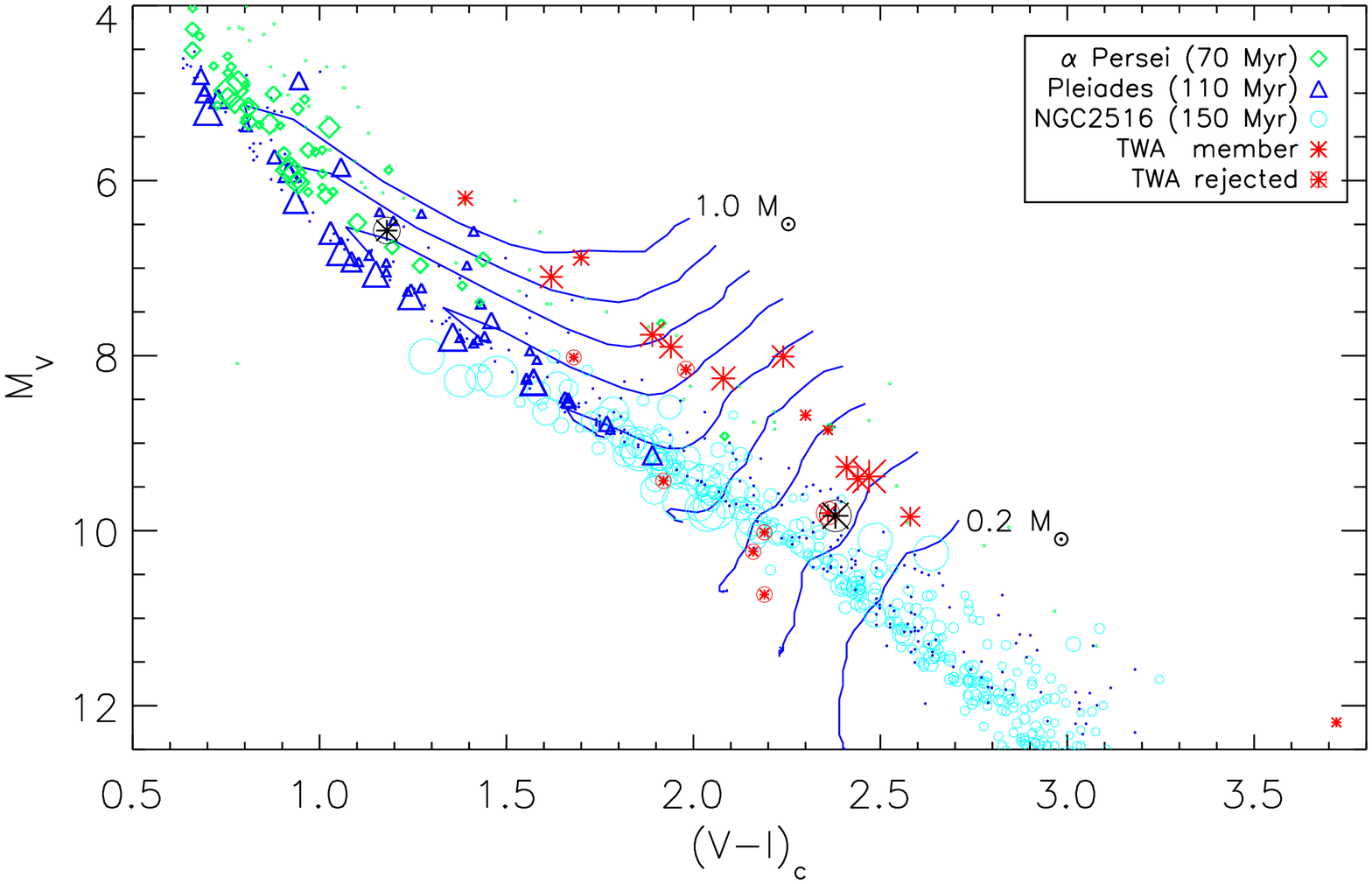,width=9.5cm,height=6.5cm,angle=90}
\psfig{file=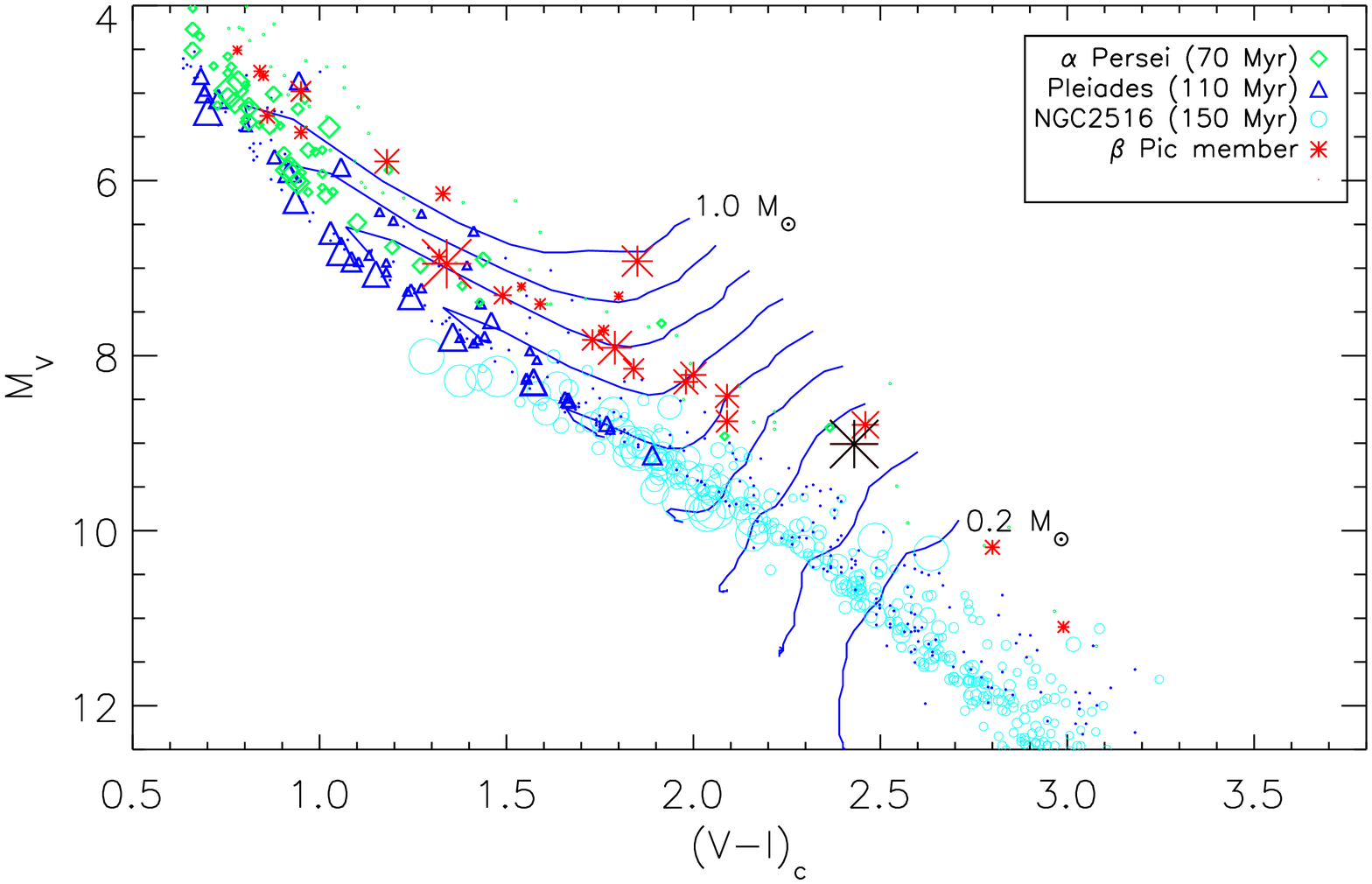,width=9.5cm,height=6.5cm,angle=90}
}
\end{minipage}

\begin{minipage}{18cm}
\centerline{
\psfig{file=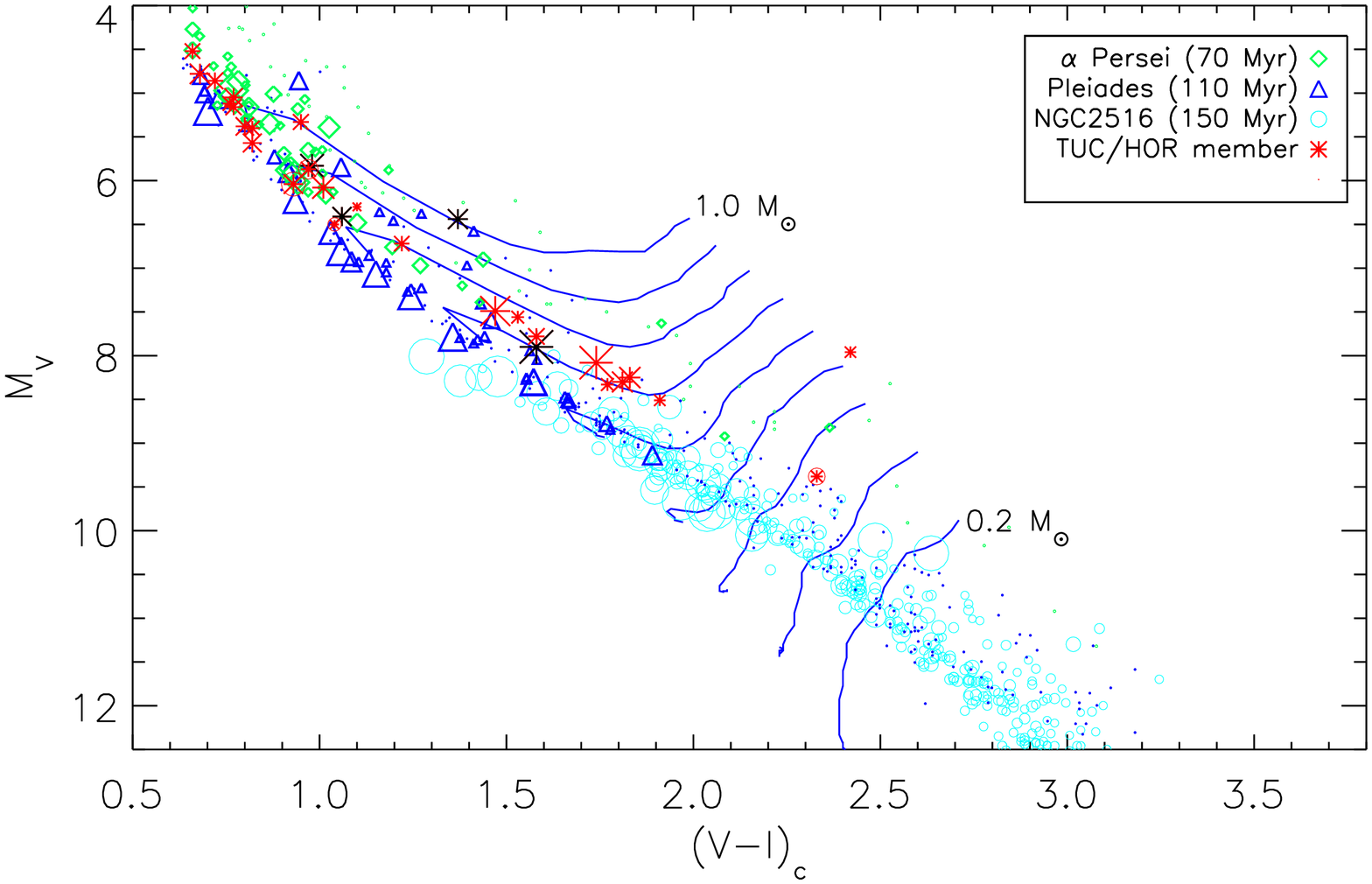,width=9.5cm,height=6.5cm,angle=90}
\psfig{file=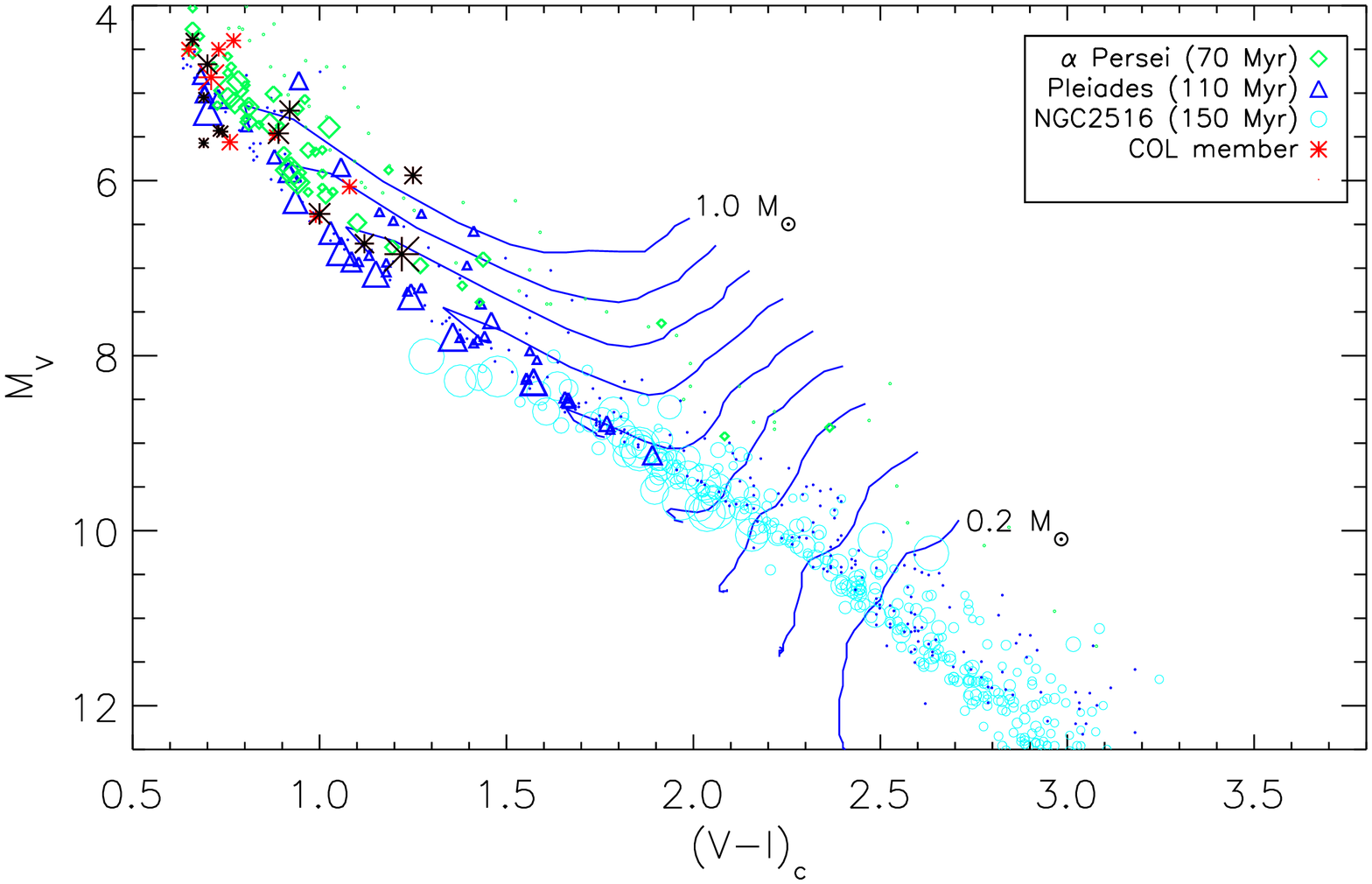,width=9.5cm,height=6.5cm,angle=90}
}
\end{minipage}

\begin{minipage}{18cm}
\centerline{
\psfig{file=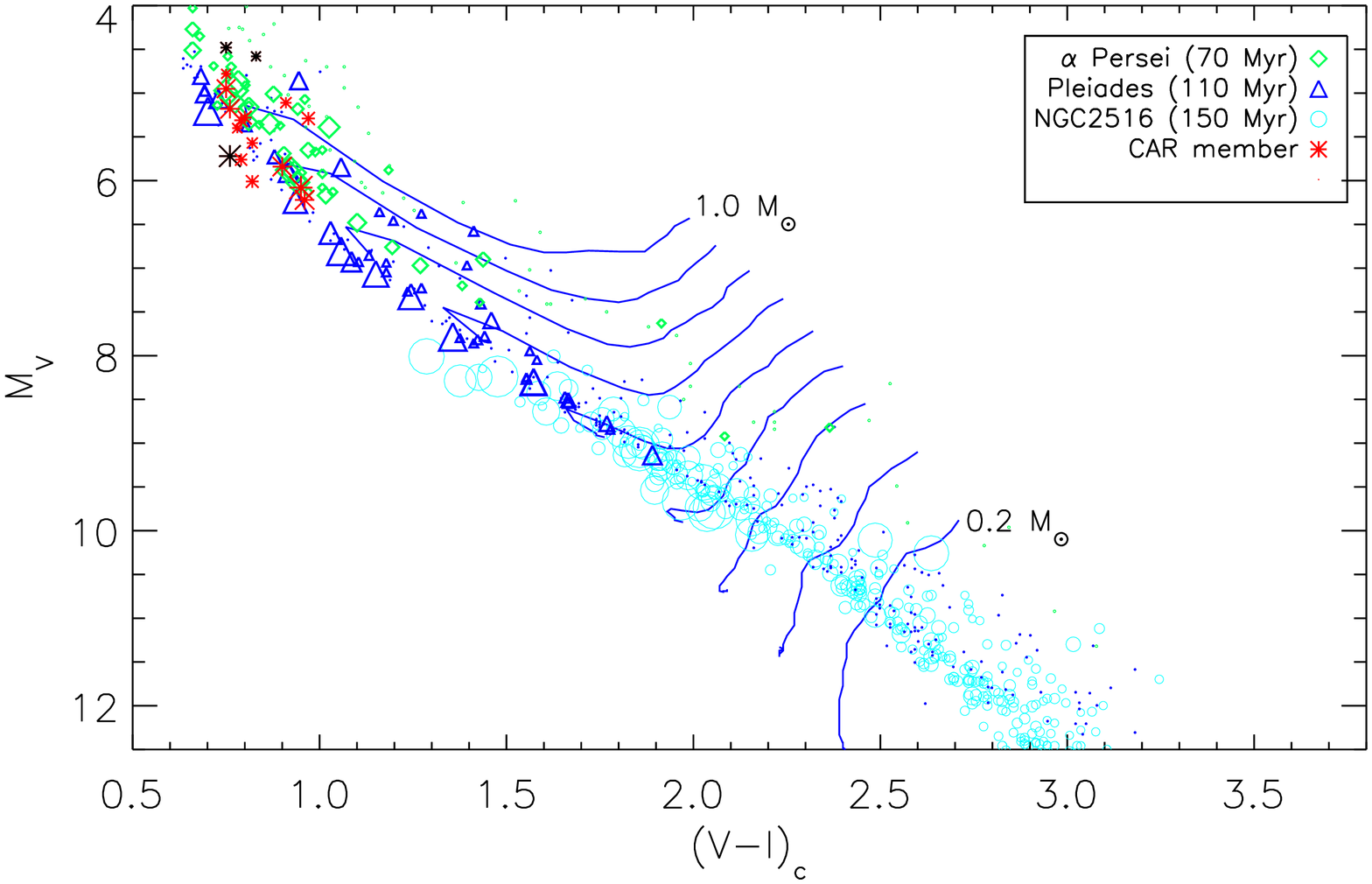,width=9.5cm,height=6.5cm,angle=90}
\psfig{file=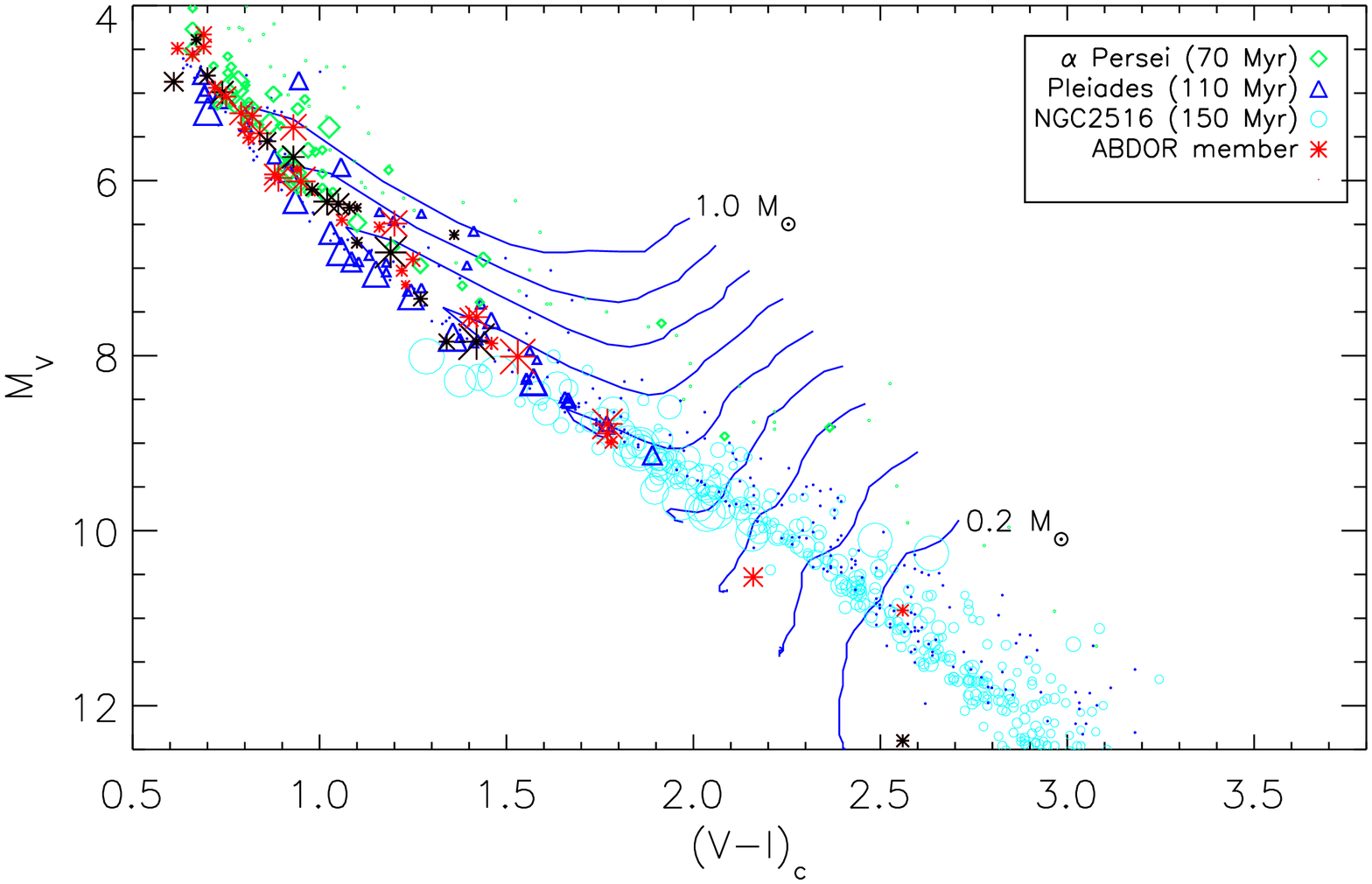,width=9.5cm,height=6.5cm,angle=90}
}
\end{minipage}
\caption{Color-Magnitude diagrams of the six associations under analysis with overplotted PMS tracks (solid lines) from Baraffe (\cite{Baraffe98}).
Different symbols indicate stars belonging to different association/clusters; the symbol size is proportional to the rotation period. 
Dots represent cluster stars with unknown rotation period. Black symbols represent stars whose V$-$I color is
derived from B$-$V.\label{cmd}}
\end{figure*}

\section{Rotation period evolution}

\subsection{Color-Magnitude diagrams}
We use the M$_{\rm V}$ vs. V$-$I CMD and a set of  low-mass PMS evolutionary tracks
to derive masses  and radii \rm and check the evolutionary stages of our targets (see Fig.\,\ref{cmd}).
Evolutionary tracks (mass range from 0.2 to 1.0M$_{\odot}$ at steps of 0.1) are taken from Baraffe et al. (1998) 
(initial metallicity [M/H]=0.0,  initial helium mass fraction Y=0.275, initial mixing length parameter H$_{\rm P}$=1.0). 

The V$-$I colors of a few stars, with the exception of the $\beta$ Pictoris members, have never been measured.
To overcome this limitation and position all the stars in the same CMD, we derived empirical relations between V$-$I and
B$-$V using all the stars belonging to the same association and having both  colors measured. 
In Fig.\,\ref{fig_fit} we give an example of the polynomial fit used to obtain V$-$I colors from B$-$V  in the case of 
Tucana/Horologium. In Table \ref{tab_fit} we list the polynomial coefficients we determined for each association.
The average error on the derived V$-$I colors is about 0.05 mag. 

Fig.\,\ref{cmd} also includes  those targets  of the TW Hya and Tuc/Hor associations that were 
rejected by Torres et al. (2008) from the high-probability  member list.  Although considered in our period search,
their rotation periods are not included in the following rotation period distribution analysis.

There are four  stars which significantly deviate from the sequence traced by the other members.
In Tucana/Horologium  it is the case of TYC\,5908-230-1 whose spectral type is unknown and whose V$-$I color is derived from B$-$V.
In Columba it is the case of BD$-$16351 whose  V$-$I color is derived from B$-$V.
In AB Dor this is the case of HIP\,17695 and  TYC\,7084\,0794\,1. Their rotational properties, however, do not deviate from the 
 average of their respective associations.

\begin{table}
\caption{Polynomial fit coefficients to derive V$-$I from B$-$V color,
for stars with no V$-$I observations.\label{tab_fit}}
\begin{tabular}{l|ccccc}

                                      &a	&b	&c	&d	&e\\
\hline
& \multicolumn{5}{c}{\bf TW Hya}\\
 B$-$V $>$ 0.5             & 3.163 & -5.014 & 3.028 & ... & ...\\
& \multicolumn{5}{c}{\bf Tucana/Horologium}\\
 B$-$V $>$ 0.5	&+0.821 & -3.550 & +9.978 & -9.190 & +3.047\\
& \multicolumn{5}{c}{\bf Columba}\\
 B$-$V $>$ 0.5	&-0.417 & 2.142 &  -0.578 &... &...\\
& \multicolumn{5}{c}{\bf Carina}\\
B$-$V $>$ 0.5 & 0.296 & 0.545 & 0.260 & ... & ...\\
& \multicolumn{5}{c}{\bf AB Dor}\\
B$-$V$<$ 1.6 &+10.19 & -44.25 & +72.94 & -50.35 & +12.66\\
B$-$V $>$ 1.6 & +0.648 & -0.494 & +0.960 &    ...   &  ...            \\
\hline

\end{tabular}
\end{table}

To better visualize the evolutionary and the rotational  stage of our targets, we considered three additional well studied open clusters
of known age: $\alpha$ Persei (70 Myr), Pleiades (110 Myr) and NGC2516 (150 Myr).
The more evolved clusters (with respect to our targets) allow us to identify in the CMD the position of the  ZAMS in the mass range of our association members.
The early-type  (more massive) members  of all three clusters have already reached the ZAMS, whereas the late-type (low mass) members are still approaching it.
The list of confirmed members of the $\alpha$ Persei and Pleiades open clusters is compiled from the WEBDA database.
The (V$-$I)$_{\rm c}$ colors and Johnson V magnitudes are from Stauffer et al. (1985, 1989) for $\alpha$ Persei, and from Stauffer (1982a, 1982b, 1984) and
Prosser et al. (1991) for  the Pleiades. The rotation periods are taken from the compilation by Messina et al. (2003, and references therein). 
The V$-$I colors, originally given in the Kron system, have been transformed into the Cousin system by using the Bessel (1979)
color-color relations.
Color excess E(B$-$V) = 0.10 and  distance modulus ($m$$-$M) = 6.60 for the $\alpha$ Persei and  E(B$-$V) = 0.04 and ($m$$-$M) = 5.60 for the Pleiades (O'Dell et al. 1994) have been used to position the cluster members in the CMD.
Photometry, rotation periods, color excess and distance modulus of NGC\,2516 members are all  taken from Irwin et al. (2009).


\subsection{Rotation period distribution}

A major goal of this paper is to look for statistically relevant differences
in the rotation rate of stars in young stellar associations that can be
ascribed to angular momentum evolution.

The main influence of stellar mass on the angular momentum evolution is to
determine the timescale of contraction towards the ZAMS and, below
approximately 1.2 M$_{\odot}$, the amount of angular momentum stored in the
radiative core (see, e.g., Allain \cite{Allain98}; Bouvier \cite{Bouvier08}; Keppens et al.\,\cite{Keppens95}).
In order to limit the range of possible variations, a binning in mass of our
sample is therefore desirable. The paucity of stars available and the
uncertainties in mass, however, allow only a rather broad mass binning. In
the following analysis we shall consider stars in the range 0.8-1.2
M$_{\odot}$ and compare the results with a larger sample in the range 0.6-1.2
M$_{\odot}$. In the former case, we limit the angular momentum evolution
timescales range still maintaining a sufficient number of stars for the
statistical analysis; in the latter we increase the number of stars at the
expenses of mixing quite different evolution timescales.\\
\indent
Colour or spectral type can be used as indicators of mass. The relationship
color or spectral type vs. mass changes,  however, with the age of the
stellar system, particularly in the PMS phase, and the MS relationship
cannot be applied to PMS stars. Without age discrimination, PMS stars of
similar colors or spectral types can belong to quite different mass ranges,
especially between $\simeq$ 0.7 and 1 M$_\odot$ where the evolutionary tracks
turn abruptly towards higher temperatures before settling to the ZAMS (see
Fig.\,\ref{cmd}). In order to take the PMS evolution into account, we derived stellar
masses and radii by comparing the position in the CMD with the Baraffe
(1998) isochrones. The uncertainties on the estimated mass and radius mostly
derive from the uncertainty on: a) V$-$I color (especially for the lowest
stellar masses); b) V magnitude (subject to variations up to a few tenths of
magnitude due to the magnetic activity); c) parallax; d) metallicity. We
estimated the cumulative uncertainty to be approximately 0.1 M$_\odot$ in
mass and  0.05 R$_\odot$ in radius, which is acceptable for the purposes of
our analysis. The derived masses and radii are listed in the online Tables\,\ref{twa_lit}-\ref{abdor_lit}.
The mass histogram of the complete sample of periodic variables
is reported in Fig. 8. About 91\% of our targets have masses between 0.6 and
1.2 M$_{\odot}$;  60\%  have masses between 0.8 and 1.2 M$_{\odot}$.\\
\indent
The uncertainties in age and the paucity of stars in each association impose
also a rather broad binning in age. In order to maintain statistical
significance, almost coeval associations are put in the same age bin. In this
way we consider TW Hya and $\beta$ Pictoris, which have estimated ages of $\sim$8
and $\sim$10 Myr, in the same age bin; Tucana/Horologium, Carina and Columba
members will be considered coeval stars with an estimated age of $\sim$30 Myr.
The age of AB Dor is estimated to be approximately 70 Myr by Torres et al.
(\cite{Torres08}), and therefore coeval to the $\alpha$ Persei cluster. We found,
however, that the period distribution vs. V$-$I is more similar to the
Pleiades than $\alpha$ Persei  (see Fig.\,\ref{abdor-age} and discussion below), and
therefore we assign AB Dor to the 110 Myr age bin  together with the
Pleiades.\\
\indent
In Fig.\,\ref{abdor-age} we compare the distribution of rotation periods of AB Dor with that of $\alpha$ Persei (top panel) and the Pleiades (bottom panel). Adopting the Barnes (\cite{Barnes03}) classification scheme, we can easily identify three different groups of stars in Fig.\,\ref{abdor-age}. The rotation upper boundary forms a sequence, populated by stars which are subject to the long timescale spin-down controlled by the stellar wind magnetic breaking. The dashed line represents the expected theoretical period distribution according to Eq. (1) and (2) of Barnes (\cite{Barnes03}). This group of stars shows an almost one-to-one correspondence between rotation period and color, which is definitively reached by the age of 500-600 Myr as shown by members of the Hyades (Radick et al.  \cite{Radick87}) and Coma Berenices (Collier Cameron et al. \cite{Cameron09}) open clusters. Very fast rotators (P $\la$ 1d) form a different sequence; the dotted line in Fig.\,\ref{abdor-age} represents their expected theoretical period distribution computed using Eq. (15) of Barnes (\cite{Barnes03}). A third intermediate group is populated, according to the Barnes (\cite{Barnes03}) scheme, by stars which are moving from the very fast rotators to the slow rotators sequence. The identification of such sequences is rather difficult in the period vs. V$-$I distribution of the younger association shown in Fig.\,\ref{distri}, TW Hya and $\beta$ Pictoris (top panel) and for Tucana/Horologium, Carina and Columba (bottom panel). In this case, stars are still contracting towards the ZAMS and are either still magnetically locked to their circumstellar disk or they just left this phase. \\
\indent
From the comparison of the rotational period  vs. V$-$I distribution AB Dor with those of $\alpha$ Persei and the Pleiades shown in Fig. \,\ref{abdor-age} we notice that the population of very fast rotators in $\alpha$ Persei in the color range 0.5 $<$ V$-$I $<$ 1.0 is missing among both AB Dor and Pleiades members. This population is expected to have migrated from the very fast rotators to the slow rotators sequence in the age range from 70 to 110 Myr. The AB Dor mean and median rotation periods are also much closer to those of the Pleiades than $\alpha$ Persei. A two-sided two-dimensional Kolmogorov-Smirnov test
confirms that the AB Dor rotation period distribution is more similar to the
Pleiades than $\alpha$ Persei,  the KS probability that the periods are
drawn from the same distribution being much lower for AB Dor / $\alpha$
Persei  (0.5\%) than for AB Dor / Pleiades (43\%). Furthermore, looking at
the AB Dor CMD (Fig.\,\ref{cmd}), we see that the $\alpha$ Persei members are
generally redder, which is consistent with a younger age than AB
Dor.  This is consistent also with the works of Luhman et al. (\cite{Luhman05}) and
Ortega et al. (\cite{Ortega07}) which suggest  a common origin of AB Dor and the
Pleiades.\\
\indent
Among slow rotators, we note two outliers: the $\alpha$ Persei member
AP\,121 and the Pleiades member HII\,2341, whose rotation period is well
above the upper boundary. We suggest that the rotation period of this two
targets, taken from the literature, is incorrect and we do not include them
in the final sample to derive rotation period histograms and distributions.\\
\indent
To put our analysis in context with earlier stellar angular momentum evolution, we considered also rotational periods for members of the Orion Nebula Cluster (ONC, $\sim$1\,Myr) and NGC\,2264 ($\sim$4 Myr). Rotation periods of ONC members were taken from Herbst et al. (\cite{Herbst02}), of NGC\,2264 members  from Rebull et al. (\cite{Rebull02}) and Lamm et al. (\cite{Lamm04}). Age-binned rotation period histograms of the full dataset are shown in Fig.\,\ref{histo}.\\
\indent
Fig.\,\ref{evolution} shows the rotation period evolution from ONC to AB Dor (plus
Pleiades) for stars with mass between 0.6 and 1.2 M$_{\odot}$ (left panel) and for the
restricted sample with mass between  0.8 and 1.2 M$_{\odot}$ (right panel). 
 Both mean and median period
decrease slowly but systematically with age from ONC to $\alpha$ Persei,
apart the mean period from ONC to NGC\,2264 in the 0.6 - 1.2 M$_{\odot}$
sample which is essentially constant. Given the paucity of the data and because
distributions are not Gaussian, we investigated the significance of such
variations by performing a two sided KS test at consecutive samples in age.
We expect that, if the variation of the mean and median are significant, the
KS probability that the period realisations in two consecutive time bins are
drawn from the same distribution will be low.  To avoid ambiguities that
arise from the color evolution at early stage, we apply only the
one-dimensional two-sided KS test to the restricted mass range 0.8 - 1.2
M$_{\odot}$ and compare the results with those obtained in the extended 0.6 -
1.2 M$_{\odot}$ range.\\
\indent
In Table \ref{tab-per} we list, for each age bin and for both the
 (0.6 - 1.2) M$_{\odot}$ and  (0.8 - 1.2) M$_{\odot}$ ranges, the mean and median rotation period, the
number of stars used in the test and the two-sided KS probability that the
realisations at consecutive sample ages are drawn from the same
distribution.\\
\indent
Considering the 0.8 - 1.2 M$_{\odot}$ range first, we see that the moderate
rotation spin-up from $\sim$1 to $\sim$9 Myr is also associated with a KS probability of  70\%
for the 1 vs. 4 Myr and 63\% for the 4 vs. 9 Myr age bins. The KS
probability decreases to 17\% for the  9 vs. 30 Myr age bins and rises up
again to 45\% for the 30 vs. 70 Myr age bins. The KS probability then
decreases for the 70 vs. 110 Myr age bins, when we also see an unambiguous
rotation spin-down. The most significant variations are then the spin-up between 9
and 30 Myr and the spin-down between 70 and 110 Myr. Between 1 and 9 Myr the
moderate spin-up is poorly supported by the KS test. The KS test on the
spin-up between 30 and 70 Myr does not allow us to draw definitive conclusions.
For the 0.8 - 1.2 M$_{\odot}$ range the analysis is therefore consistent with
a considerable disk-locking before 9 Myr, followed by a moderate but
unambiguous spin-up from 9 to 30 Myr, consistent with stellar contraction
towards the ZAMS. Variations between 30 and 70 Myr are rather doubtful,
despite the median indicates a significant spin-up. In fact the mean rotation period does
not indicate any spin-up at all and the KS-probability does not support a
strong difference in the period distribution at these two age bins. This
situation may be due to the heterogeneity of the sample: all stars with
masses above 1 M$_{\odot}$  are expected to complete their contraction
toward the ZAMS at ages earlier than about 30 Myr, but stars with lower mass
will end the contraction towards the ZAMS later (around 50 Myr for a star of 0.8
M$_{\odot}$). The unambiguous spin-down from 70 to 110 Myr is consistent
with the fact that, starting from the 70 Myr age bin, all stars in the 0.8 -
1.2 M$_{\odot}$  mass range  have entered the MS phase and therefore the
angular momentum evolution is dominated by wind-braking.\\
\indent
The same considerations can be applied to the extended (0.6 - 1.2) M$_{\odot}$
range, despite all KS-probabilities are lower than in the (0.8 - 1.2)
M$_{\odot}$ range. At ages earlier than 9 Myr, evidences for a moderate
spin-up are rather poor, the two-sided test giving a probability around 60\%
that the distributions in the 1 and 4 Myr and in the 4 and 9 age bins are the
same. The spin-up from 9 to 30 Myr remains unambiguous, with a probability
of only 9\% that the period distributions in these two age bins are the
same. The moderate spin-up from 30 to 70 Myr is somewhat more significant,
which is likely  due to a higher number of stars ending their
contraction towards the ZAMS at ages later than 30 Myr (85 Myr for a star of 0.6
M$_{\odot}$). The spin-down between 70 and 110 Myr remains also
unambiguous, the KS-probability for these two bins being only 11\%.

\begin{figure}
\begin{minipage}{10cm}
\includegraphics[scale = 0.3, trim = 0 0 50 0, clip, angle=90]{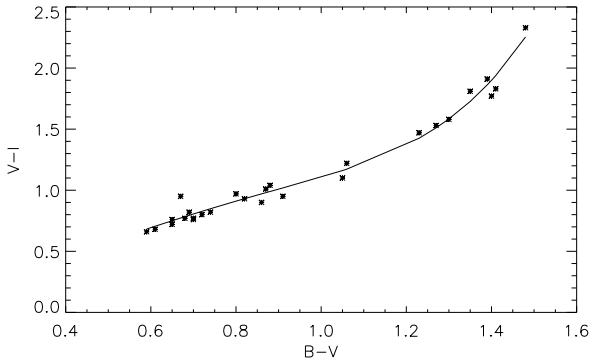}
\end{minipage}
\caption{Empirical relation to derive V$-$I  from B$-$V colors for the Tucana/Horologium members lacking
V$-$I measurements.\label{fig_fit}}
\end{figure}

\rm
\begin{figure}
\begin{minipage}{10cm}
\includegraphics[scale = 0.5, trim = 0 0 0 0, clip, angle=0]{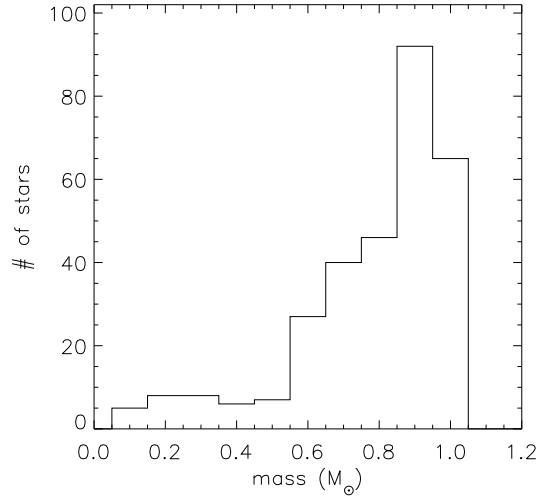}
\end{minipage}
\caption{Histogram of masses for the complete sample of periodic stars inferred from comparison with Baraffe et al.  (\cite{Baraffe98}) evolutionary tracks. \label{fig-histo-mass}}
\end{figure}

\begin{figure}
\begin{minipage}{18cm}
\includegraphics[scale = 0.5, trim = 0 0 0 0, clip, angle=0]{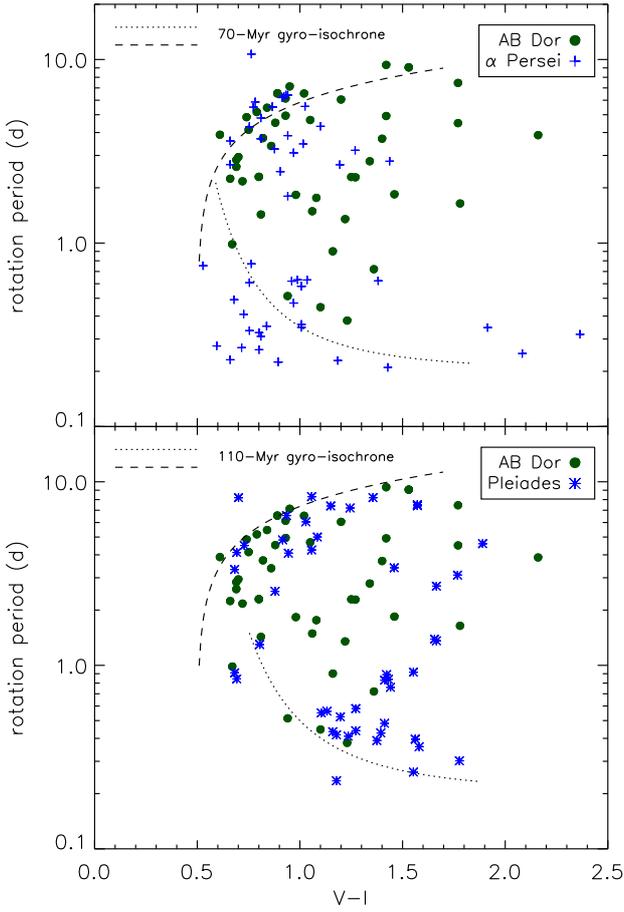}
\end{minipage}
\caption{Rotation period distribution of AB Dor members as compared to $\alpha$ Persei (top panel) and Pleiades members (bottom panel). Dashed and dotted lines represent the gyro-isochrones from Barnes (2003).\label{abdor-age}}
\end{figure}

\begin{figure}
\begin{minipage}{10cm}
\includegraphics[scale = 0.5, trim = 0 0 0 0, clip, angle=0]{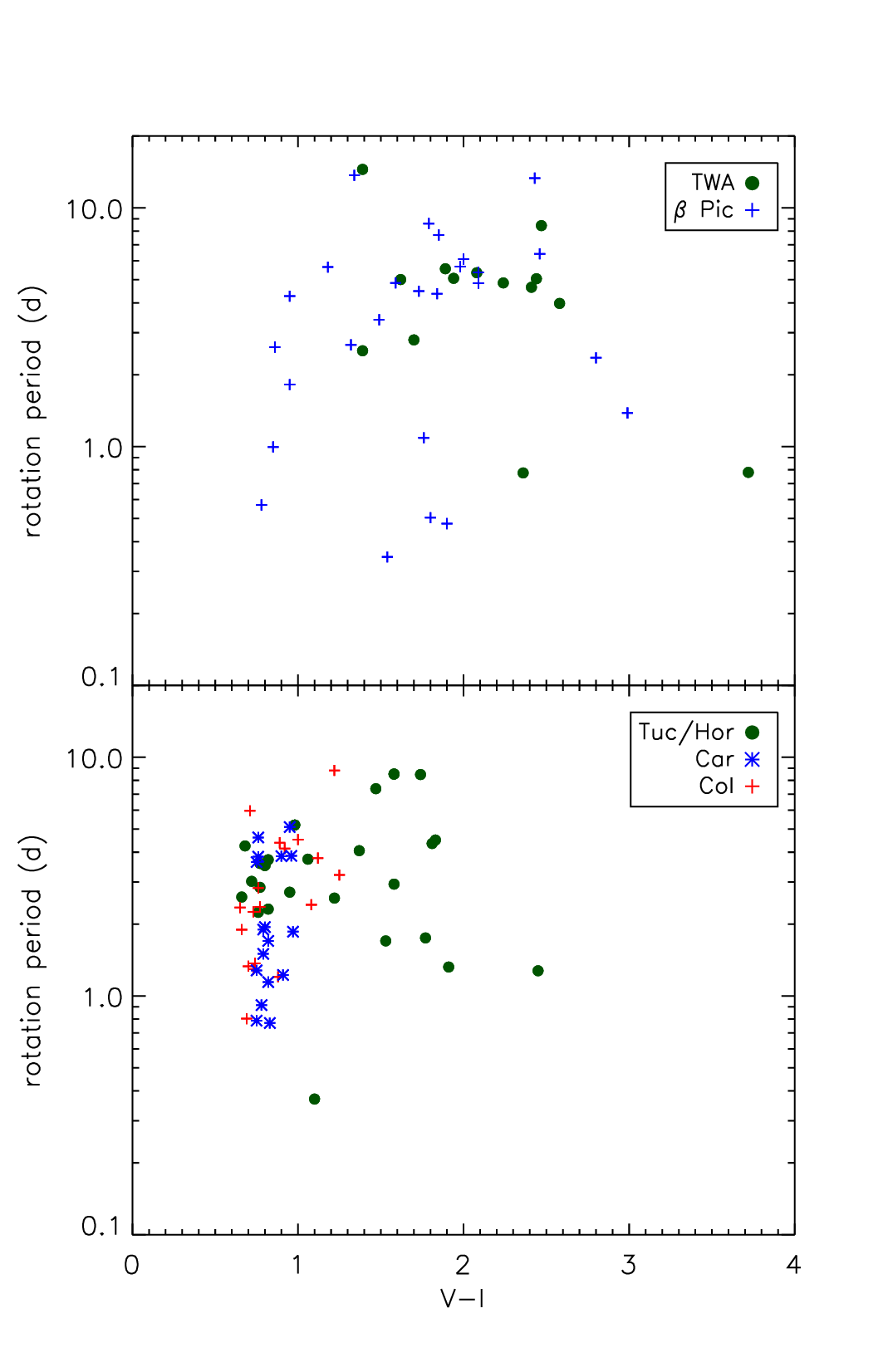}
\end{minipage}
\caption{Rotation period distribution of TWA and $\beta$ Pic (top panel), and Tucana/Horologium, Columba and Carina (bottom panel)\label{distri}}
\end{figure}

\indent
The most recent work on rotation and activity in PMS stars was carried out by Scholz et al. (\cite{Scholz07}) and based on data 
of four associations in the age range from $\sim$4 to $\sim$30 Myr ($\eta$ Chamaleontis, TW Hya, $\beta$ Pictoris and Tucana/Horologium). It is based 
on $v\sin i$ measurements and the stellar mass is inferred from the spectral type, differently than our comparison
with evolutionary tracks.   Their study shows a monotonic increase of $v\sin i$ (decrease of rotation period) until an age of about 30 Myr, which is the oldest age considered in their analysis.\\
Despite some difference with respect to the Scholz et al. (\cite{Scholz07})  analysis, we substantially confirm their results, however on a firmer basis thanks to the use of rotation periods instead of $v\sin i$ values and of a more numerous sample of associations and members.\\

\begin{figure}
\begin{minipage}{10cm}
\includegraphics[scale = 0.5, trim = 0 100 0 0, clip, angle=0]{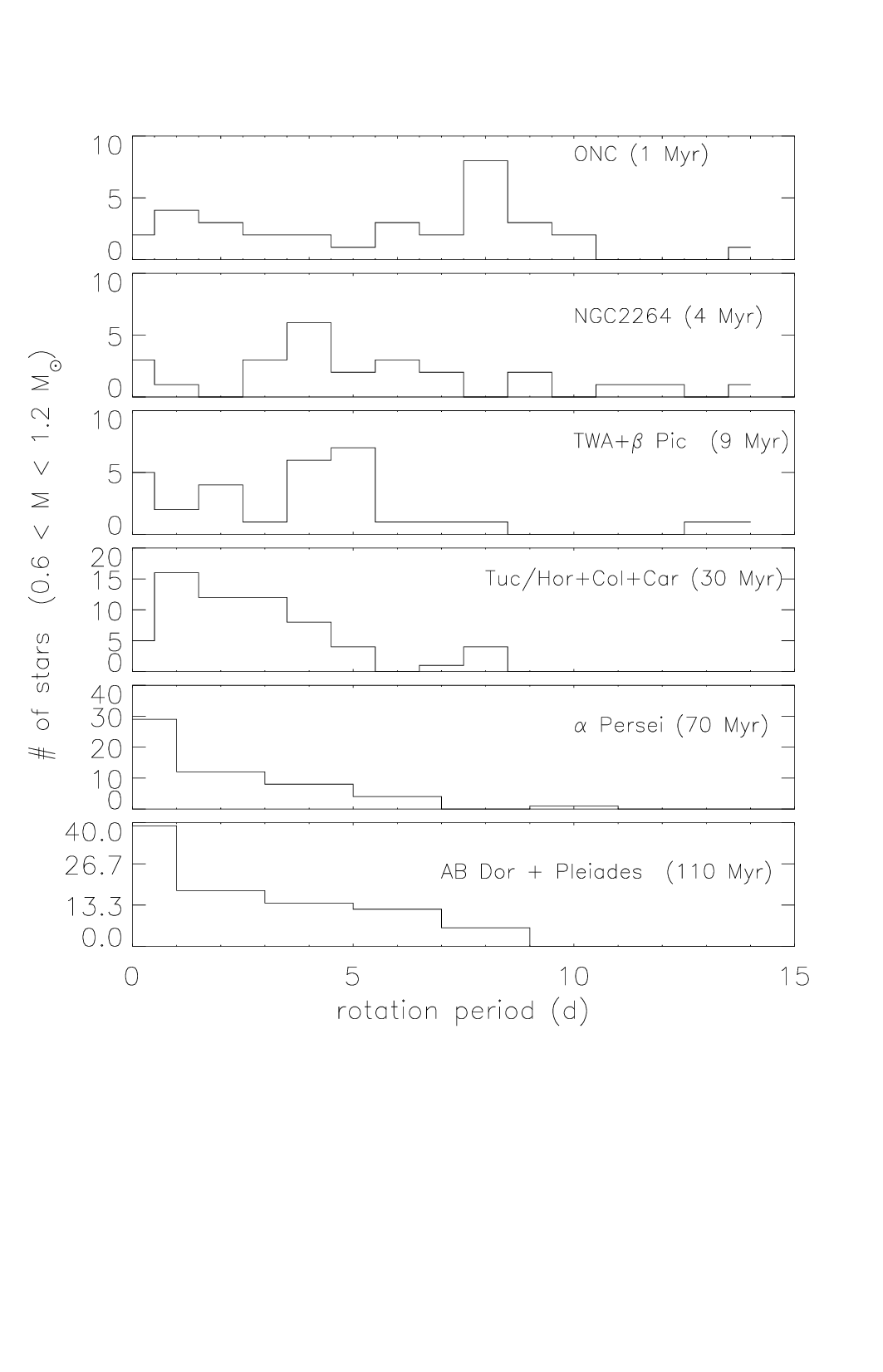}
\end{minipage}
\caption{Histograms of the rotation period distribution at the ages considered in the present study in the mass range 0.6$<$M/M$_{\odot}$$<$1.2. \label{histo}}
\end{figure}

\begin{figure*}
\begin{minipage}{18cm}
\centerline{
\psfig{file=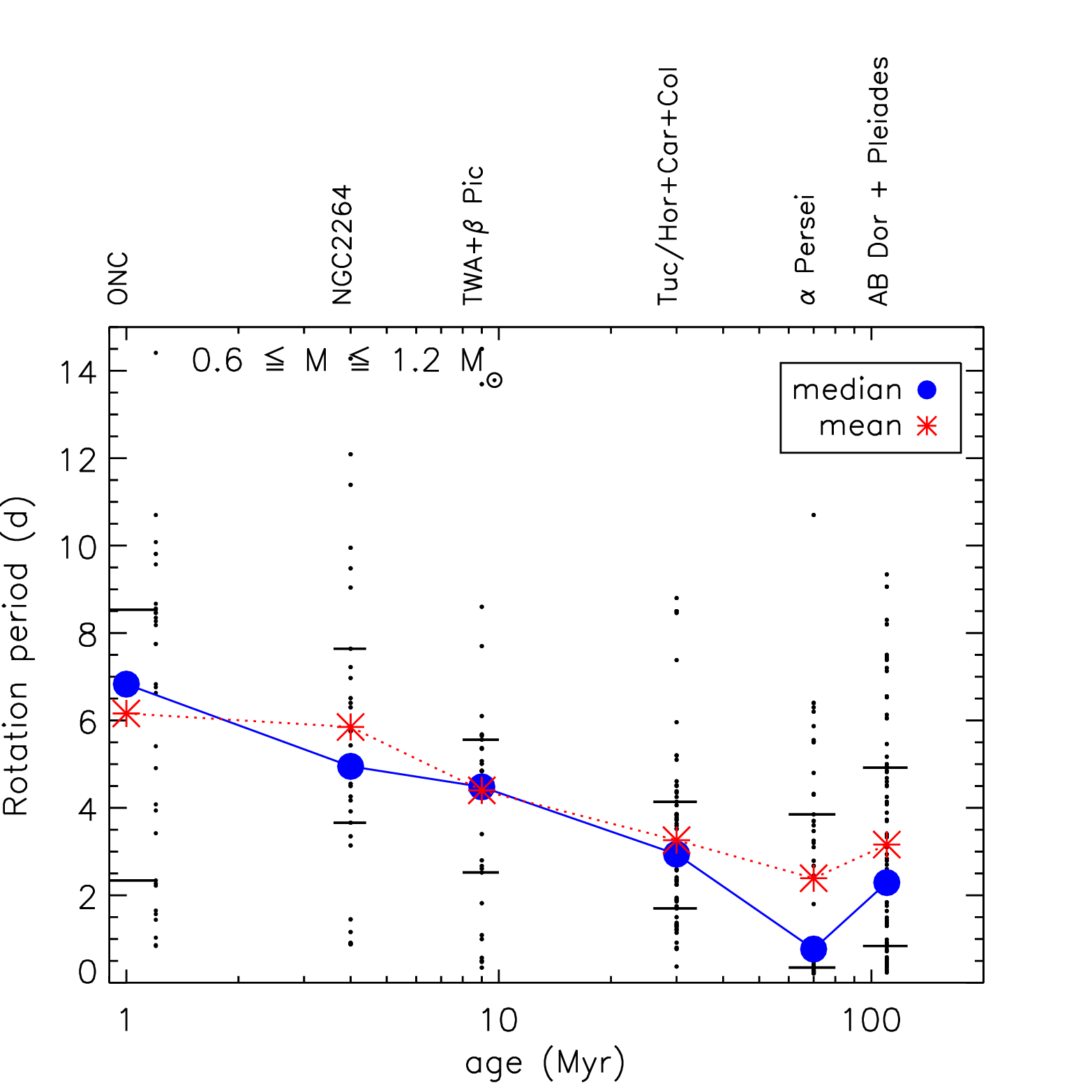,angle=0,width=9cm,height=10cm}
\psfig{file=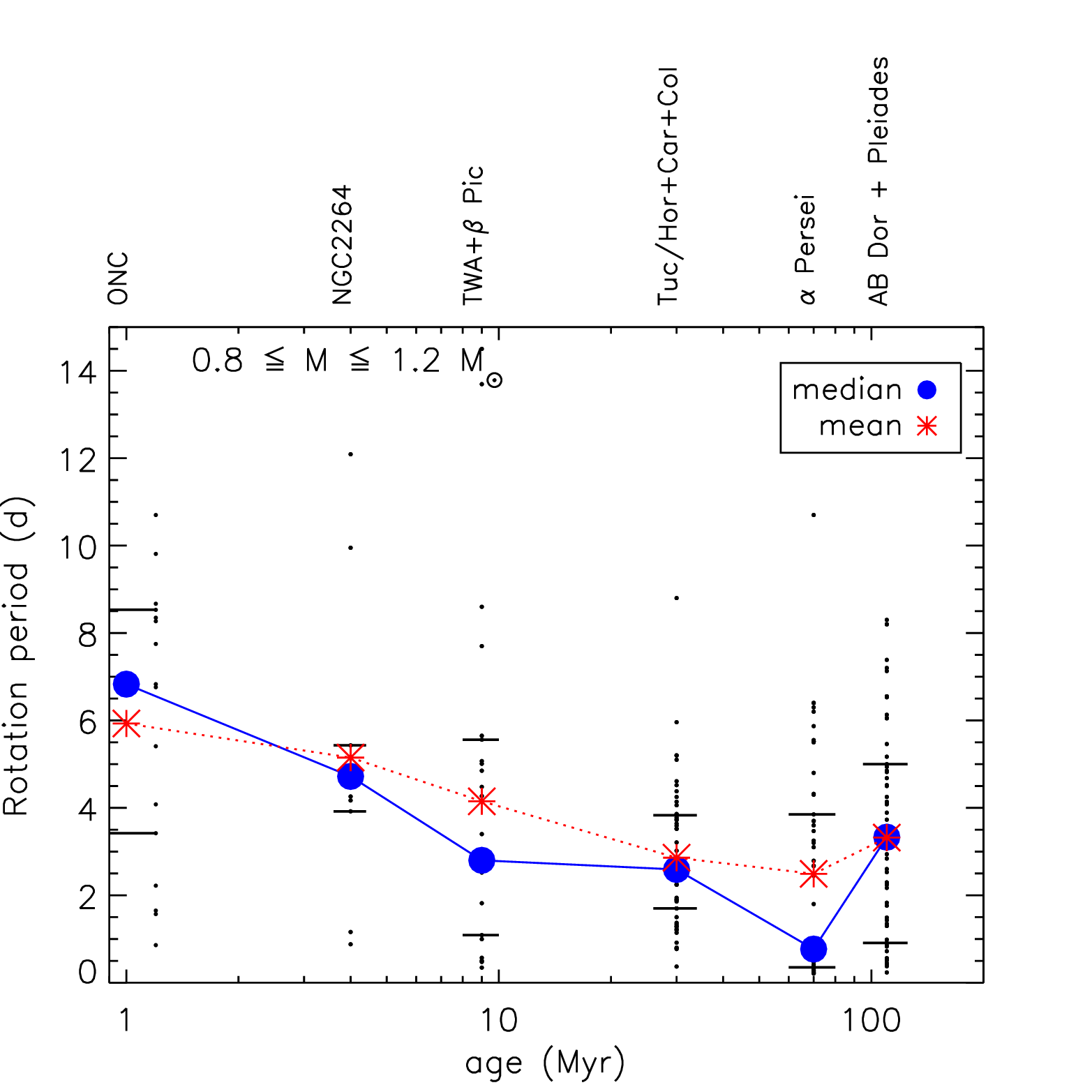,angle=0,width=9cm,height=10cm}
}
\end{minipage}
\caption{Rotation period evolution vs. time in the  0.6-1.2 (left panel) and in the 0.8-1.2 (right panel)  solar mass ranges. Small dots represent the individual rotation period measurements 
Bullets connected by solid lines are the median periods, whereas asterisks connected by dotted lines are mean periods.
Short horizontal lines represent the 25th and 75th percentiles of rotation period.
\label{evolution}}
\end{figure*}

\subsection{Rotation-photospheric activity connection}
The periodic light modulations shown by our stars arise from the presence of temperature inhomogeneities (i.e. starspots) on the stellar photosphere. 
Possibly, similar to the Sun, such inhomogeneities originate from photospheric magnetic fields whose total filling factor and distribution depend on the properties
of the dynamo mechanism operating in the stellar interior.
The amplitude of the light curve provides a lower limit on the amount of magnetic fields asymmetrically distributed along the stellar longitude,
which is in turn proportional to the total  magnetic field filling factor. As shown by Messina et al. (\cite{Messina01}, \cite{Messina03}), the upper bound of the light curve amplitude distribution  is observed to decrease with increasing rotation period, when the dynamo becomes less efficient. 
In Fig.\,\ref{ampiezza} we plot the  maximum \rm V-band peak-to-peak light curve amplitude vs. rotation period of stars in the associations under analysis.
 Such values represent the largest amplitude ever measured in all time sections in which the complete data time series of each target was divided, as explained in Sect.\,4.1. \rm
Bullets represent stars with masses M $>$ 0.6 M$_{\odot}$, whereas open triangles stars with masses  M $<$ 0.6 M$_{\odot}$.
Circled symbols are those stars whose measured $v\sin i$ values are inconsistent with the equatorial velocity v$_{\rm eq}$=2$\pi$R/P.
 In four  out of six associations, the upper bounds of the light curve amplitude distributions do not show any evident correlation with the rotation period. The only two exceptions are $\beta$ Pic and AB Dor associations, whose upper lightcurve amplitude bound decreases with increasing rotation period. To improve our statistics, we combined data from coeval clusters as well data from $\alpha$ Persei and Pleiades clusters, 
as earlier done for the rotation period distributions. In Fig.\,\ref{ampiezza_bis} we note that  the maximum light curve amplitude upper bound (solid line) 
begins to clearly correlate with the rotation periods starting from an age of $\sim$70 Myr.
Therefore, the photospheric activity behaviour of our young members of loose associations older than 70 Myr seem to be similar to those
observed in older stars where an $\alpha$-$\Omega$ dynamo operates. 
\rm Unfortunately, we have no data of very low mass (VLM) stars in our associations to study differences in magnetic activity with respect to higher-mass stars. 
Low-mass dM stars have a deep convection zone and stars with masses $<$0.3 M$_{\odot}$
(i.e. later that dM3/4) are expected to be fully convective and may gererate magnetic fields via a turbulent $\alpha^2$ dynamo.\\
\indent
 We note  evidence of a dependence of the light curve amplitude  also on  age. Considering stars of similar mass and rotation period, the light amplitudes are observed to be largest in the youngest associations TW Hya and $\beta$ Pic, where the variability is dominated by hot/cool spots and disk-accretion phenomena. Then, the amplitudes are observed to  decrease
until an age of $\sim$30 Myr, where we expect that the variability arises chiefly from cool spots. Then, the light curve amplitudes are observed to increase again reaching a maximum 
level at $\sim$110 Myr, which remains  constant untill an age of $\sim$230 Myr, as shown by a similar study among the members of the intermediate age open cluster M\,11 (Messina et al. \cite{Messina09})  
After this age, stars appear to show periodic light modulations of smaller amplitudes (see, e.g., Messina et al. \cite{Messina09}; Hartman et a.\,\cite{Hartman09}; Radick et al. \cite{Radick87}). \rm
In other words, stars with similar rotation, mass and internal
structure but with different ages, produce on average different light curves amplitude and, consequently, either different amount of magnetic fields or different
surface distribution of magnetic fields. To understand which unknown, yet age-dependent, parameters  play a role in the activity level is a challenge

\begin{table*}
\caption{List of target associations/clusters, age, number of stars used to make the statistics, median and mean rotation period, and significance level
that consecutive period distributions are drawn from the same distribution.\label{tab-per}}
\begin{tabular}{lc|cccc|cccc|l}
\hline
					&	& \multicolumn{4}{c|}{\bf  0.6-1.2 M$_{\odot}$} & \multicolumn{4}{c|}{\bf 0.8-1.2 M$_{\odot}$} &  \\
\hline
Target    & age  &  \#     &     P$_{\rm median}$     &     P$_{\rm mean}$ &  KS  & \#     &     P$_{\rm median}$     &     P$_{\rm mean}$ &  KS  &  \\
              & (Myr) & stars  &	(d)					& (d)			& & stars  &	(d)					& (d)			&\\
\hline
ONC       					&  1  	&   33   &   6.83   &    6.17 & 0.57 &  16    & 6.83   & 5.93 & 0.70  & 1 vs. 4 Myr\\
NGC\,2264     				&  4    &   26   &   5.43   &    6.19 & 0.59 &  10    & 4.72   & 5.15 & 0.63  & 4 vs. 9 Myr\\
TW Hya + $\beta$ Pic    	        &  9  	&   30   &   4.83   &    4.52 & 0.09 &  23    & 3.40   & 4.31 & 0.17  & 9 vs. 30 Myr\\
Tuc/Hor + Car + Col     		&  30 	&   62   &   2.85   &    3.22 & 0.36 &  53    & 2.60   & 2.87 & 0.45  & 30 vs. 70 Myr\\
$\alpha$ Persei    			&  70  	&   54   &   0.77   &    2.39 & 0.11 &  48    & 0.77   & 2.49 & 0.26  &70 vs. 110 Myr\\
AB\,Dor + Pleiades    		&  110 	&   89   &   2.29   &    3.16 & ...     & 48    & 3.33   & 3.32 & ...     & ... \\

\hline

\end{tabular}
\end{table*}

\section{Conclusions}

We have analysed the rotational properties of late-type members of six young associations within 100 pc 
and with age in the range 8-110 Myr. Our period search was based on  photometric time series taken from the ASAS 
catalog. Our analysis has allowed us to obtain the following results:

\begin{itemize}
\item We newly discovered the rotation period of 93 stars, confirmed the period already known from the 
literature of 41 stars, revised the period of 10 stars, and finally we retrieved from the literature the period of 21 additional stars.
After excluding all the stars rejected by Torres et al. (\cite{Torres08}) from the high-probability member list,
our final sample consists of  150 periodic confirmed members.

\item We determined for the first time the rotation periods of a number of confirmed members  in
$\beta$ Pictoris (10 stars), Tucana/Horologium (17 stars), Columba (15 stars), and Carina (16 stars),
as well we increased  the number of known periodic members of AB Doradus (+150\%) and  TWA (+15\%).

\item  A two-dimensional two-sided Kolmogorov-Smirnov test
applied to period-color distributions allowed us to confirm that the AB Dor
association is older than 70 Myr (as reported by Torres et al. \cite{Torres08}) and
likely coeval of the Pleiades cluster, i.e. 110 Myr old. 
\rm

\item
 Comparing the $v\sin i$ values from the literature with the calculated equatorial velocity v$_{\rm eq}$ = 2$\pi$R/P, where P and R are rotation period and stellar radius,
we found that the average inclination of the stellar rotation axis in each association is generally higher than expected from a random distribution of stellar axes. 

\rm
\item  About 91\% of our stars have mass in the  0.6 $<$ M $<$ 1.2 M$_{\odot}$ range. We could determine in this mass range 
the rotation period distributions and derive their median and mean rotation period.
Such values are the first available at ages of 8, 10, and 30 Myr in this mass range.
\rm 

\item 
In the 0.8 - 1.2 M$_{\odot}$ range, we found the most significant variations of
the rotation period distribution to be the spin-up between 9
and 30 Myr and the spin-down between 70 and 110 Myr. Two sided KS tests confirm the 
significativity of such variations. Between 1 and 9 Myr the
moderate spin-up is poorly supported by the KS test. The KS test on the
spin-up between 30 and 70 Myr does not allow us to draw definitive conclusions.
Our analysis is therefore consistent with
a considerable disk-locking before 9 Myr, followed by a moderate but
unambiguous spin-up from 9 to 30 Myr, consistent with stellar contraction
towards the ZAMS. Variations between 30 and 70 Myr are rather doubtful,
despite the median indicates a significant spin-up. 
The unambiguous spin-down from 70 to 110 Myr is consistent
with the fact that, starting from the 70 Myr age bin, all stars in the 0.8 -
1.2 M$_{\odot}$  mass range  have entered the MS phase and therefore the
angular momentum evolution is dominated by wind-braking.\\
\indent
The same considerations can be applied to the extended (0.6 - 1.2) M$_{\odot}$
range, despite all KS-probabilities are lower than in the (0.8 - 1.2)
M$_{\odot}$ range.  The moderate spin-up from 30 to 70 Myr is somewhat more significant,
which is likely to be due to a higher number of stars ending their
contraction towards the ZAMS at ages later than 30 Myr.

\item We found that the photospheric magnetic activity, as described by the upper bound of the light curve amplitude distribution,
correlates with the rotation period starting from an age of about 70 Myr.
Moreover, stars of similar mass and rotation show evidence of an age dependence of the activity level.
 It is highest at ages younger than 10 Myr, where hot spots and accretion processes are dominant. Then, the light curve amplitude is observed to be at  minimum level
at an age of 30 Myr, when only dynamo-generated cool spots are expected to dominate the variability. 
Then the level of spot activity again increases reaching its maximum level at the age of Pleiades. This behaviour suggests the existence of some age-dependent parameter which, apart from rotation and mass,  also plays a role in driving the  level of photospheric magnetic activity.
\end{itemize}

\rm

\begin{figure}
\begin{minipage}{10cm}
\centerline{
\psfig{file=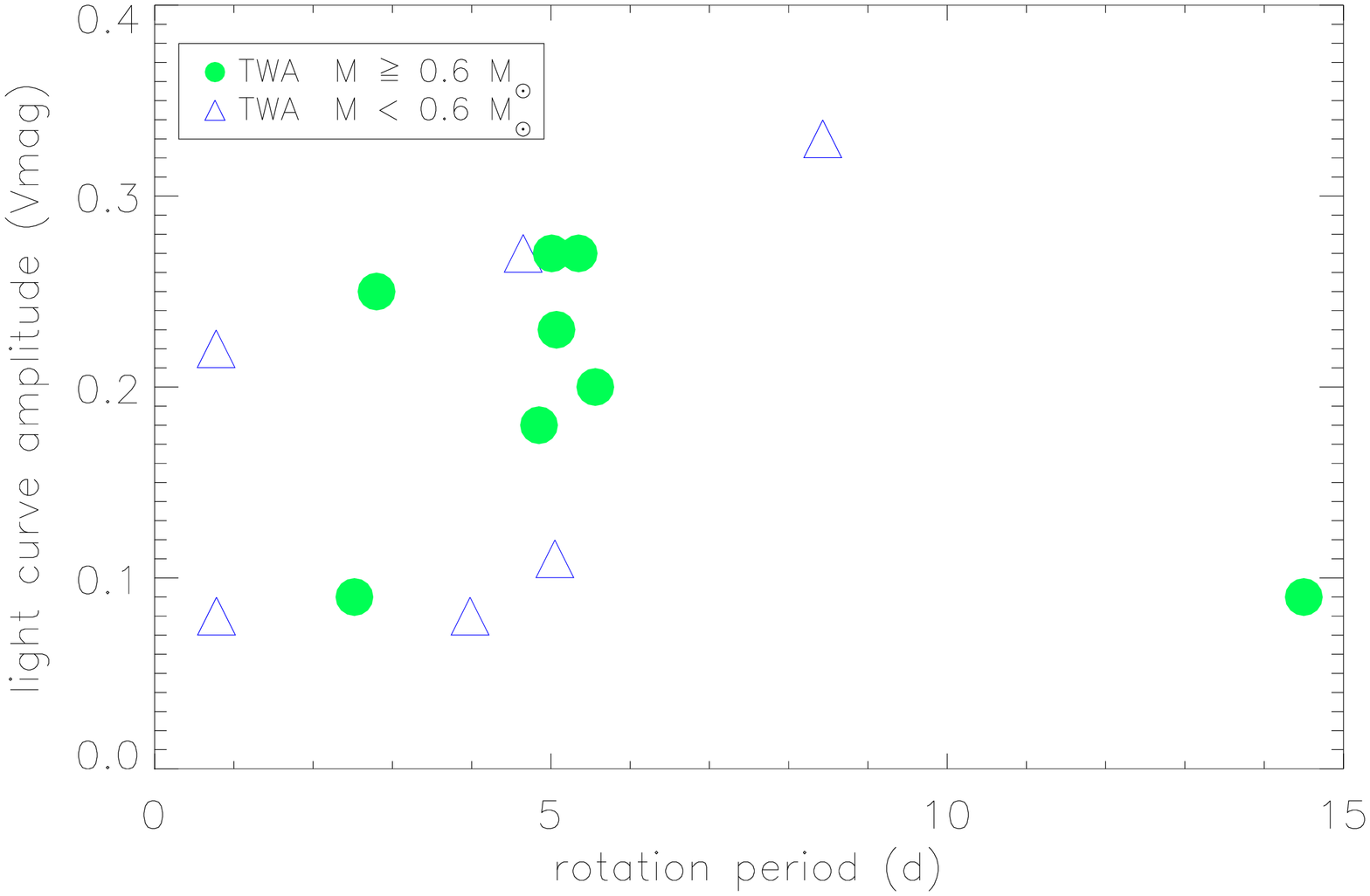,width=4.5cm,height=3.7cm,angle=90}
\psfig{file=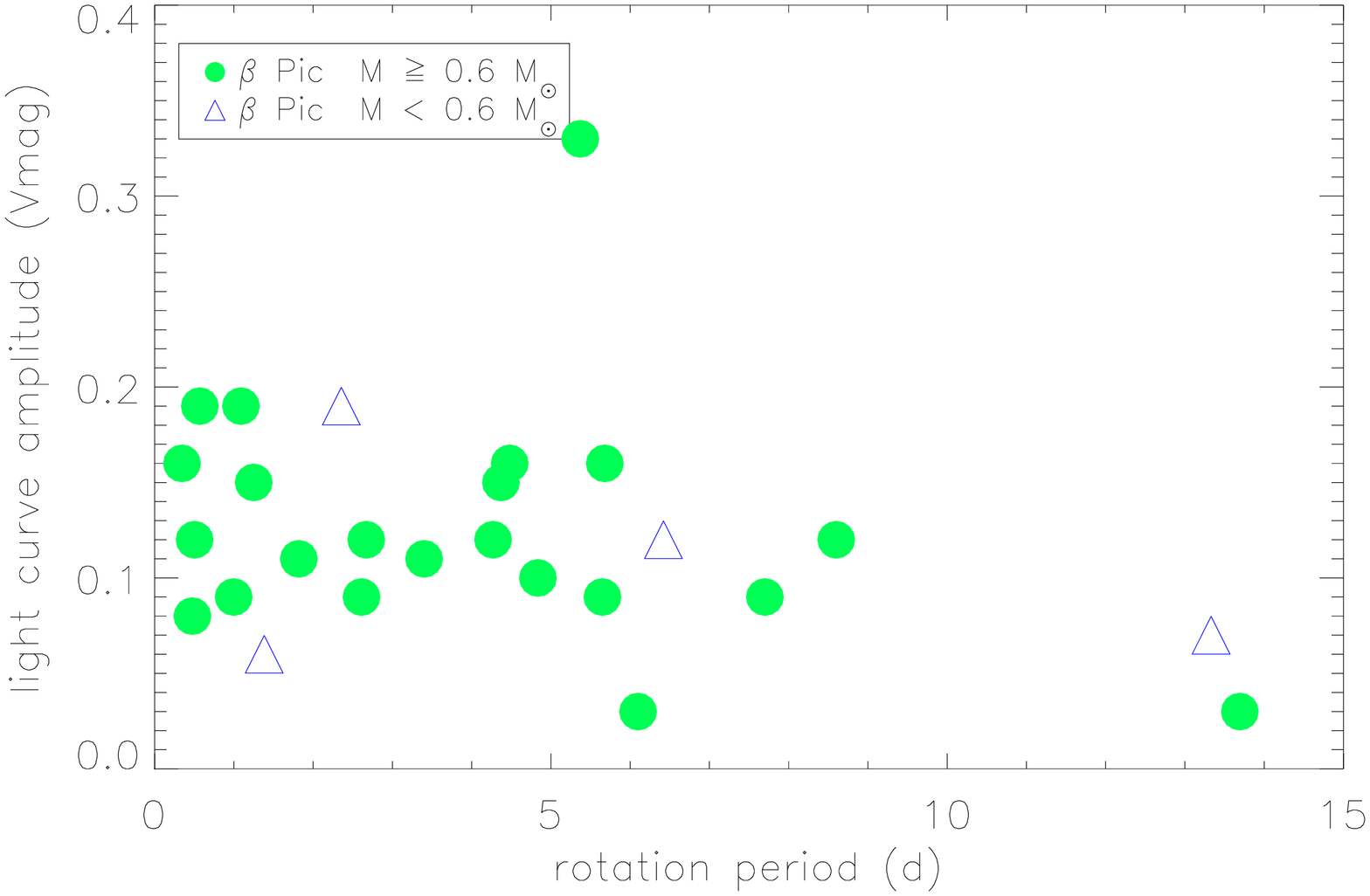,width=4.5cm,height=3.7cm,angle=90}
}
\end{minipage}

\begin{minipage}{10cm}
\centerline{
\psfig{file=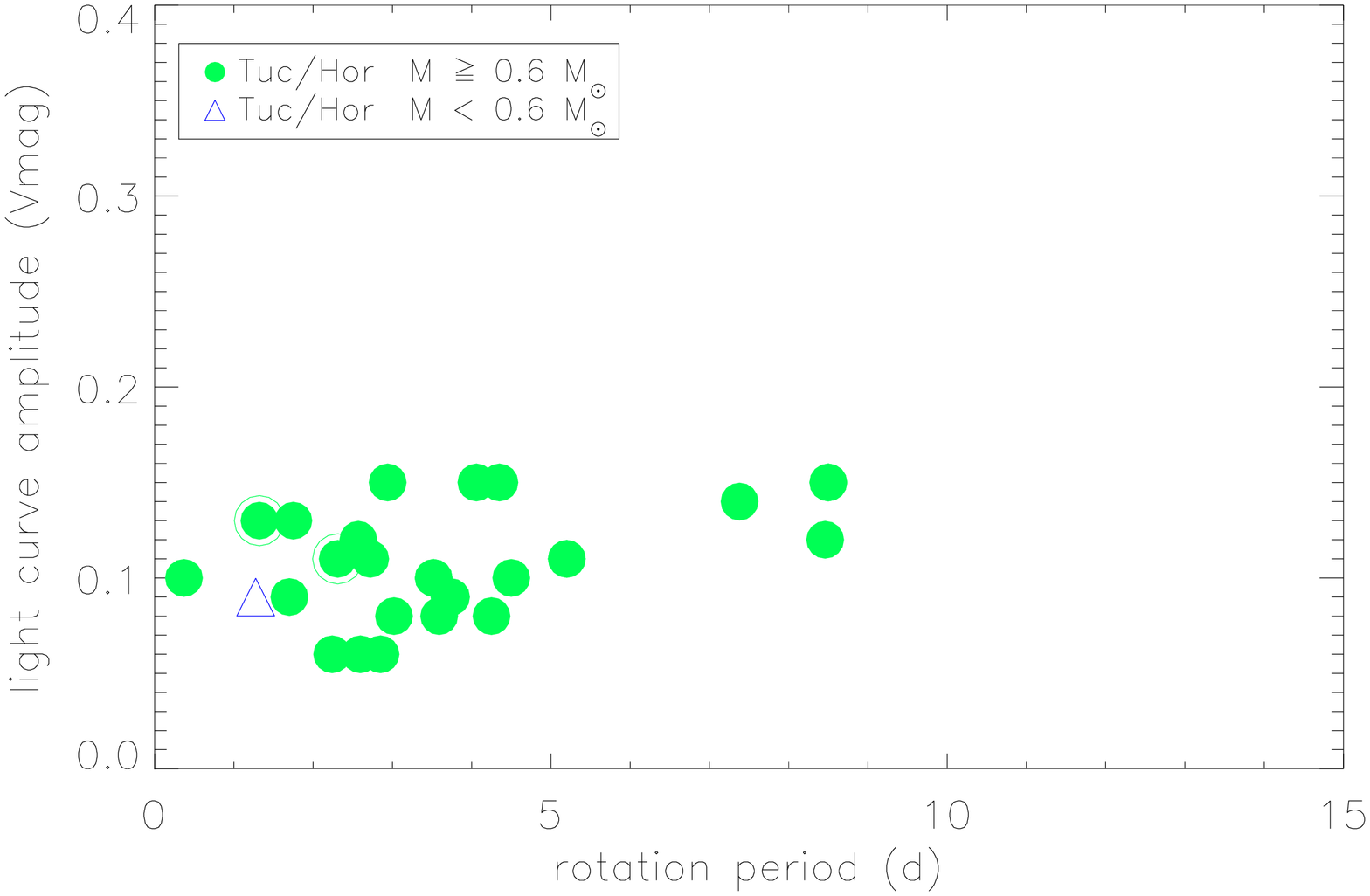,width=4.5cm,height=3.7cm,angle=90}
\psfig{file=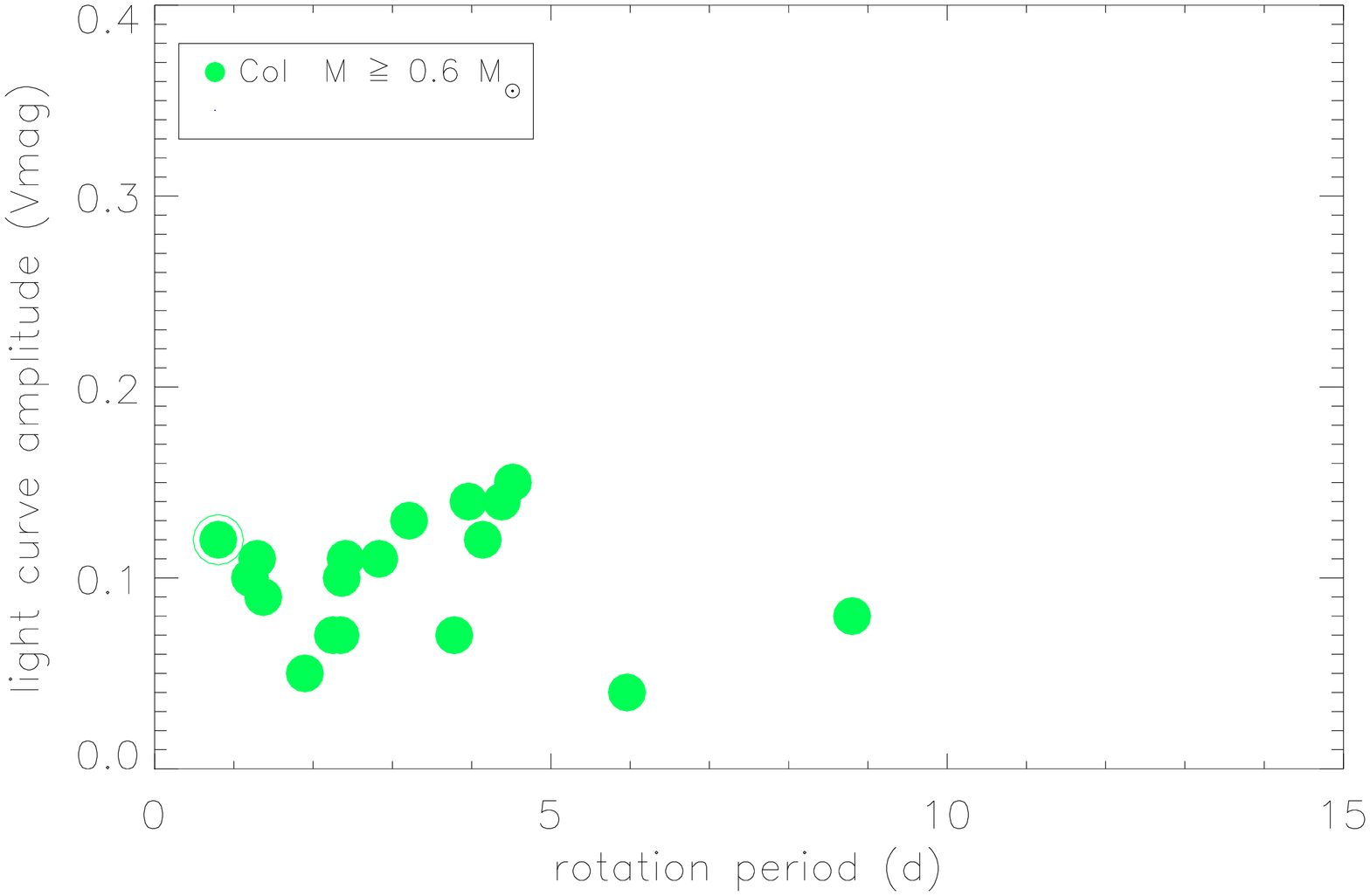,width=4.5cm,height=3.7cm,angle=90}
}
\end{minipage}
\begin{minipage}{10cm}
\centerline{
\psfig{file=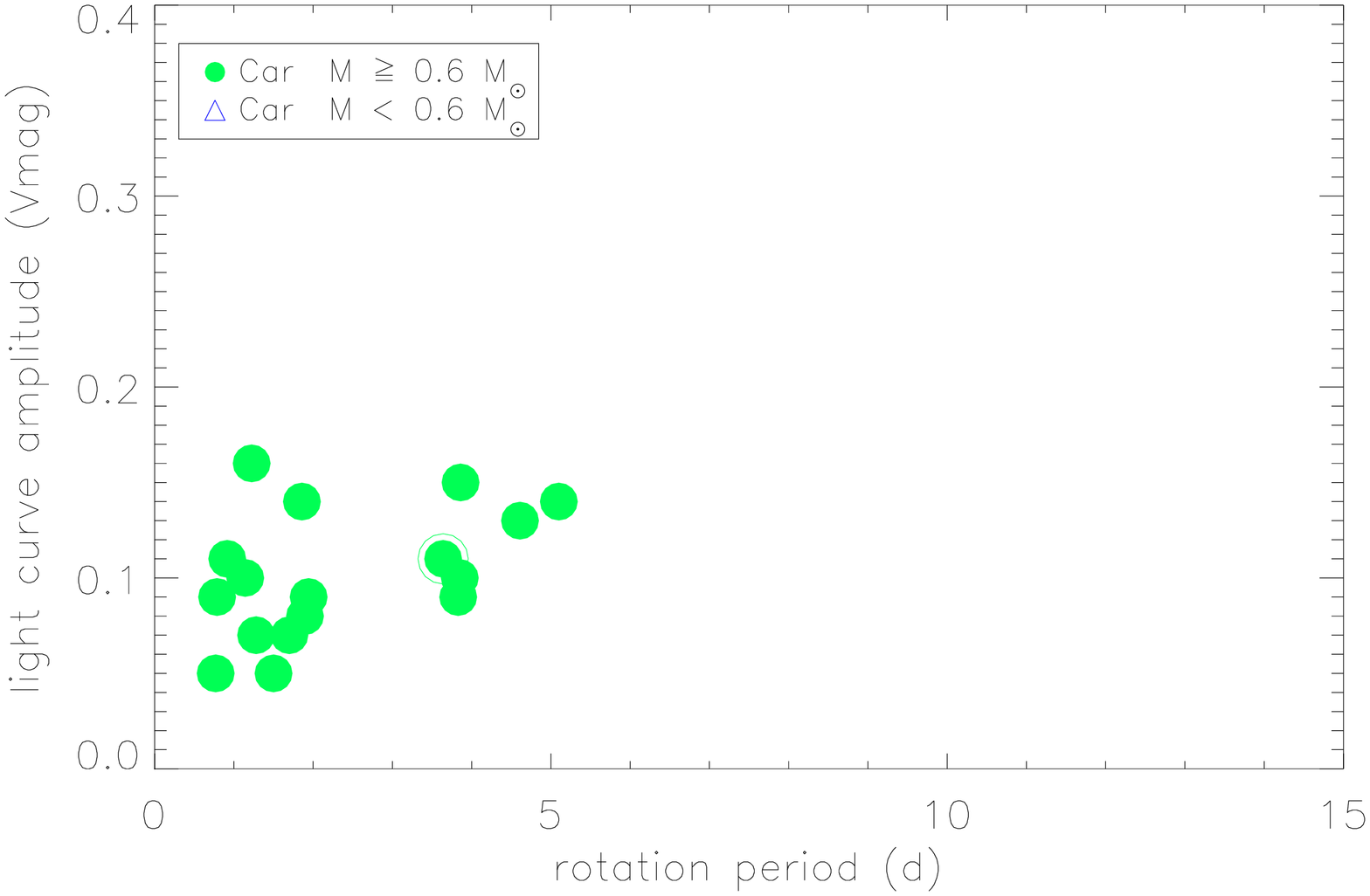,width=4.5cm,height=3.7cm,angle=90}
\psfig{file=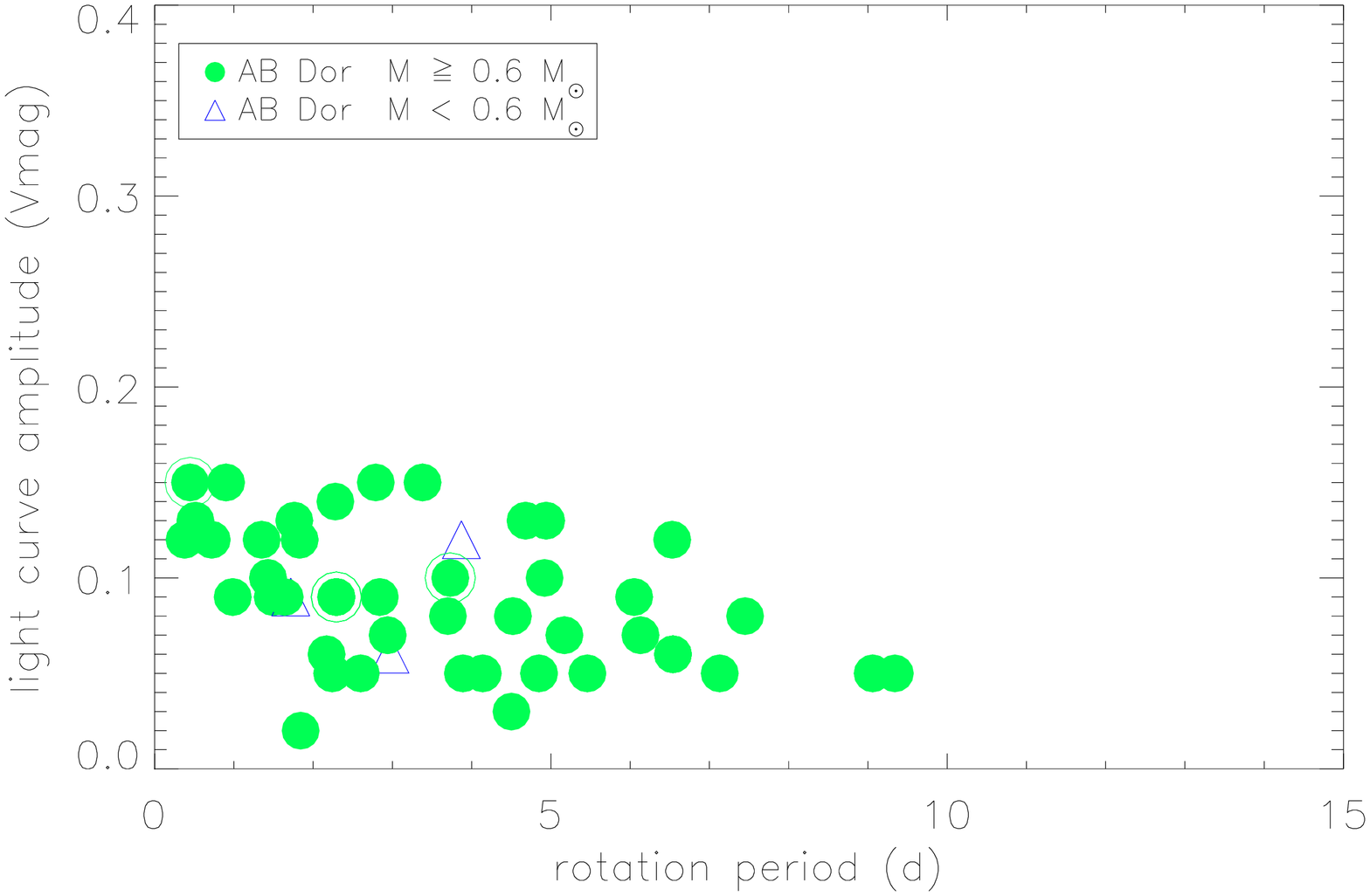,width=4.5cm,height=3.7cm,angle=90}
}
\end{minipage}
\caption{V-band peak-to-peak light curve amplitude vs. rotation period. Circled symbols are
stars whose $v\sin i$ is inconsistent with the equatorial velocity v$_{\rm eq}$=2$\pi$R/P.\label{ampiezza}}

\end{figure}

\begin{figure}
\begin{minipage}{10cm}
\centerline{
\psfig{file=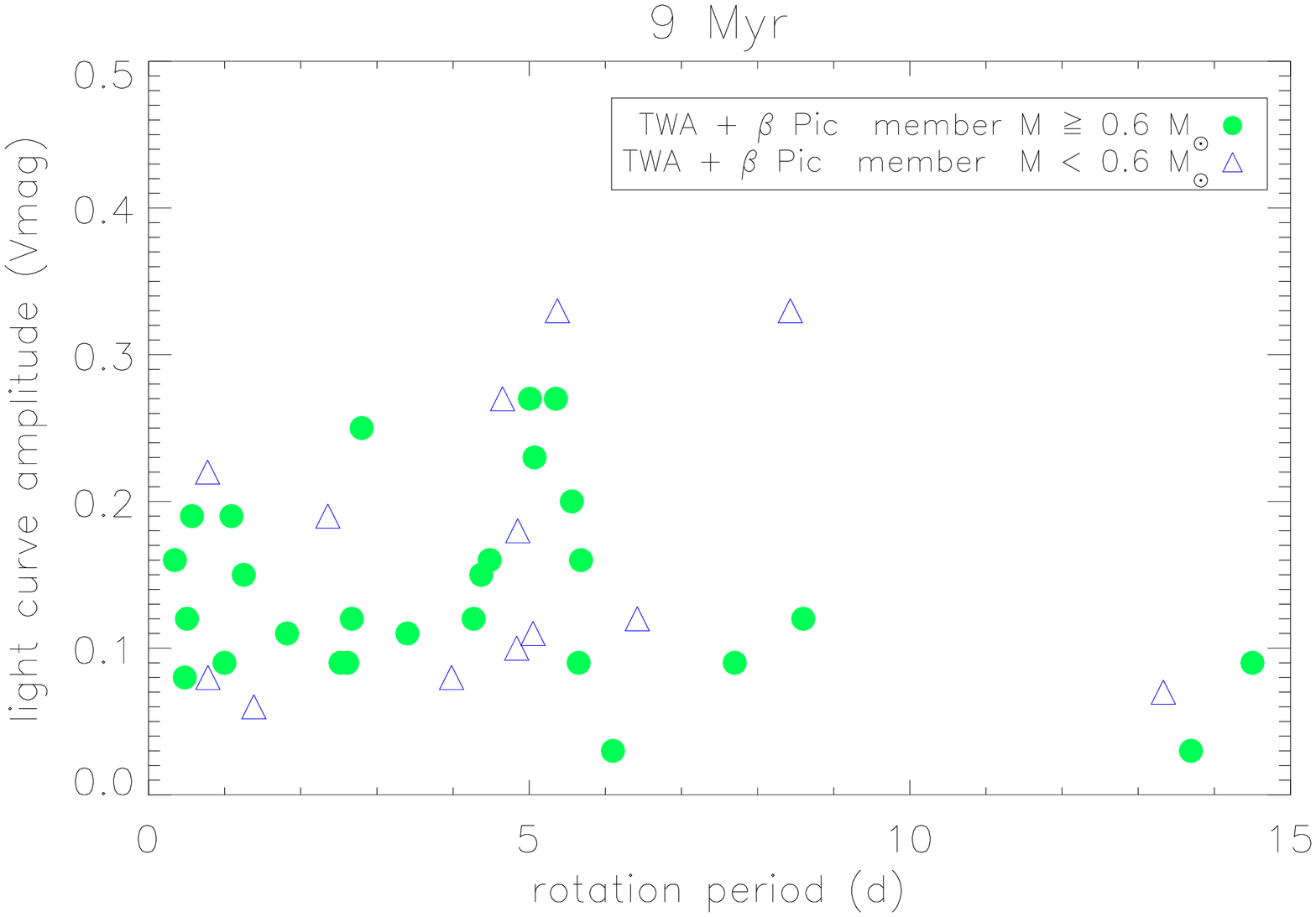,width=4.5cm,height=3.7cm,angle=90}
\psfig{file=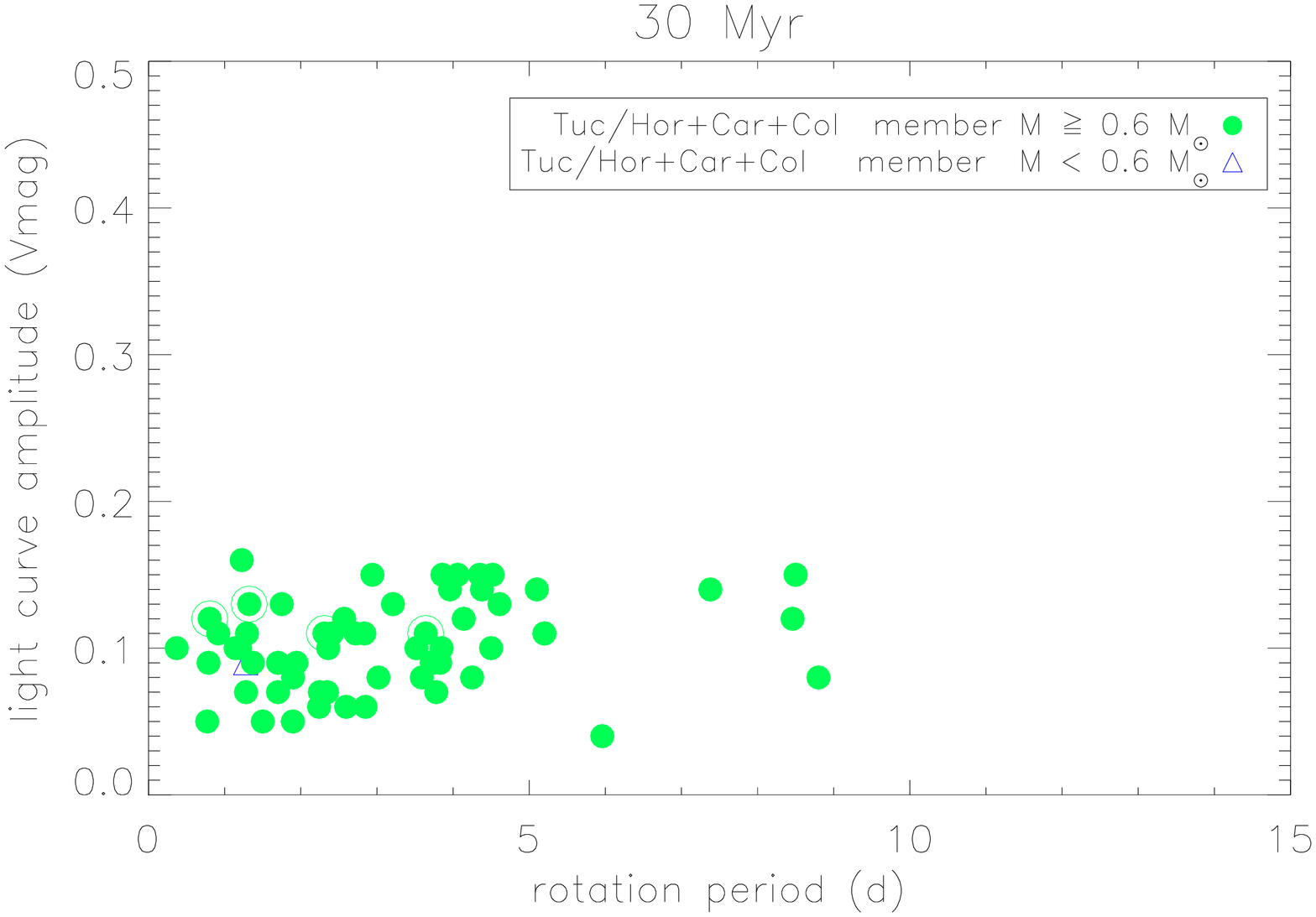,width=4.5cm,height=3.7cm,angle=90}
}
\end{minipage}

\begin{minipage}{10cm}
\centerline{
\psfig{file=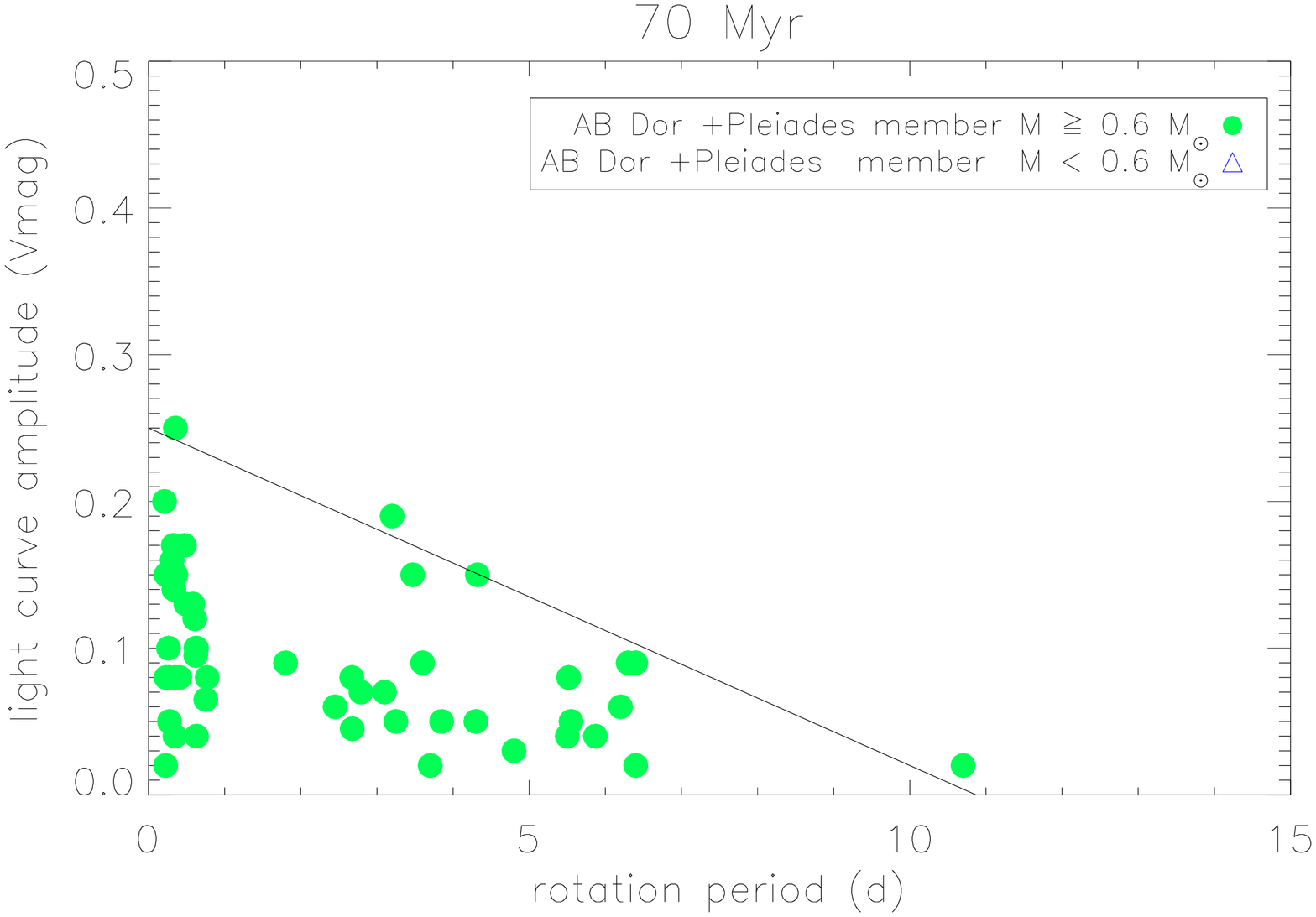,width=4.5cm,height=3.7cm,angle=90}
\psfig{file=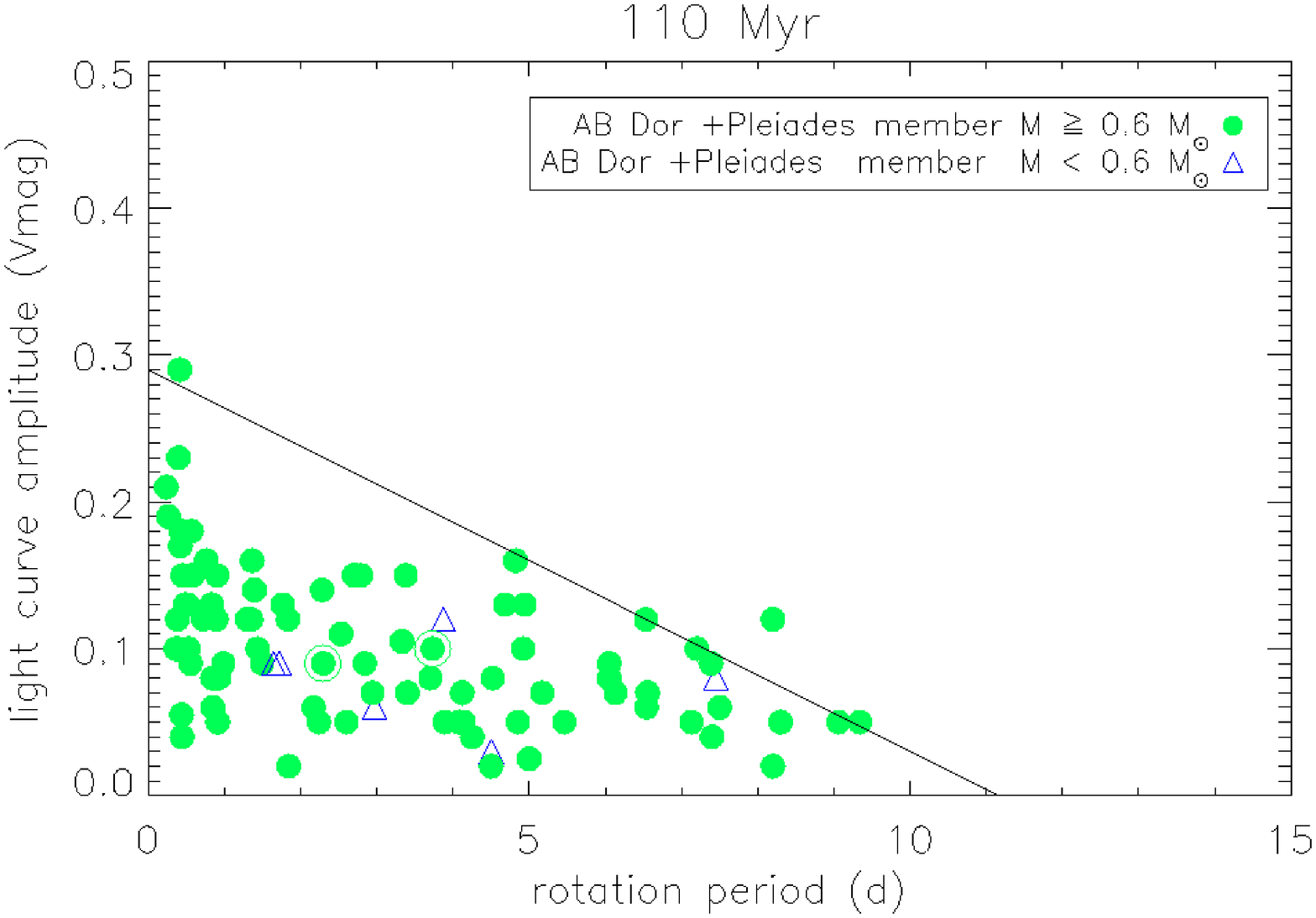,width=4.5cm,height=3.7cm,angle=90}
}
\end{minipage}
\caption{\label{ampiezza_bis} Similar to Fig.\,\ref{ampiezza} but with data of coeval stars plotted together. Solid lines mark the decreasing upper bound
of the amplitude distribution.}

\end{figure}

\appendix

\section{Individuals}
\subsection{TW Hydrae}

\indent

 \bf TW\,Hya: \rm the most recent and extensive study of the variability of TW\,Hya  was carried out by Rucinsky et al. (2008). 
It is based on the MOST satellite high-precision photometry and the contemporaneous  ASAS photometry.
Their analysis reveals a number of periodicities probably arising  from different mechanisms either operating
 in the photosphere (starspot activity) or related to accretion processes from its disk (veiling, accretion). 
The highest power peak detected in our periodograms is at P=6.86d which is probably not related to rotation.
For the aim of the present study, we adopt the $\Delta$V light curve amplitude  and the P=2.80d period
determined by Lawson \& Crause (\cite{Lawson05}), which is in good agreement with earlier determinations by Koen \& Eyer (\cite{Koen02})
and by Alencar \& Bathala (\cite{Alencar02}).

{\bf TWA\,2}: is a visual binary  whose components are separated by 0.55 arcsec and differ by $\Delta$V=1.0 mag.

{\bf TWA\,3AB}: is a visual binary  whose components are separated by 1.5 arcsec, differ by $\Delta$V=0.9 mag, and are not resolved by ASAS system. We could  determine the rotation period of neither he A nor the B component.

\bf TWA\,4:  \rm  is a quadruple system formed by two pairs of SB at 0.8 arcsec and with $\Delta$V=0.5.
TWA\,4A is a SB1 with  P$_{orb}$=262d (Torres et al. \cite{Torres95}); TWA\,4B is a  SB2 with  P$_{orb}$=315d (Torres et al. \cite{Torres95}).
Aa, Ba, and Bb have similar brightness. One of the components is a clasiscal T Tauri star (CTTS).
Our periodogram analysis found a high confidence level rotation period of P=14.29d   in 9 out of the 14 analysed time intervals in which we divided the complete magnitude series. This period is different from the \rm rotation period P=2.521d 
reported by Koen \& Eyer (\cite{Koen02}) and derived from the Hipparcos photometry. Since in both cases the system components are not resolved 
and have similar brightness, they may be both responsible for the observed photometric variability.
 Assuming that P=2.521d is the correct period of TWA4A, which is consistent with the $v\sin i$ = 8.9 kms$^{-1}$ and the derived stellar radius, the P=14.29d period may be attributed to TWA4B, which may have dominated the observed variability during the ASAS observations. \rm

\bf TWA\,5: \rm  is a triple system consisting of a binary with period 5.94 yr (Konopacky \cite{Konopacky07}), $\Delta$mag$\sim$0.1 (JHK)
plus a brown dwarf at 2 arcsec (Lowrance et al. \cite{Lowrance99}). Our analysis shows the highest power peak at P=0.776d, which is in good agreement with  the value of P=0.77d reported in the  ACVS.

{\bf TWA\,6}:  we adopt the rotation period from the literature (Lawson \& Crause \cite{Lawson05}), since the available ASAS photometry did not provide any periodicity.
TWA\,6 was not included in the high-probability member list by Torres et al. (\cite{Torres08}).
The $v\sin i$  is from Skelly et al. (\cite{skelly08}).

{\bf TWA\,7}: we found a rotation period in good agreement with the literature value (Lawson \& Crause \cite{Lawson05}).

{\bf TWA\,8AB}: is a  visual binary  whose components have a separation of 13.2 arcsec, a magnitude difference $\Delta$V=3.2,
and are not resolved by ASAS system. Our rotation period agrees with that reported by Lawson \& Crause (\cite{Lawson05})
for the brighter component TWA8A. For TWA\,8B we adopt the rotation period of Lawson \& Crause  (\cite{Lawson05}).

{\bf TWA\,9AB}: is a visual binary  whose components have a separation of  5.8 arcsec and $\Delta$V=2.7.
For the brighter component TWA9A our period agrees with that reported by Lawson \& Crause  (\cite{Lawson05}), whereas it disagrees
with the P=0.83d in the  ACVS. The ASAS photometry could not resolve the fainter TWA9B whose rotation period is taken from Lawson \& Crause  (\cite{Lawson05}). The $v\sin i$ of both components are from Scholz et al. (\cite{Scholz07}).

{\bf TWA\,10} and {\bf TWA\,12}: our rotation periods agree with those reported by Lawson \& Crause  (\cite{Lawson05}).
TWA\,12 was not included in the high-probability member list by Torres et al. (\cite{Torres08}).

\bf TWA\,13AB:  \rm  is visual binary whose components are separated by 5.1 arcsec and have $\Delta$V=0.5. The ASAS systems does not resolve the components of the binary. However, the observed variability is likely due to both components. In fact, our analysis revealed two rotation periods:  P=5.56d which is in agreement with the determination by Lawson \& Crause  (\cite{Lawson05}), and P=5.35d which is also in good 
agreement with the rotation period that Lawson \& Crause  (\cite{Lawson05}) report for TWA\,13B.

\bf TWA\,14:  \rm  our analysis did not reveal any significant periodicity. We adopt the rotation period determined by Lawson \& Crause  (\cite{Lawson05}).
It was not included in the high-probability member list by Torres et al. (\cite{Torres08}).

{\bf TWA\,15AB}: is visual binary rejected as member of the TW Hya group by Torres et al. (\cite{Torres08}). Since our analysis did not reveal the rotation period, we adopt the rotation periods from  Lawson \& Crause  (\cite{Lawson05}). The $v\sin i$ are from Scholz et al. (\cite{Scholz07}).

{\bf TWA\,16}: is a binary with components separated by 0.7 arcsec. Its membership to the TW Hya group must be confirmed yet according to Torres et al. (\cite{Torres08}).

\bf TWA\,17:  \rm  our analysis did not reveal any significant periodicity. We adopt the rotation period of Lawson \& Crause  (\cite{Lawson05}) and the 
$v\sin i$ of  Reid (\cite{Reid03}). It was not included in the high-probability member list by Torres et al. (\cite{Torres08}).

\bf TWA\,18: \rm  our analysis did not reveal any significant periodicity. We adopt the  P=1.11d period and light curve amplitude of  \rm  Lawson \& Krause  (\cite{Lawson05}).  The v$\sin i$ is from Scholz et al. (\cite{Scholz07}).  It was not included in the high-probability member list by Torres et al. (\cite{Torres08}).

{\bf TWA\,19AB}:  is a visual  binary  with a separation between the components of  37 arcsec and $\Delta$V=2.8. It was rejected as member by Torres et al. (\cite{Torres08}).
The ASAS photometry is probably contaminated by the light contribution by the fainter component. We could not determine any periodicity.

{\bf TWA\,20}: We could not determine any significant periodicity.

\bf TWA\,21 \rm and \bf TWA\,24: \rm were rejected as members by Torres et al. (\cite{Torres08}). 

\bf TWA\,23: \rm were not included in  the high-probability member list by Torres et al. (\cite{Torres08}) due to lack of complete
kinematic data.

\subsection{$\beta$ Pictoris }

\indent

\bf TYC\,1186\,706\,1: \rm this star was recently discovered to be a $\beta$ Pic member by Lepine \& Simon (\cite{Lepine09}). 
Our period is in good agreement with the literature value (Norton et al. \cite{Norton07}).

\bf HIP\,12545: \rm is classified as SB1 by  Torres et al. (\cite{Torres08}). The period P = 0.5569d reported in the ACVS  is not confirmed
by our period analysis  (see the periodogram in Fig.\,\ref{bpic_fig1}). The $v\sin i$=9.5 km s$^{-1}$ is from Scholz et al. (\cite{Scholz07}).

{\bf GJ 3305:} is a close binary (Kasper et al.~\cite{kasper07}) and has a probable further 
wide companion (51 Eri=HIP 21547, F0V; Feigelson et al. 2006) at an angular distance of  66 arcsec. 
Our analysis did not reveal any significant periodicity. In the following analysis we adopt the period
P=6.1d of Feigelson et al. (\cite{Feigelson06}). 
The $v \sin i$ is from Scholz et al.~(\cite{Scholz07}).

{\bf HIP 23200:} our period is in good agreement with both the literature (Alekseev \cite{Alekseev96}) and the ACVS values. 
The $v \sin i$ is from Favata et al.~(\cite{favata95}).

\bf HIP 23418: \rm  is a triple system consisting of a SB2 binary (the primary is a M3V) with orbital period P$_{orb}$=11.96d and 
eccentricity  e=0.323, and of a visual companion with $\Delta$V$\sim$ 1 at an angular distance of 
0.7-1.0 arcsec (Delfosse et al. \cite{Delfosse99}). The 
$v \sin i$ is from Scholz et al.~(\cite{Scholz07}).

\bf BD-21 1074: \rm  is a triple system consisting of a M2V (V=10.29) star and a binary companion at separation of  23 arcsec
 (M3, V=11.61).  We detected a periodicity of P=13.3d in 4 out of 14 time intervals as well as when the whole time series was analysed.
Unfortunately, no $v\sin i$ value is at present available to check the consistency between $v\sin i$ and equatorial velocity.\rm

\bf HIP 76629: \rm  is a triple system consisting of a binary system (the primary is a K0V) and of a fainter visual component 
(at a separation of 10.0 arcsec and $\Delta$V=6.8).
The RV trend by Gunther \& Esposito (\cite{Gunther07a}) and the Hipparcos acceleration imply the presence of an additional closer companion with period of
several years.  Our rotation period agrees with the period found by Cutispoto (\cite{Cutispoto98a}).

\bf TWA\,22: \rm  it was an unconfirmed member of TWA according to Torres et al. (\cite{Torres08}) 
because of incomplete kinematic information.
The revised  kinematic data by Teixeira et al. (\cite{Teixeira09}) indicate its membership
to $\beta$ Pic moving group, that we adopt here.

{\bf HIP\,84586:} is an SB2 system (G5 IV + K0 IV)  with an additional visual companion HD\,155555C, 5.6 mag fainter in V at
33 arcsec. Our rotation period is in agreement with the literature values, e.g.  by Cutispoto (\cite{Cutispoto98b}), Pasquini et
al. (\cite{Pasquini91}), Strassmeier \& Rice (\cite{Strassmeier00}) and with the orbital period. The system is tidally-locked.

\bf TYC 8742\,2065\,1: \rm is classified as SB2 (the primary is a K0IV) by Torres et al. (\cite{Torres06}). It has a very close optical
companion (Torres et al. \cite{Torres08}) of similar brightness ($\Delta$V=0.2).   We detected two
periodicities of P=2.61d and P=1.61d of comparable power in almost each of about half of the selected time intervals. 
Probably, they represent the rotation period of either the SB2 or the optical companion, respectively. We are not in the position
to assign to the SB2 system the corresponding rotation period. However, due the relative small difference, an incorrect assignment
it would not imply significant difference in the results of the rotation period distribution analysis. \rm

\bf HIP88399: \rm  has a brighter F6V companion at a distance of 6.5 arcsec, which falls within the aperture radius used to extract the ASAS photometry.

\bf V4046 Sgr:  \rm is an SB2 (the primary is a K5) tidally-locked binary
accreting object (CTTS). Our rotation period agrees  with the P=2.42d found by  Quast et al. (\cite{Quast00}). The
$v\sin i$ is from Quast et al. (\cite{Quast00}).

\bf UCAC2 18035440: \rm  GSC7396-0759 is a probable comoving system at 169 arcsec (Torres et al. \cite{Torres08}). We note in its
 periodogram  numerous peaks at confidence level higher than 99\%. However, we could not identify which peak is related to stellar rotation.

\bf TYC 9077\,2489\,1: \rm is a triple system consisting of a binary (the primary is a K5Ve)   with a separation of 0.18 arcsec (= 5.2 AU) 
and  $\Delta$K=2.3 (Chauvin et al. \cite{Chauvin09}), and a wide companion (HIP 92024, A7V) at a 70 arcsec distance. This distance is
sufficiently large to allow the ASAS system to observe the only close visual binary system. 

\bf HIP\,92680: \rm our rotation period agrees with the period reported by  Innis et al. (\cite{Innis07}).


\bf TYC 6878\,0195\,1: \rm is reported by Torres et al. (\cite{Torres06}) as visual binary system (the brighter component is a K4Ve) 
whose components have a separation of  1.10 arcsec and  $\Delta$V=3.50.

\bf HIP 102141: \rm is a binary system formed by two very similar M dwarfs ($\Delta$V$<$0.05)
at a separation of about 3 arcsec. Our analysis did not reveal the rotation period.

\bf HIP 102409: \rm  our rotation period is in good agreement with literature values of, e.g., Hebb et al.  (\cite{Hebb07}), Rodon\`o et al.  (\cite{Rodono86}).

\bf TYC 6349\,0200\,1: \rm is reported by Neuhauser et al. (\cite{Neuhauser03}) as a visual binary (the brighter star is a K6Ve) with a separation of 2.2 arcsec and
 $\Delta$K=1.6.

\bf TYC 2211\,1309\,1: \rm this star was recently discovered to be a $\beta$ Pic member by Lepine \& Simon (\cite{Lepine09}). 
Our period is in good agreement with the literature value (Norton et al. \cite{Norton07}).

\bf HIP 112312: \rm its  companion TX PsA  is sufficiently distant  (36 arcsec) to not contribute
to the observed variability.

\bf HIP 11437: \rm  is at declination +30 and no ASAS photometry exists. The rotation period is taken from Norton et al. (\cite{Norton07}).
It has a companion at an angular distance of 22 arcsec  and $\Delta$V=2.4.
The $v \sin i$ is from Cutispoto et al. (\cite{Cutispoto00})

\bf HIP10679: \rm has a nearby  F5V companion (HIP\,10680) at a distance of 13.8 arcsec.


\subsection{Tucana/Horologium}

\indent

{\bf HIP 490:} the $v \sin i$ is from Cutispoto et al. (\cite{Cutispoto02}).

\bf HIP 1910: \rm is a binary system, its components having a separation of 0.7 arcsec and  $\Delta$V=2.4.

\bf HIP\,2729:  \rm our analysis did not find any significant periodicity. We adopt the P=0.37d rotation period found by Koen \& Eyer (\cite{Koen02}) using the Hipparcos photometry. 

\bf TYC\,8852\,0264\,1: \rm  our analysis revealed a rotation period P=4.8d that we detected with high confidence level in 8 out of 11 time intervals and which is about half the value reported in the ACVS. The latter has no power peak in our periodogram. Nonetheless, the 4.8d period, together with computed stellar radius, give an equatorial velocity inconsistent with the  
three independent measurements of $v\sin i$ which agree within errors ($30.22\pm1.09$ km s$^{-1}$ Scholz et al. \cite{Scholz07}; $32.7\pm1.8$ km s$^{-1}$, Torres et al.
 \cite{Torres06}; 32 km s$^{-1}$ De La Reza \& Pinzon \cite{Delareza04}). In any case it is not included when determining  the rotation period distribution
because it was  rejected as member of Tuc/Hor by Torres et al. (\cite{Torres08}).

\bf HIP\,6485: \rm our rotation period agrees with the period found by Koen \& Eyer (\cite{Koen02}) using the Hipparcos photometry.

\bf HIP 9141: \rm is a binary system, its components having a separation of 0.15 arcsec, and $\Delta$H=0.1 (Biller et al. \cite{Biller07}).
The $v \sin i$ is from Nordstrom et al. (\cite{Nordstrom04}).

\bf HIP\,9892: \rm is a SB1 system with long period but no other orbital elements determined (Gunther \& Esposito \cite{Gunther07a}). Our rotation period is about half the period from the literature P=4.3215d (Koen \& Eyer \cite{Koen02}). As shown in the online Fig.\,\ref{tuc_fig1}, the latter period is not present at all in our periodogram.

\bf TYC\,8489\,1155\,1: \rm is the wide companion of the F7V star HIP 9902. The components are sufficiently separated
to be resolved by the ASAS photometry. However, we could not determine the rotation period.

\bf TYC\,8497\,0995\,1: \rm our analysis revealed a rotation period of P=7.38d which is about half the period reported in the ACVS. The latter is not present
in our periodogram as shown in the online Fig.\,\ref{tuc_fig2}

\bf AF\,Hor \rm and \bf TYC\,8491\,0656\,1 \rm are an M2V and K6V stars, respectively, separated by about 22 arcsec. They are too close to be separated by the ASAS photometry.  Our analysis revealed a period of P=1.275d which likely represents the rotation period of the brighter star (TYC\,8491\,0656\,1)
dominating the observed variability. \rm 

\bf TYC\,7026\,0325\,1: \rm in the  ACVS is reported with a rotation period of P=2.2613d.  However, our analysis revealed the P=8.48d to be the only significant periodicity.

{\bf TYC\,8060\,1673\,1:} the $v \sin i$ is from Viana Almeida et al. (\cite{Viana09}).

\bf HIP 16853\rm: is a binary system with astrometric orbit (P=200 d). 

\bf HIP\,21632:  \rm we adopt  the rotation period from Koen \& Eyer (\cite{Koen02}), since the analysis of ASAS photometry did not reveal any significant periodicity.

\bf TYC\,5907\,1244\,1:  \rm it is classified as  SB2 by Torres et al. (\cite{Torres08}). However, neither its orbital elements nor the spectral type are known. In the  ACVS  this star is reported  with a rotation period of P=1.10473d, whereas our analysis gives the P=5.21d as the most significant periodicity.

{\bf TYC\,7600\,0516\,1: } it is rejected as member of Tuc/Hor association by Torres et al.~(\cite{Torres08}).  Our period is in good agreement with the P=2.47d  reported by Cutispoto et al. (\cite{Cutispoto99}).  \rm

\bf TYC\,7065\,0879\,1: \rm it is a close visual binary with very similar components according to Torres et al.~(\cite{Torres06}). It is rejected as member of Tuc/Hor association by Torres et al.~(\cite{Torres08})

\bf HIP\,105404: \rm is the only eclipsing binary star in our sample. It is a triple system composed by a very short-period eclipsing binary and a long-period SB (P=3y, Guenther et al. \cite{Gunther07a}). 
 Since the Lomb-Scargle periodogram is best suited to search for single sinusoidal flux variations, it fails to detect the right orbital period. However, using as first guess the rotation period available from the literature and the phase dispersion minimization the ASAS extended time series
has allowed us to improve the estimation of the orbital period. It is rejected as member of Tuc/Hor association by Torres et al.~(\cite{Torres08})

\bf TYC 9344 0293 1: \rm is binary system whose components have a separation of  0.2 arcsec (Torres et al. \cite{Torres08}). The rotation period, 
that we detected with high confidence level in 8 out of 12 time intervals, when combined
with the stellar radius, gives an equatorial velocity inconsistent with the only available measurements of  $v\sin i$=61km s$^{-1}$ (Torres et al. \cite{Torres06}). We can guess that the vsini value was overestimated due to line blending, since the binary components 
have a very small separation.

\bf TYC\,9529\,0340\,1: \rm we have  information neither on spectral type nor on binarity. The 
P=2.31d rotation period, that we detected with high confidence level in 11 out of 13 time intervals,  in combination with the computed stellar radius provides an equatorial velocity inconsistent with the only available measurement of  $v\sin i$=73.90 km s$^{-1}$ (Torres et al. \cite{Torres06}).

\bf HIP 116748AB: \rm is a binary system whose components have a separation of 5.3 arcsec and $\Delta$V=1.3. 


\subsection{Columba }

\indent

\bf TYC 8047\,0232\,1: \rm  has a brown dwarf companion at 3.2 arcsec (Chauvin et al. \cite{Chauvin03}). 

\bf HIP 16413: \rm  is a binary system whose components have a separation of 0.90 arcsec and  $\Delta$V=1.90.

\bf TYC 5882\,1169\,1: \rm although previously classified as member of the Tuc/Hor association, it is proposed by Torres et al. (\cite{Torres08}) as high-probability member of Columba association. The  $v \sin i$ is from Scholz et al. (\cite{Scholz07}).

{\bf TYC 6457\,2731\,1:} the $v \sin i$ is from Viana Almeida et al. (\cite{Viana09}).

\bf TYC\,7584\,1630\,1: \rm the rotation period we found is half the value reported in the ACVS, which is absent in our periodogram (see online Fig.\,\ref{col_fig1}).

\bf TYC 8077\,0657\,1: \rm is a binary system whose components have a separation of 21.3 arcsec and $\Delta$V=3.20 (Torres et al. \cite{Torres06}).
The ASAS photometry does not resolve this star from UCAC2\,11686780.

\bf HIP 25709: \rm is classified as  SB2 by Torres et al. (\cite{Torres08}), but no orbital elements  were derived. 


 \bf TYC\,7617\,0549\,1: \rm in the  ACVS is reported with a rotation period of P=1.3038d. However,  in the periodogram no significant power peak is 
evident other than that at P=4.1395d.

\bf TYC\,7100\,2112\,1: \rm the rotation period, that we detected with high confidence level in 11 out of 15 time intervals, when combined
 with the stellar radius, gives an equatorial velocity inconsistent with the only available measurements of  $v\sin i$=170km s$^{-1}$ (Torres et al. \cite{Torres06}).

\bf AG Lep: \rm the rotation period is retrieved from the literature (Messina et al. \cite{Messina01}), no ASAS photometry being available.
The $v \sin i$ is from Cutispoto et al. (\cite{Cutispoto99}).

\subsection{Carina }

\indent

\bf  HIP 30034: \rm is a member of Tuc/Hor according to Zuckerman \& Song (\cite{Zuckerman04}). In the present study is considered as
member of Carina according to Torres et al. (\cite{Torres08}).
It has a brown dwarf companion at wide separation  (Chauvin et al. \cite{Chauvin05}).

\bf  HIP\,32235 \rm and \bf HIP\,33737: \rm  are  member of Tuc/Hor according to Zuckerman \& Song (\cite{Zuckerman04}). In the present study are considered as
member of Carina according to Torres et al. (\cite{Torres08}).

\bf TYC 8559\,1016\,1: \rm is a visual binary whose components have a  separation of  5.8 arcsec and  $\Delta$V=3.0 (Torres et al. \cite{Torres06}).

\bf  TYC\,8929\,0927\,1: \rm we find two peaks of comparable power and very high confidence level. However, only the P=0.73d period combined the computed stellar radius gives an equatorial velocity consistent with the measured $v\sin i$.

\bf TYC 8569\,3597\,1: \rm is a SB2, with orbital period P$_{orb}$=24.06d (Torres et al. \cite{Torres08}).

\bf TYC\,8160\,0958\,1: \rm the rotation period we found is half the period reported in the ACVS, which is absent in our periodogram.

\bf TYC\,8586\,2431\,1: \rm the rotation period, that we detected with high confidence level in 9 out of 14 time intervals, when combined with the stellar radius computed from PMS tracks, gives an equatorial velocity inconsistent with the only available measurements of  $v\sin i$=128km s$^{-1}$ (Torres et al. \cite{Torres06}). The luminosity class  IV assigned to this star may indicate that the star has already left the MS and its radius has already started increasing. However, a stellar radius larger than about 6R$_{\odot}$ would reconcile $v\sin i$ and v$_{eq}$.

\subsection{AB Doradus}

\indent

{\bf HIP 5191}: is a visual binary whose components are separated by  23 arcsec. 


{\bf TYC 8042\,1050\,1}: is a visual binary whose components are separated by 21.7 arcsec.

{\bf HIP 10272}:  is a binary whose components are separated by  1.8 arcsec and have $\Delta$V=1.6.

{\bf HIP 13027}:  is a binary whose components are separated by 3.6 arcsec and with $\Delta$V=0.8.

{\bf HIP 14809}: \rm together with HIP\,14807 represent a binary whose components are separated by  33.2 arcsec and with  $\Delta$V=2.0.

{\bf HIP 22738AB}: is a binary whose components are separated by  7.8 arcsec and have $\Delta$V=0.9. The ASAS system could not resolve the components.

{\bf HIP 25647}: is a quadruple system (very low mass star at about 1 AU + close pair of M dwarfs at 9 arcsec). 
Our rotation period agrees with periods from the literature, e.g., Cutispoto \& Rodon\`o (\cite{Cutispoto88}). The
$v \sin i$ is from Wichmann et al. (\cite{Wichmann03}).

{\bf TYC\,7059\,1111\,1:} \rm our period is in good agreement with the literature value (Cutispoto et al. \cite{Cutispoto03}).
However, the derived v$_{\rm eq}$=2$\pi$R/P is iconsistent with two independent measurements of $v\sin i$
which agree within errors ($41.5\pm3.5$ km s$^{-1}$, Torres et al. \cite{Torres06}; 40 km s$^{-1}$, Tagliaferri et al. \cite{Tagliaferri94}).

{\bf HIP 26373 - HIP 26369} is a binary system whose components are separated by  18.3 arcsec and with $\Delta$V=1.9. 
Our rotation period agrees with that from Cutispoto et al. (\cite{Cutispoto99}). The ASAS photometry does not resolve the component of this K0+K6 system.

{\bf HIP 27727}: is a possible binary system, but to be confirmed yet. Our rotation period agrees with the period from the literature (Strassmeier et al. \cite{Strassmeier97}).


\bf TYC\,7598\,1488\,1:  \rm  our period is in good  agreement with the literature value reported by Cutispoto et al. (\cite{Cutispoto01}). However, when it is combined with the stellar radius, the derived equatorial velocity is inconsistent with two independent measurements (Torres et al. \cite{Torres06}; Tagliaferri et al. \cite{Tagliaferri94}) which give the same value of $v\sin i$ (55 km s$^{-1}$). A larger stellar radius of about 3-4 R$_{\odot}$, as reported by Cutispoto (\cite{Cutispoto98b}) and based on a photometric distance d$>$86 pc, would partly solve the disagreement of  v$_{eq}$ with $v\sin i$. \rm

{\bf HIP 30314}: is a binary whose components are separated by 16.2 arcsec.

{\bf HIP 31711}: is a binary whose components are separated by  0.8 arcsec and with  $\Delta$V=2.3. Since our periodogram shows several peaks of similar power, we adopt the period from Cutispoto et al. (\cite{Cutispoto99}) which gives a reasonably smooth light curve.

\bf TYC\,1355\,214\,1: \rm  our rotation period agrees with that determined by Norton et al. (\cite{Norton07}).

{\bf HIP 36108}: is a binary whose components are separated by  1.20 arcsec and with $\Delta$V=1.2

{\bf HIP 36349}: is a binary whose components are separated by  0.3 arcsec and with $\Delta$V=1.9. 
 Our period agrees with  the period P=1.642d of Koen \& Eyer (\cite{Koen02}) derived from the Hipparcos photometry.\rm


{\bf HIP 76768}: is a binary whose components are separated by  0.9 arcsec and with $\Delta$V=1.3.
Our period P=3.70d differs from the period P=0.336d reported in the  ACVS, which is absent in our periodogram.
The latter period, if correct, together with the  $v\sin i$=8.0 kms$^{-1}$ 
would imply a pole-on orientation of the rotation axis. 

\bf BD-13 4687: \rm  the  $v\sin i$=140.0 km s$^{-1}$ is from da Silva et al. (\cite{daSilva09}).

{\bf HIP 93375}: is a binary whose components are separated by  11.2 arcsec (Torres et al. 2008). The 
$v \sin i$ is from Nordstrom et al. (\cite{Nordstrom04}).

{\bf HIP 94235}: the $v \sin i$ is from Nordstrom et al. (\cite{Nordstrom04}).

{\bf  TYC 1090-0543}: our analysis gives a rotation period in agreement with the period  P=2.2374d found by Norton et al. (\cite{Norton07}).

\bf HIP\,106231:  \rm our period analysis revealed the most significant periodicity to be P=0.42312, which is in agreement with the known LO Peg rotation period 
from either ACVS and literature (e.g., Jeffries et al. 1994). The $v \sin i$ is from Barnes et al. ( \cite{Barnes05}).

{\bf HIP 113597}: is a binary whose components are separated by  1.8 arcsec and with $\Delta$V=0.6.

{\bf HIP 114530}: is a binary whose components are separated by  19.6  arcsec and with $\Delta$V=4.2.

\bf PW\,And: \rm the rotation period is taken from the literature (Strassmeier \& Rice \cite{Strassmeier06}), no ASAS photometry being 
available. The  $v \sin i$ is from Strassmeier \& Rice (\cite{Strassmeier06}).

{\bf HIP 26401}: is a binary whose components are separated by  3.9 arcsec and with  $\Delta$V=1.1

{\bf HIP 63742}: is an  astrometric (Hipparcos) and spectroscopic (Gunther \& Esposito \cite{Gunther07a}) binary. 
We adopt the rotation period from  Gaidos et al. (2000), no ASAS photometry being available.
The $v \sin i$ is from Zuckerman et al. (\cite{Zuckerman04}).

{\bf HIP 86346}: is a triple system whose brightest component has a close companion at 0.2 arcsec 
(Hortmuth et al. 2007) and a  companion at 19.1 arcsec. We adopt the rotation period from Henry et al. (\cite{Henry95}), 
 no ASAS photometry being available. Both B$-$V and V$-$I colors are taken from Weis (\cite{Weis93}). The
$v \sin i$ is from Zuckerman et al. (\cite{Zuckerman04}).

{\bf HIP 114066}: we adopt the rotation period of  Koen \& Eyer (2002), no ASAS photometry being available. The
$v \sin i$ is from Zuckerman et al. (\cite{Zuckerman04}).

{\bf HIP 16563}: is a binary whose components are separated by  9.5 arcsec and with   $\Delta$V=2.9.  We adopt the rotation period of Messina (\cite{Messina98}), no ASAS photometry being available.

{\bf HIP 12635-12638}:  is a binary whose components are separated by 14.6 arcsec and with $\Delta$V=1.5.

{\bf HIP 110526AB}: is a binary whose components are separated by  1.8 arcsec and with $\Delta$V=0.1.

\section*{Acknowledgements}
This work was supported  by the  Italian Ministero dell'Universit\`a, dell'Istruzione e della Ricerca (MIUR)
and the Istituto Nazionale di Astrofisica (INAF).
The extensive use  of the SIMBAD  and ADS  databases  operated by  the  CDS center,  Strasbourg,
France,  is gratefully  acknowledged. The Authors would like to thank Dr.\,G. Pojma\'nski for the extensive use
we made of the ASAS database. The Authors would like to thank the Referee for helpful comments.

\Online

\begin{table*}
\scriptsize
\caption{\label{twa_lit} TW Hydrae association. Summary data from the literature and mass derived from evolutionary tracks. Note n1=1 indicates that V$-$I is derived from B$-$V; note n2=m indicates a confirmed member. }
\centering
\begin{tabular}{l@{\hspace{.1cm}}l@{\hspace{.3cm}}r@{\hspace{.1cm}}r@{\hspace{.1cm}}r@{\hspace{.3cm}}r@{\hspace{.1cm}}r@{\hspace{.1cm}}l@{\hspace{.1cm}}r@{\hspace{.1cm}}r@{\hspace{.1cm}}r@{\hspace{.1cm}}r@{\hspace{.1cm}}r@{\hspace{.1cm}}c@{\hspace{.1cm}}c@{\hspace{.1cm}}c@{\hspace{.1cm}}l@{\hspace{.1cm}}l@{\hspace{.1cm}}l@{\hspace{.1cm}}l@{\hspace{.1cm}}}
\hline
  name      &     &  \multicolumn{3}{c}{RA } & \multicolumn{3}{c}{DEC} &   \multicolumn{1}{c}{V}      & B$-$V  & V$-$I   &  M$_V$ & \multicolumn{1}{c}{d}      &  \multicolumn{1}{c}{$v\sin i$} & Mass & Radius & Sp.T  &  mgr & n1 & n2\\ 
		&	& \multicolumn{3}{c}{(2000.0)}                    &	 \multicolumn{3}{c}{(2000.0)}                    &  \multicolumn{1}{c}{(mag)}  & (mag)  & (mag) & (mag)   & \multicolumn{1}{c}{(pc)}  &  \multicolumn{1}{c}{(km s$^{-1}$)} & (M$_{\odot}$) & (R$_{\odot}$) & &   &      &                 \\
\hline
TWHya  &  HIP53911  &  11  &  1  &  51.90  &  -34  &  42  &  17.0  &  10.50  &  0.97  &  1.70  &  6.88  &  53.00  &  6.00   &      1.0  &  1.4  & K6Ve  &    TWA  &  --
  &  m  \\
TWA2  &  CD-298887  &  11  &  9  &  13.80  &  -30  &  1  &  40.0  &  11.07  &  1.48  &  2.24  &  8.01  &  41.00  &  12.60    &      0.6  &  1.5 & M2Ve  & TWA  &  --
  &  m  \\
TWA4  &  HIP55505  &  11  &  22  &  5.30  &  -24  &  46  &  40.0  &  9.42  &  1.17  &  1.39  &  6.20  &  44.00  &  8.90     &      1.0  &  1.2 &  K5V  &    TWA  &  --
  &  m  \\
TWA5A  &  CD-337795  &  11  &  31  &  55.30  &  -34  &  36  &  27.0  &  12.12  &  1.47  &  2.36  &  8.85  &  45.00  &  53.50    &      0.4  &  1.1 & M2Ve    &  TWA
&  --  &  m  \\
TWA7  &  GSC7190-2111  &  10  &  42  &  30.10  &  -33  &  40  &  17.0  &  11.65  &  1.46  &  2.44  &  9.41  &  28.00  &  4.40    &      0.3  & 1.0 &  M2Ve   &  TWA  &
  --  &  m  \\
TWA8A  &  CD-268623A  &  11  &  32  &  41.20  &  -26  &  51  &  56.0  &  12.23  &  1.46  &  2.41  &  9.27  &  39.00  &  6.60    &      0.4  &  1.0 & M3Ve   &  TWA  &
  --  &  m  \\
\hline
\end{tabular}

\end{table*}

 \begin{table*}
\scriptsize
\caption{\label{bpic_lit} $\beta$ Pictoris association. Summary data from the literature and mass derived from evolutionary tracks. Note n1=1 indicates that V$-$I is derived from B$-$V;  note n2=m indicates a confirmed member. }
\centering
\begin{tabular}{l@{\hspace{.1cm}}l@{\hspace{.3cm}}r@{\hspace{.1cm}}r@{\hspace{.1cm}}r@{\hspace{.3cm}}r@{\hspace{.1cm}}r@{\hspace{.1cm}}l@{\hspace{.1cm}}r@{\hspace{.1cm}}r@{\hspace{.1cm}}r@{\hspace{.1cm}}r@{\hspace{.1cm}}r@{\hspace{.1cm}}c@{\hspace{.1cm}}c@{\hspace{.1cm}}c@{\hspace{.1cm}}l@{\hspace{.1cm}}l@{\hspace{.1cm}}l@{\hspace{.1cm}}l@{\hspace{.1cm}}}
\hline
  name      &     &  \multicolumn{3}{c}{RA } & \multicolumn{3}{c}{DEC} &   \multicolumn{1}{c}{V}      & B$-$V  & V$-$I   &  M$_V$ & \multicolumn{1}{c}{d}      &  \multicolumn{1}{c}{$v\sin i$} & Mass & Radius & Sp.T  &  mgr & n1 & n2\\ 
		&	& \multicolumn{3}{c}{(2000.0)}                    &	 \multicolumn{3}{c}{(2000.0)}                    &  \multicolumn{1}{c}{(mag)}  & (mag)  & (mag) & (mag)   & \multicolumn{1}{c}{(pc)}  &  \multicolumn{1}{c}{(km s$^{-1}$)} & (M$_{\odot}$) & (R$_{\odot}$) & &   &      &                 \\
\hline
HIP12545  &  BD+05378  &  2  &  41  &  25.90  &  5  &  59  &  18.0  &  10.37  &  1.21  &  1.59  &  7.41  &  39.00  &  40.00 &        0.8  &  0.9 & K6Ve   &  BPIC
  &  --  &  m  \\
HIP23200  &  V1005Ori  &  4  &  59  &  34.83  &  1  &  47  &  0.68  &  10.05  &  1.39  &  1.84  &  8.15  &  24.00  &  14.00 &        0.7  & 0.8 &  M0.5Ve   &
BPIC  &  --  &  m  \\
HIP23309  &  CD-571054  &  5  &  0  &  47.10  &  -57  &  15  &  25.0  &  10.00  &  1.40  &  1.79  &  7.91  &  26.20  &  5.80 &        0.8  & 1.0 &  M0Ve   &  BPIC
&  --  &  m  \\
HIP23418  &  GJ3322  &  5  &  1  &  58.80  &  9  &  59  &  0.00  &  11.45  &  1.54  &  2.46  &  8.79  &  34.00  &  7.67 &        0.4  &  1.2 & M3V  &   BPIC  &
--  &  m  \\
HIP29964  &  HD45081  &  6  &  18  &  28.20  &  -72  &  2  &  41.0  &  9.80  &  1.13  &  1.32  &  6.87  &  38.60  &  16.40 &        0.8  & 0.8 &  K4Ve  &   BPIC
  &  --  &  m  \\
HIP76629  &  HD139084A  &  15  &  38  &  57.50  &  -57  &  42  &  27.0  &  7.97  &  0.81  &  0.95  &  4.98  &  39.70  &  16.60 &         1.0  &  1.2 & K0V  & 
BPIC  &  --  &  m  \\
\hline
\end{tabular}

\end{table*}

\begin{table*}
\scriptsize
\caption{\label{tuc_lit} Tucana/Horologium association. Summary data from the literature and mass derived from evolutionary tracks. Note n1=1 indicates that V$-$I is derived from B$-$V; note n2=m indicates a confirmed member; n2=n indicates a rejected member. }
\centering
\begin{tabular}{l@{\hspace{.1cm}}l@{\hspace{.3cm}}r@{\hspace{.1cm}}r@{\hspace{.1cm}}r@{\hspace{.3cm}}r@{\hspace{.1cm}}r@{\hspace{.1cm}}l@{\hspace{.1cm}}r@{\hspace{.1cm}}r@{\hspace{.1cm}}r@{\hspace{.1cm}}r@{\hspace{.1cm}}r@{\hspace{.1cm}}c@{\hspace{.1cm}}c@{\hspace{.1cm}}c@{\hspace{.1cm}}l@{\hspace{.1cm}}l@{\hspace{.1cm}}l@{\hspace{.1cm}}l@{\hspace{.1cm}}}
\hline
  name      &     &  \multicolumn{3}{c}{RA } & \multicolumn{3}{c}{DEC} &   \multicolumn{1}{c}{V}      & B$-$V  & V$-$I   &  M$_V$ & \multicolumn{1}{c}{d}      &  \multicolumn{1}{c}{$v\sin i$} & Mass & Radius & Sp.T  &  mgr & n1 & n2\\ 
		&	& \multicolumn{3}{c}{(2000.0)}                    &	 \multicolumn{3}{c}{(2000.0)}                    &  \multicolumn{1}{c}{(mag)}  & (mag)  & (mag) & (mag)   & \multicolumn{1}{c}{(pc)}  &  \multicolumn{1}{c}{(km s$^{-1}$)} & (M$_{\odot}$) & (R$_{\odot}$) & &   &      &                 \\
\hline
HIP1113  &  HD987  &  0  &  13  &  53.00  &  -74  &  41  &  18.0  &  8.78  &  0.74  &  0.82  &  5.57  &  43.90  &  7.30 &         1.0  & 1.0 &  G8V    &  TUC  &  --
&  m  \\
HIP1910  &  GSC8841-0065  &  0  &  24  &  9.00  &  -62  &  11  &  4.00  &  11.55  &  1.40  &  1.77  &  8.33  &  44.00  &  20.90 &        0.7  & 0.8 &  M0Ve    &
TUC  &  --  &  m  \\
HIP1993  &  GSC8841-1145  &  0  &  25  &  14.70  &  -61  &  30  &  48.0  &  11.47  &  1.35  &  1.81  &  8.30  &  43.00  &  7.30 &        0.7  & 0.8 &  M0Ve    &  TUC
  &  --  &  m  \\
HIP2729  &  HD3221  &  0  &  34  &  51.20  &  -61  &  54  &  58.0  &  9.61  &  1.05  &  1.10  &  6.30  &  45.90  &  122.80 &        0.9  & 0.8 &  K4Ve  &    TUC
&  --  &  m  \\
TYC935111101  &  CD-7824  &  0  &  42  &  20.30  &  -77  &  47  &  40.0  &  10.21  &  1.06  &  1.22  &  6.72  &  50.00  &  19.50 &        0.8  & 0.8 &  K3Ve    &
TUC  &  --  &  m  \\
HIP6485  &  HD8558  &  1  &  23  &  21.30  &  -57  &  28  &  51.0  &  8.51  &  0.68  &  0.77  &  5.05  &  49.30  &  13.80 &         1.0  & 1.1 &  G7V  & TUC  &
--  &  m  \\
\hline
\end{tabular}

\end{table*}

\begin{table*}
\scriptsize
\caption{\label{col_lit} Columba association. Summary data from the literature and mass derived from evolutionary tracks. Note n1=1 indicates that V$-$I is derived from B$-$V; note n2=m indicates a confirmed member. }
\centering
\begin{tabular}{l@{\hspace{.1cm}}l@{\hspace{.3cm}}r@{\hspace{.1cm}}r@{\hspace{.1cm}}r@{\hspace{.3cm}}r@{\hspace{.1cm}}r@{\hspace{.1cm}}l@{\hspace{.1cm}}r@{\hspace{.1cm}}r@{\hspace{.1cm}}r@{\hspace{.1cm}}r@{\hspace{.1cm}}r@{\hspace{.1cm}}c@{\hspace{.1cm}}c@{\hspace{.1cm}}c@{\hspace{.1cm}}l@{\hspace{.1cm}}l@{\hspace{.1cm}}l@{\hspace{.1cm}}l@{\hspace{.1cm}}}
\hline
  name      &     &  \multicolumn{3}{c}{RA } & \multicolumn{3}{c}{DEC} &   \multicolumn{1}{c}{V}      & B$-$V  & V$-$I   &  M$_V$ & \multicolumn{1}{c}{d}      &  \multicolumn{1}{c}{$v\sin i$} & Mass & Radius & Sp.T  &  mgr & n1 & n2\\ 
		&	& \multicolumn{3}{c}{(2000.0)}                    &	 \multicolumn{3}{c}{(2000.0)}                    &  \multicolumn{1}{c}{(mag)}  & (mag)  & (mag) & (mag)   & \multicolumn{1}{c}{(pc)}  &  \multicolumn{1}{c}{(km s$^{-1}$)} & (M$_{\odot}$) & (R$_{\odot}$) & &   &      &                 \\
\hline
TYC804702321  &  CD-52381  &  1  &  52  &  14.60  &  -52  &  19  &  33.0  &  10.89  &  0.95  &  1.08  &  6.07  &  92.00  &  19.80 &        0.9  & 0.9 &  K2V(e)  &
  COL  &  --  &  m  \\
BD-16351  &  xxx  &  2  &  1  &  35.60  &  -16  &  10  &  0.00  &  10.40  &  1.10  &  1.25  &  5.94  &  78.00  &  999.00 &         1.0  & 1.2 &  K5  &  COL  &  1  &
  m  \\
TYC755806551  &  CD-44753  &  2  &  30  &  32.40  &  -43  &  42  &  23.0  &  10.42  &  1.07  &  1.22  &  6.84  &  52.00  &  7.00 &        0.8  & 0.8 &  K5V(e)   &
COL  &  1  &  m  \\
HIP16413  &  HD21955  &  3  &  31  &  20.80  &  -30  &  30  &  59.0  &  9.93  &  0.64  &  0.73  &  4.50  &  122.00  &  22.30 &         1.0  &1.3 &   G7IV  &   COL  &
  --  &  m  \\
HIP19775  &  HD26980  &  4  &  14  &  22.60  &  -38  &  19  &  2.00  &  9.08  &  0.58  &  0.65  &  4.50  &  82.60  &  13.90 &         1.0  & 1.3 &  G3V  &  COL  &
--  &  m  \\
TYC645727311  &  HD27679  &  4  &  21  &  10.30  &  -24  &  32  &  21.0  &  9.43  &  0.62  &  0.69  &  5.05  &  75.00  &  11.00 &         1.0  & 1.3 &  G2V   &  COL
  &  1  &  m  \\
\hline
\end{tabular}

\end{table*}

\begin{table*}
\scriptsize
\caption{\label{car_lit} Carina association. Summary data from the literature and mass derived from evolutionary tracks. Note n1=1 indicates that V$-$I is derived from B$-$V; note n2=m indicates a confirmed member. }
\centering
\begin{tabular}{l@{\hspace{.1cm}}l@{\hspace{.3cm}}r@{\hspace{.1cm}}r@{\hspace{.1cm}}r@{\hspace{.3cm}}r@{\hspace{.1cm}}r@{\hspace{.1cm}}l@{\hspace{.1cm}}r@{\hspace{.1cm}}r@{\hspace{.1cm}}r@{\hspace{.1cm}}r@{\hspace{.1cm}}r@{\hspace{.1cm}}c@{\hspace{.1cm}}c@{\hspace{.1cm}}c@{\hspace{.1cm}}l@{\hspace{.1cm}}l@{\hspace{.1cm}}l@{\hspace{.1cm}}l@{\hspace{.1cm}}}
\hline
  name      &     &  \multicolumn{3}{c}{RA } & \multicolumn{3}{c}{DEC} &   \multicolumn{1}{c}{V}      & B$-$V  & V$-$I   &  M$_V$ & \multicolumn{1}{c}{d}      &  \multicolumn{1}{c}{$v\sin i$} & Mass & Radius & Sp.T  &  mgr & n1 & n2\\ 
		&	& \multicolumn{3}{c}{(2000.0)}                    &	 \multicolumn{3}{c}{(2000.0)}                    &  \multicolumn{1}{c}{(mag)}  & (mag)  & (mag) & (mag)   & \multicolumn{1}{c}{(pc)}  &  \multicolumn{1}{c}{(km s$^{-1}$)} & (M$_{\odot}$) & (R$_{\odot}$) & &   &      &                 \\
\hline
TYC939003221  &  HD42270  &  5  &  53  &  29.30  &  -81  &  56  &  53.0  &  9.14  &  0.79  &  0.97  &  5.29  &  59.00  &  30.30 &         1.0  & 1.2 &  K0V+VI    &
CAR  &  --  &  m  \\
TYC855910161  &  CD-571709  &  7  &  21  &  23.70  &  -57  &  20  &  37.0  &  10.72  &  0.64  &  0.76  &  5.72  &  100.00  &  11.70 &        0.9  & 1.0 &  K0V+D    &
  CAR  &  1  &  m  \\
TYC892909271  &  CD-63408  &  8  &  24  &  6.00  &  -63  &  34  &  3.00  &  9.87  &  0.63  &  0.75  &  4.78  &  104.00  &  73.00 &         1.0  & 1.2 &  G5V  &   CAR
  &  --  &  m  \\
TYC856917611  &  CP-531875  &  8  &  45  &  52.70  &  -53  &  27  &  28.0  &  10.29  &  0.63  &  0.75  &  4.48  &  145.00  &  45.00 &         1.0  & 1.0 &  G2V   &
CAR  &  1  &  m  \\
TYC939521391  &  CD-75392  &  8  &  50  &  5.40  &  -75  &  54  &  38.0  &  10.59  &  0.76  &  0.82  &  5.57  &  101.00  &  43.80 &         1.0  & 1.0 &  G9V  &  
  CAR  &  --  &  m  \\
TYC856935971  &  CD-532515  &  8  &  51  &  56.40  &  -53  &  55  &  57.0  &  11.06  &  0.69  &  0.79  &  5.31  &  141.00  &  29.00 &         1.0  & 1.0 &  G9V  & 
  CAR  &  --  &  m  \\
\hline
\end{tabular}

\end{table*}

\begin{table*}
\scriptsize
\caption{\label{abdor_lit}AB Doradus association. Summary data from the literature and mass derived from evolutionary tracks. Note n1=1 indicates that V$-$I is derived from B$-$V; note n2=m indicates a confirmed member. }
\centering
\begin{tabular}{l@{\hspace{.1cm}}l@{\hspace{.3cm}}r@{\hspace{.1cm}}r@{\hspace{.1cm}}r@{\hspace{.3cm}}r@{\hspace{.1cm}}r@{\hspace{.1cm}}l@{\hspace{.1cm}}r@{\hspace{.1cm}}r@{\hspace{.1cm}}r@{\hspace{.1cm}}r@{\hspace{.1cm}}r@{\hspace{.1cm}}c@{\hspace{.1cm}}c@{\hspace{.1cm}}c@{\hspace{.1cm}}l@{\hspace{.1cm}}l@{\hspace{.1cm}}l@{\hspace{.1cm}}l@{\hspace{.1cm}}}
\hline
  name      &     &  \multicolumn{3}{c}{RA } & \multicolumn{3}{c}{DEC} &   \multicolumn{1}{c}{V}      & B$-$V  & V$-$I   &  M$_V$ & \multicolumn{1}{c}{d}      &  \multicolumn{1}{c}{$v\sin i$} & Mass & Radius & Sp.T  &  mgr & n1 & n2\\ 
		&	& \multicolumn{3}{c}{(2000.0)}                    &	 \multicolumn{3}{c}{(2000.0)}                    &  \multicolumn{1}{c}{(mag)}  & (mag)  & (mag) & (mag)   & \multicolumn{1}{c}{(pc)}  &  \multicolumn{1}{c}{(km s$^{-1}$)} & (M$_{\odot}$) & (R$_{\odot}$) & &   &      &                 \\
\hline

HIP5191  &  HD6569  &  1  &  6  &  26.15  &  -14  &  17  &  47.1  &  9.50  &  0.91  &  0.95  &  6.01  &  50.00  &  10.00 &        0.9  & 0.9 &  K1V  &    ABDOR  &
--  &  m  \\
TYC804210501  &  CD-46644  &  2  &  10  &  55.40  &  -46  &  3  &  59.0  &  11.24  &  0.91  &  1.16  &  6.53  &  87.40  &  36.00 &        0.9  & 0.8 &   K3IVe    &
ABDOR  &  --  &  m  \\
HIP10272  &  HD13482  &  2  &  12  &  15.41  &  23  &  57  &  29.5  &  7.94  &  1.13  &  0.93  &  5.39  &  32.30  &  6.00 &         1.0  & 1.1 &  K1  &  ABDOR  &
--  &  m  \\
HIP14684  &  ISEri  &  3  &  9  &  42.30  &  -9  &  34  &  47.0  &  8.48  &  0.81  &  0.84  &  5.46  &  40.20  &  7.00 &         1.0  & 1.0 &  G0  &   ABDOR  &  --
&  m  \\
HIP17695  &  EUVEJ0347-01.9  &  3  &  47  &  23.34  &  -1  &  58  &  19.9  &  11.59  &  1.51  &  2.16  &  10.53  &  16.30  &  18.00 &        0.4  & 0.4 &  M3   &
ABDOR  &  --  &  m  \\
HIP22738AB  &  CD-561032B  &  4  &  53  &  30.50  &  -55  &  51  &  32.0  &  11.16  &  1.56  &  2.56  &  10.91  &  11.20  &  999.00 &        0.2  & 0.6 &  M3Ve   &
  ABDOR  &  --  &  m  \\
\hline
\end{tabular}

\end{table*}

\clearpage

\begin{deluxetable}{lrcccccc}
\tablecolumns{8}
\tablecaption{\label{tab_period} Summary of period search.
}
\tablewidth{0pt}
 \tablehead{
\colhead {Target}   & \colhead {ASAS}    & \colhead {JD}       & \colhead {JD}    & \colhead {Period} & \colhead {$\Delta$P}  & \colhead {Normalized       } & \colhead {Power at 99\%}\\   
\colhead {          }   & \colhead {number} & \colhead {initial} & \colhead {final} & \colhead {(d)}      & \colhead {(d)           }  & {power} & {confidence level}  \\
}
\startdata
\multicolumn{8}{c}{}\\
\multicolumn{8}{c}{\bf TW Hydrae}\\
\multicolumn{8}{c}{}\\
\hline
                          TWA2 &                  110914-3001.7 &  1869.8491 &  1971.7092 &   4.845 &   0.107 & 13.06 &  6.07  \\
                          TWA2 &                  110914-3001.7 &  1971.7092 &  2080.4703 &   4.860 &   0.072 &  5.49 &  5.00  \\
                          TWA2 &                  110914-3001.7 &  2577.8675 &  2683.7281 &   4.932 &   0.148 &  6.57 &  5.14  \\
                          TWA2 &                  110914-3001.7 &  2683.7281 &  2784.5825 &   4.845 &   0.107 & 12.63 &  7.49  \\
                          TWA2 &                  110914-3001.7 &  2971.8337 &  3072.8017 &   4.833 &   0.106 &  7.04 &  5.41  \\
\enddata
\end{deluxetable}


\begin{thebibliography}{}

\bibitem[1996]{Alekseev96} Alekseev, I.Y. 1996,  Astron. Rep., 40, 74
\bibitem[2002]{Alencar02} Alencar, S.H.P., \& Bathala, C. 2002, ApJ, 571, 378
\bibitem[1998]{Allain98} Allain, S. 1998, A\&A, 333, 629 
\bibitem[1998]{Baraffe98} Baraffe, I., Chabrier, G., Allard, F., \& Hauschildt, P. 1998, A\&A, 337,403
\bibitem[2003]{Barnes03} Barnes, S., 2003, ApJ, 586, 464
\bibitem[2005]{Barnes05} Barnes, J.R.; Collier Cameron, A., Lister, T.A., Pointer, G.R., Still, M.D. 2005, MNRAS, 356, 1501
\bibitem[1979]{Bessel79} Bessel, M.S., 1979, PASP, 91, 589
\bibitem[2008]{Beuzit08} Beuzit, J.-L., Feldt, M.,  Dohlen, K., et al. 2008, Ground-based and Airborne Instrumentation for Astronomy II. 
Ed. McLean, I. S., \& Casali, M. M. Proceedings of the SPIE, Volume 7014,  701418
\bibitem[2007]{Biller07} Biller, B.A., Close, L.M., Masciadri, E., et al. 2007, ApJS, 173, 143
\bibitem[2008]{Bouvier08}Bouvier, J. 2008, A\&A,489, 53
\bibitem[1997]{burrows97} Burrows, A., Marley, M., Hubbard, W. B et al. 1997, ApJ, 491, 856 
\bibitem[2003]{Chauvin03}  Chauvin, G., Thomson, M., Dumas, C., et al. 2003, A\&A, 404, 157
\bibitem[2005]{Chauvin05} Chauvin G., Lagrange A.-M., Zuckerman B., et al.  2005, A\&A, 438, 29C
\bibitem[2009]{Chauvin09} Chauvin, G., Lagrange, A. -M., Bonavita, M., et al. 2009,, A\&A, 2010, 509, 52
\bibitem[2009]{Cameron09} Collier Cameron, A., Davidson, V.A., Hebb, L., et al. 2009, MNRAS, 
\bibitem[1988]{Cutispoto88} Cutispoto, G., \& Rodon\`o, M. 1988, IBVS, 3232
\bibitem[1998a]{Cutispoto98a} Cutispoto, G. 1998a, A\&AS 127, 207
\bibitem[1998b]{Cutispoto98b} Cutispoto, G. 1998b, A\&AS 131, 321
\bibitem[1999]{Cutispoto99} Cutispoto, G.; Pastori, L.; Tagliaferri, G.; Messina, S.; Pallavicini, R. 1999, A\&AS, 138, 87
\bibitem[2000]{Cutispoto00} Cutispoto, G., Pastori, L., Guerrero, A. et al. 2000 A\&A, 364, 205 
\bibitem[2001]{Cutispoto01} Cutispoto, G., Messina, S., \& Rodon\`o, M. 2001, A\&A, 367, 910
\bibitem[2002]{Cutispoto02} Cutispoto, G., Pastori, L., Pasquini, L. et al. 2002 A\&A, 384, 491
\bibitem[2003]{Cutispoto03} Cutispoto, G., Messina, S., Rodon\`o, M. 2003, A\&A, 400, 659
\bibitem[2009]{daSilva09} da Silva, L., Torres, C.A.O., de la Reza, R. et al. 2009, A\&A,  508, 833
\bibitem[2004]{Delareza04}  De La Reza, R. \& Pinzon, G. 2004, AJ, 128, 1812 
\bibitem[1999]{Delfosse99} Delfosse. X., Forveille, T., Beuzit, J.-L., et al. 1999, A\&A, 344, 897
\bibitem[1995]{favata95} Favata, F., Barbera, M., Micela, G., Sciortino, S. 1995 A\&A 295, 147
\bibitem[2006]{Feigelson06} Feigelson E.D., Lawson W.A., Stark M., Townsley L. \& Garmire G.P. 2006, AJ, 131, 1730 
\bibitem[2000]{Gaidos00} Gaidos, E. J.; Henry, G. W.; Henry, S. M., 2000, AJ, 120, 1006
\bibitem[2007a]{Gunther07a} Guenther, E. W., Esposito, M., Mundt, R., et al. 2007, A\&A, 467,1147
\bibitem[2007b]{Gunther07b} Guenther, E. W. \& Esposito, M. arXiv:0701293
\bibitem[1994]{Jeffries94} Jeffries, R. D., Byrne, P. B., Doyle, J. G., et al.  1994, MNRAS, 270, 153 
\bibitem[2009]{Hartman09} Hartman, J. D., Gaudi, B. S., Pinsonneault, M. H., et al. 2009, ApJ, 691,342
\bibitem[2007]{Hebb07} Hebb, L., Petro, L., Ford, H.C., et al. 2007, MNRAS, 379, 63 
\bibitem[1995]{Henry95} Henry, Gregory W.; Fekel, Francis C.; Hall, Douglas S., 1995, AJ, 110, 2926
\bibitem[1996]{Herbst96} Herbst, W., \& Wittenmyer, R. 1996, BAAS, 28, 1338
\bibitem[2002]{Herbst02} Herbst, W., Bailer-Jones, C. A. L., Mundt, R., Meisenheimer, K., \& Wackermann, R. 2002, A\&A, 396, 513
\bibitem[2005]{Herbst05} Herbst, W., \& Mundt, R. 2005, ApJ, 633, 967
\bibitem[2007]{Herbst07} Herbst, W., Eisl\"offel, J., Mundt, R., \& Scholz, A.2007, Protostars and Planets V, B. Reipurth, D. Jewitt, and K. Keil (eds.), 
University of Arizona Press, Tucson, 951 pp., 2007., p.297-311
\bibitem[2006]{Hodgkin06} Hodgkin, S.T., Irwin, J.M., Aigrain, S., et al. 2006, AN, 327, 9 
\bibitem[2007]{Hormuth07}  Hormuth, F., Brandner, W., Hippler, S., Janson, M., \&  Henning, T., 2007, A\&A, 463, 707
\bibitem[1986]{Horne86} Horne, J.H., \& Baliunas, S.L. 1986, ApJ, 302, 757
\bibitem[2007]{kasper07} Kasper, M., Apai, D., Janson, M., Brandner, W. 2007 A\&A, 472, 321
\bibitem[1988]{Kawaler88} Kawaler, S.D., 1988, ApJ, 333,236
\bibitem[1995]{Keppens95} Keppens, R., MacGregor, K. B., \& Charbonneau, P. 1995, A\&A, 294, 469
\bibitem[2002]{Koen02} Koen, C., Eyer, L, 2002, MNRAS, 331,45
\bibitem[2007]{Konopacky07} Konopacky, Q. M., Ghez, A. M., Duchêne, G., McCabe, C., Macintosh, B. A. 2007, AJ, 133, 2008
\bibitem[1981]{Kovacs81} Kovacs, G. 1981, Ap\&SS, 78, 175
\bibitem[1997]{Krishnamurthi97}Krishnamurthi, A., Pinsonneault, M.H., Barnes, S. \& Sofia, S. 1997, ApJ, 480, 303
\bibitem[2007]{Innis07} Innis, J.; Coates, D. W.; Kaye, T. G.; Borisova, A.; Tsvetkov, M.	, 2007, Peremennye Zvezdy, vol.27, no. 4.
\bibitem[2009]{Irwin09} Irwin, J., Aigrain, S., Bouvier, J., et al. 2009, MNRAS, 392, 1456
\bibitem[2004]{Lamm04} Lamm, M. H., Bailer-Jones, C. A. L., Mundt, R., Herbst, W., \& Scholz, A. 2004, A\&A, 417, 557
\bibitem[2010a]{Lanza10a} Lanza, A.F., Bonomo, A.S., Moutou, C., et al. 2010, A\&A, submitted
\bibitem[2010b]{Lanza10b} Lanza, A.F. 2010, A\&A, in press, 2009arXiv0912.4585L

\bibitem[2005]{Lawson05} Lawson, W. A., \& Crause, L.A., 2005, MNRAS, 357,1399
\bibitem[2009]{Lepine09} Lepine, S., \& Simon, M., 2009, AJ,137, 3632
\bibitem[1999]{Lowrance99} Lowrance, P. J., McCarthy, C., Becklin, E. E.,  1999, ApJ, 512, 69
\bibitem[2005]{Luhman05} Luhman, K.L., Stauffer, J.R., \& Mamajek, E.E., 2005, ApJ, 628, L69
\bibitem[1991]{Macgregor91} MacGregor, K.B., \& Brenner, M. 1991, ApJ, 376, 204
\bibitem[2008]{marois08} Marois, C., Macintosh, B., Barman, T. et al. 2008, Sci 322, 1348
\bibitem[1998]{Messina98} Messina, S., 1998, PhD Thesis, University of Catania
\bibitem[2001]{Messina01} Messina, S., Rodon\'o, M., \& Guinan, E. F. 2001, A\&A, 366,215
\bibitem[2003]{Messina03} Messina, S., Pizzolato, N., Guinan, E. F., \& Rodon\'o, M 2003, A\&A, 410,671
\bibitem[2004]{Messina04} Messina, S., Rodon\'o, M., \& Cutispoto, G. 2004, AN, 325, 660
\bibitem[2007]{Messina07} Messina, S., 2007, Memorie Societ\`a Astron, It., 78, 628
\bibitem[2008]{Messina08} Messina, S., Distefano, E., Parihar, P., et al. 2008, A\&A, 483, 253
\bibitem[2009]{Messina09} Messina, S.,  Parihar, P., Koo, J.-R.  et al. 2009, A\&A, in press
\bibitem[2003]{Neuhauser03} Neuhäuser, R., Guenther, E. W., Alves, J., et al. 2003, AN, 324, 535
\bibitem[2009]{Nielsen09} Nielsen, E.L. \& Close, L.M. 2009, arXiv:0909.4531
\bibitem[2004]{Nordstrom04} Nordstrom, B., Mayor, M., Andersen, J. et al. 2004, A\&A 418, 989 
\bibitem[2007]{Norton07} Norton, A. J.; Wheatley, P. J.; West, R. G., et al. 2007, A\&A, 467, 785
\bibitem[1994]{Odell94} O'Dell, M. A., Hendry, M. A., \&Collier Cameron, A. 1994, MNRAS, 268,  181
\bibitem[2007]{Ortega07} Ortega, V.G., Jilinski, E., de la Reza, R., \& Bazzanella, B. 2007, MNRAS, 377, 441
\bibitem[2009]{Parihar09} Parihar, P., Messina, S., Distefano, E.,  Shantikumar N., S., \& Medhi, B.J. , 2009, MNRAS, 400, 603
\bibitem[1991]{Pasquini91} Pasquini, L., Cutispoto, G., Gratton, R., Mayor M. 1991, A\&A 248, 72, 80
\bibitem[1997]{Perryman97} Perryman, M. A. C., Lindegren, L., Kovalevsky, J., et al. 1997, A\&A, 323, 49
\bibitem[1997]{Pojmanski97} Pojmanski G., 1997, Acta Astronomica, 47, 467
\bibitem[1998]{Pojmanski98} Pojmanski G., 1998, Acta Astronomica, 48, 35
\bibitem[2002]{Pojmanski02} Pojmanski G., 2002, Acta Astronomica, 52, 397
\bibitem[2003]{Pojmanski03} Pojmanski G., 2003, Acta Astronomica, 53, 341
\bibitem[2009]{Pont09} Pont,F. 2009, MNRAS, 396, 1789
\bibitem[1992]{Press92} Press, W.H., Teukolsky, S.A., Vetterling, W.T., \& Flannery, B.P., 1992, Numerical Recipes, Cambridge University
\bibitem[1991]{Prosser91} Prosser, C.F., Stauffer, J., \& Kraft R.P. 1991. AJ, 101, 1361 
\bibitem[2000]{Quast00} Quast, G. R., Torres, C,A. O., de La Reza, R. et al. 2000, IAU Symp. 200 'The Formation of Binary Stars'
Ed. Bo Reipurth and Hans Zinnecker,  28.
\bibitem[1987] {Radick87}Radick, R.R., Thompson, D.T., Lockwood, G.W., Duncan, D.K., \& Bagget, W.E., 1987, ApJ, 321, 459
\bibitem[2002]{Rebull02} Rebull, L.M., Madikon,R.B., Strom,S.E., et al. 2002, AJ, 123, 1528 
\bibitem[2004]{Rebull04} Rebull, L.M., Wolff, S.C., \& Strom, S.E. 2004, AJ, 127, 1029
\bibitem[2003]{Reid03} Reid, N. 2003 MNRAS 342, 837
\bibitem[2001]{Rhode01} Rhode,K.L., Herbst, W, \& Mathieu, R.D., 2001, AJ, 122, 3258
\bibitem[1986]{Rodono86} Rodon\`o, M., Cutispoto, G., Pazzani, V., et al. 1986, A\&A, 165, 135
\bibitem[1983]{Rucinski83} Rucinski, S. M., \&  Krautter, J., 1983, A\&A, 121, 217
\bibitem[2008]{Rucinski08} Rucinski, S. M., Slavek, M., Matthews, J.M., et al. 2008, MNRAS, 391, 1913
\bibitem[1982]{Scargle82} Scargle, J.D., 1982, ApJ, 263, 835
\bibitem[1989]{Schwarnemberg89} Schwarnemberg-Czerny, A. 1989, MNRAS, 241, 153
\bibitem[2007]{Scholz07} Scholz, A., Coffey, J., Brandeker, A., \& Jayawardhana, R. 2007, ApJ, 662,1254
\bibitem[2009]{Scholz09} Scholz, A., Eisloffel, J., \& Mundt, R., 2009, MNRAS.tmp 1474
\bibitem[2008]{setiawan08} Setiawan, J., Weise, P., Henning, Th., et al. 2008, in Precision Spectroscopy in Astrophysics, Edited by N.C. Santos, L. Pasquini, A.C.M. Correia, and M. Romaniello, p. 201 (arxiv 0704.2145)
\bibitem[2008]{skelly08} Skelly M.B., Unruh, Y.C., Collier Cameron, A. et al. 2008, MNRAS 385, 708 
\bibitem[1999]{Stassun99} Stassun, K.G., Mathieu, R.D., Mazeh, T., \& Vrba, F. 1999, AJ, 117, 2941
\bibitem[1982a]{Stauffer82a} Stauffer, J.R. 1982a, AJ 87, 899 
\bibitem[1982b]{Stauffer82b} Stauffer, J.R. 1982b AJ 87, 1507 
\bibitem[1984]{Stauffer84} Stauffer, J.R. 1984, ApJ, 280, 189 
    Optical and infrared photometry of late type stars in the Pleiades 
    1984ApJ...280..189S 
\bibitem[1985]{Stauffer85} Stauffer, J.R., Hartmann, L.W., Burnham, J.N., \& Jones B.F. 1985, ApJ, 289, 247
\bibitem[1989]{Stauffer89} Stauffer, J.R., Hartmann, L.W., \& Jones B.F. 1989, ApJ, 346, 160
\bibitem[1997]{Strassmeier97} Strassmeier K.G., Bartus J., Cutispoto G. \& Rodon\`o M., 1997, A\&AS, 125, 11 
\bibitem[2000]{Strassmeier00} Strassmeier, K.G.  \& Rice J.B  2000 A\&A 360, 1019 
\bibitem[2006]{Strassmeier06} Strassmeier, K. G.; Rice, J. B., 2006, A\&A, 460, 715
\bibitem[1994]{Tagliaferri94} Tagliaferri,, G., Cutispoto, G., Pallavicini, R., Randich, S., Pasquini, L. 1994, A\&A 285, 272 
\bibitem[2009]{Teixeira09} Teixeira, R., Ducourant, C., Chauvin, G. et al. 2009, A\&A 503, 281 
\bibitem[1995]{Torres95} Torres, G., Stefanik, R. P.. Latham, D.W.. \& Mazeh, T. 1995, ApJ, 452, 870
\bibitem[2006]{Torres06} Torres, C.A.O., Quast, G.R., da Silva, L. et al. 2006, A\&A, 460, 695
\bibitem[2008]{Torres08} Torres, C.A.O., Quast, G.R., Melo, C.H.F., Sterzik, M.F. 2008, 
  Handbook of Star Forming Regions, Volume II: The Southern Sky ASP Monograph 
  Publications, Vol. 5. Edited by Bo Reipurth, p.757  (arXiv:0808.3362)
\bibitem[1998]{Urban98} Urban, S. E., Corbin, T. E., \& Wycoff, G. L. 1998, AJ, 115, 2161
\bibitem[2009]{Viana09} Viana Almeida, P., Santos, N.C., Melo, C. et al. 2009, A\&A, 501, 965
\bibitem[2005]{vonbraun05} von Braun, Lee, B.L., Seager, S., et al. 2005, PASP, 117, 141
\bibitem[1993]{Weis93} Weis, E.W. 1993, AJ, 195, 1962
\bibitem[2003]{Wichmann03} Wichmann, R., Schmitt, J.H.M.M., Hubrig, S. 2003, A\&A 399, 983
\bibitem[2004]{Zuckerman04} Zuckerman, B.\&  Song, I. 2004, Ann. Rev. Astron. Astr. 42, 685 
\bibitem[2004]{Zuckerman04b} Zuckerman, B., Song, I., Bessell, M.S 2004, ApJ 613, L65
\bibitem[2006]{Zuckerman06} Zuckerman, B., Bessell, M.S., Song, I., Kim, S. 2006, ApJ 649, L11

\end{thebibliography}
\end{document}